\documentclass[review]{elsarticle}
\usepackage{lineno,hyperref}
\usepackage{float}

\usepackage{tikz}
\usetikzlibrary{patterns,arrows}
\usepackage{amsmath}
\usepackage{amssymb}
\usepackage{amsthm}

\usepackage{exscale}
\usepackage{mathrsfs}
\usepackage[mathscr]{eucal}
\usepackage{bm}
\usepackage{eqlist} 
\usepackage{subfig}
\usepackage[margin=2.5cm]{geometry}
\usepackage{verbatim}
\usepackage{tikz} 
\usetikzlibrary{arrows.meta}
\usetikzlibrary{calc}
\usepackage[nameinlink, capitalise]{cleveref}	
\usepackage{pdflscape}	
\usepackage{todonotes}	

\newcommand{\bea}{ \begin{eqnarray*}}
\newcommand{\eea}{\end{eqnarray*} }
\newcommand{\beq}{ \begin{center}\begin{equation*}}
\newcommand{\eeq}{\end{equation*} \end{center}}
\newcommand{\be}{ \begin{equation}}
\newcommand{\ee}{\end{equation}}

\newcounter{exnum}

\journal{Acta Astronautica}

\biboptions{numbers,sort&compress}
\hypersetup{pdfauthor={Gabriel Surina}}

\usepackage{xcolor}
\newcommand*{\missingreference}{\colorbox{red}{?reference?}}
\newcommand*{\missingcitation}{\colorbox{red}{?citation?}}
\makeatletter
\def\@setref#1#2#3{%
  \ifx#1\relax
  \protect\G@refundefinedtrue
  \nfss@text{\reset@font\missingreference}%
  \@latex@warning{Reference `#3' on page \thepage \space
       undefined}%
  \else
  \expandafter#2#1\null
  \fi}
\def\@citex[#1]#2{\leavevmode
  \let\@citea\@empty
  \@cite{\@for\@citeb:=#2\do
   {\@citea\def\@citea{,\penalty\@m\ }%
   \edef\@citeb{\expandafter\@firstofone\@citeb\@empty}%
   \if@filesw\immediate\write\@auxout{\string\citation{\@citeb}}\fi
   \@ifundefined{b@\@citeb}{\hbox{\reset@font\missingcitation}%
    \G@refundefinedtrue
    \@latex@warning
     {Citation `\@citeb' on page \thepage \space undefined}}%
    {\@cite@ofmt{\csname b@\@citeb\endcsname}}}}{#1}}
\makeatother
\usepackage{xcolor}
\usepackage{soul}

\newlength{\tempheight}
\newlength{\tempwidth}

\newcommand{\rowname}[1]
{{\makebox[\tempheight][c]{\textbf{#1}}}}

\newcommand{\columnname}[1]
{\makebox[\tempwidth][c]{\textbf{#1}}}

\usepackage{amsmath,amssymb,enumitem,mathtools}

\usepackage[most]{tcolorbox}
\newtcolorbox{highlighted}{colback=yellow,coltext=black,breakable}
\usepackage{lipsum}
\begin{document}

\begin{titlepage}
   \begin{center}
       \vspace*{1cm}
        \large
       \textbf{Measurement of Hybrid Rocket Solid Fuel Regression Rate for a Slab Burner using Deep Learning}
        \normalsize
       \vspace{1.0cm}

       Gabriel Surina III$^a$, Georgios Georgalis$^b$, Siddhant S. Aphale$^a$, Abani Patra$^b$, Paul E. DesJardin$^{a,*}$ \\
    
    
       $^a$Department of Mechanical and Aerospace Engineering, University at Buffalo,\\ the State University of New York, Buffalo, NY $14260$-$4400$, USA \\
       
       
       $^b$Data Intensive Studies Center, Tufts University, Medford, MA $02155$, USA \\
       
        
        $^{*}$Corresponding author, \textit{Email address:} ped3@buffalo.edu
        
        \vspace{1.5cm}
    \end{center}    
        
        
        
        

\end{titlepage}

\begin{frontmatter}

\title{Measurement of Hybrid Rocket Solid Fuel Regression Rate for a Slab Burner using Deep Learning}

\author[add1]{Gabriel Surina III}
\author[add2]{Georgios Georgalis}
\author[add1]{Siddhant S. Aphale}
\author[add2]{Abani Patra}
\author[add1]{Paul E. DesJardin\corref{mycorrespondingauthor}}

\cortext[mycorrespondingauthor]{Corresponding author}
\ead{ped3@buffalo.edu}
	
\address[add1]{Department of Mechanical and Aerospace Engineering, 
University at Buffalo, the State University of New York, Buffalo, 
NY 14260-4400, USA \\
}
\address[add2]{Data Intensive Studies Center, Tufts University, Medford, MA 02155, USA \\
 }

\begin{abstract}
This study presents an imaging-based deep learning tool to measure the fuel regression rate in a $2$D slab burner experiment for hybrid rocket fuels. The slab burner experiment is designed to verify mechanistic models of reacting boundary layer combustion in hybrid rockets by the measurement of fuel regression rates. A DSLR camera with a high intensity flash is used to capture images throughout the burn and the images are then used to find the fuel boundary to calculate the regression rate. A U-net convolutional neural network architecture is explored to segment the fuel from the experimental images. A Monte-Carlo Dropout process is used to quantify the regression rate uncertainty produced from the network. The U-net computed regression rates are compared with values from other techniques from literature and show error less than $10$\%. An oxidizer flux dependency study is performed and shows the U-net predictions of regression rates are accurate and independent of the oxidizer flux, when the images in the training set are not over-saturated. Training with monochrome images is explored and is not successful at predicting the fuel regression rate from images with high noise. The network is superior at filtering out noise introduced by soot, pitting, and wax deposition on the chamber glass as well as the flame when compared to traditional image processing techniques, such as threshold binary conversion and spatial filtering. U-net consistently provides low error image segmentations to allow accurate computation of the regression rate of the fuel. 
\end{abstract}


\begin{keyword}
Hybrid fuel, regression rate measurement, convolutional neural network, Monte-Carlo Dropout, model uncertainty
\end{keyword}

\end{frontmatter}

\section{Introduction}
Hybrid rockets are recently gaining attention and becoming a competitive propulsion system because they have the benefits of solid and liquid rockets in one platform \cite{Jens2016,Gallo2021}. The fuel and oxidizer are in two different states: typically the fuel is a solid and the oxidizer is a gas/liquid, providing a highly dense fuel source similar to a bi-propellant solid motor but with the throttle control of a liquid motor \cite{Chiaverini2000, Karabeyoglu98, BAKSD20}. The fuel regression rate ($\dot{r}$) corresponds to the rate that the fuel recedes over the course of a burn and is an important measurement since it directly defines the motor thrust and geometry (e.g., a higher regression rate results in smaller combustion chambers) \cite{Zilliac2006}. Low fuel regression rates correspond to low fuel mass flow rates, leading to excess oxidizer passing through the motor, which reduces the available thrust and therefore reduces the overall efficiency \cite{Hirata2011}. From a prediction perspective, the regression rate is a simple measurement that can be used to verify model estimates.

The simplest method to measure temporally and spatially averaged regression rate is to measure the fuel grain mass before and after the burn, then divide by the time of the burn, as performed in \cite{DGEDJ18, Cai2013, Knuth2002}. Measuring the regression rate in this way may be subject to error caused by part of the fuel mass not combusting and melting away (e.g., for solid fuels that have a low melting point such as paraffin wax). Another drawback to weighing the fuel grain is that it does not provide data of local and instantaneous regression rates which is useful for validating regression rate models \cite{Zilliac2006}. Other methods of regression rate measurement include using combustion chamber pressure \cite{Kumar2014}, x-ray radiography \cite{Chiaverini2000}, initial and final port diameter \cite{Karabeyoglu2003, Zilliac2006, Shin2005}, and ultrasonic transducers \cite{Korting1987, Carmicino2006, Chiaverini2001}. More recently, \cite{BAKSD20} developed an imaging-based regression rate measurement for a $2$D slab burner experiment. They collected images of the fuel profiles captured continuously throughout the burn and manually traced the fuel boundary to segment the fuel from the background (binary masks where a value of ``$0$" corresponds to background and a value of ``$1$" corresponds to fuel). The regression rate was then computed by measuring the fuel height variation over the burn time.

Even if manually segmenting the fuel and noise from experimental images is accurate, it is certainly not the most efficient and repeatable process for a large number of images and experimental dataset. The task of solid fuel image segmentation for measuring the fuel regression rate can benefit from advances in deep learning processes. Image segmentation refers to the process of classifying objects and identifying their location in an image. Modern image processing tools based on deep learning, such as convolutional neural networks (CNNs) have been used to segment images for autonomous driving \cite{Ghosh2019} and medical applications \cite{Chowdhary2020}. CNNs are primarily made up of convolutional layers, intermediate activation function (or transfer function) layers, and pooling layers. Convolutional layers apply kernels (or filters) of weights that convolve with the image creating a feature map to extract important learning features from the images. The activation function layers apply a non-linear function, a Rectified Linear Unit (ReLU), that determines the output of previous convolution nodes in the neural network and whether they will be activated. The pooling layers reduce the size of the feature map and also help retain the useful information for the segmentation task \cite{MBPPKT20}. The pooled feature maps result in smaller number of parameters in the network and reduced computational expense. Training of a CNN refers to the process of computing the values of the kernel weights. After initialization, the weights are updated using the gradient descent optimization method after each pass through the training data (or epoch) \cite{brainsci}. 

U-net is a widely used CNN architecture that was initially created for biomedical image segmentation with a low number of training data \cite{RFB15}. Examples of U-net based architecture applications include infant brain segmentation \cite{QJZAU20}, nuclei segmentation for cancer grading and cell detection / classification \cite{ZXZL19}, chronic stroke lesion segmentation \cite{ZHDXW19}, liver and tumor segmentation \cite{LCDFH18}, retinal vessel segmentation for discover and treatment of ocular diseases \cite{XLLL18}, and even cloud and shadow segmentation \cite{JHHT20}. Although, U-net has mostly been applied in the medical field, more recent studies have demonstrated applications specifically in combustion, such as aluminum droplet combustion detection in solid propellant from shadowgraph images \cite{DNCBP19}, flame segmentation to locate the hottest region for fire suppression \cite{BK21}, and segment coolant film and hot exhaust gases in a rocket combustion chamber \cite{MZHTH20}. Neural networks in general have also been recently utilized in the combustion field, including maximizing power density in proton exchange membrane fuel cells \cite{WXXJ20}, creating a fuel consumption model to predict the most efficient route for truck transportation \cite{PPN17}, and to classify solid fuels based on ash content, volatile matter, and fixed carbon \cite{EBYM20}. Similar machine learning processes have been used in hybrid rockets to recognize the different burning phases of solid fuel in a similar $2$D slab burner experiment \cite{RPK20}. The most relevant work to the goal of this study is the particle shape and size regression detection, an example, BubCNN, presented by \cite{HSEP20}, where neural networks were used to detect the shape and size of bubbles in a gas-liquid multiphase flow while segmenting them from the background. Another example would be \cite{LSH21} where the U-net was used to segment bubble images and classify particle shapes and sizes. The work from \cite{HSEP20} and \cite{LSH21} however do not have noise from soot, pitting, wax, and flame interfering with the imaging. 

The overall research focus of this study is on automating the manual regression rate measurement technique in \cite{BAKSD20} using U-net to produce the fuel segmentation masks instead of manually tracing them. The manual approach is used as the ground truth and serves as a baseline to validate the U-net results. The goal of this paper is to evaluate whether the U-net architecture is a viable and self-consistent way to produce segmented fuel masks to measure the fuel regression rate. To accomplish this goal, a series of U-net models of similar architecture are built. Each model is trained on all or a subset of the data collected in \cite{BAKSD20}, a $2$D slab burner experiment with four different oxidizer fluxes ($ $5.91$,~9.58,~18.59,~22.19~[kg/m^2-s]$). The data included, in total, $150$ RGB images and their corresponding manually segmented fuel masks. The U-net models are then used to predict the segmented fuel masks from the input images together with estimations of model uncertainty. The fuel masks and uncertainty information are then used to measure the fuel regression rate and are compared with values reported from the manual process. Lastly, the network robustness and limitations are investigated through an oxidizer flux independency and monochromatic data study. 

The rest of this paper is organized as follows. Section \ref{sec:exset} presents the experimental setup. Section \ref{sec:meth} describes various methods to produce the fuel segmentation masks including the U-net architecture. Results and associated discussion are presented in Section \ref{sec:results}. Conclusions of this study are in Section \ref{sec:con}.

\section{Experimental Setup}
\label{sec:exset}

\begin{figure}[H]
	\centering
	\subfloat[\label{fig:setup}]{%
	  \includegraphics[width=0.7\textwidth]{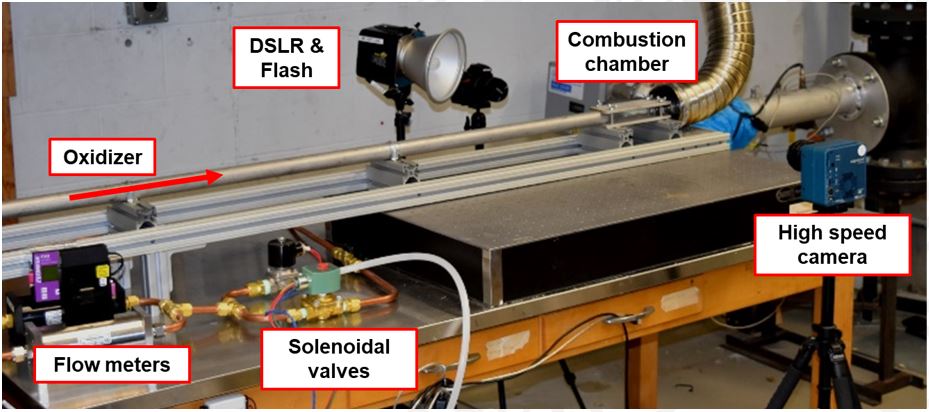}
	}\\
	\subfloat[\label{fig:cc}]{%
		\includegraphics[width=0.7\textwidth]{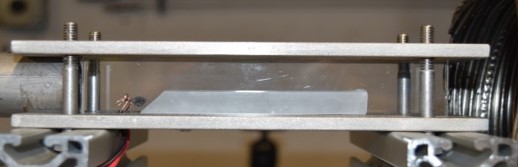}
	}
	\caption{The $2$D slab burner \protect\subref{fig:setup} overall experimental setup, and \protect\subref{fig:cc} side view of the combustion chamber showing paraffin wax fuel \cite{BAKSD20}.}
	\label{fig:Figure1}
\end{figure}

\Cref{fig:setup} shows the $2$D slab burner experiment for paraffin wax-gaseous oxygen hybrid combustion. The experiments are conducted in the setup that is based on \cite{BAKSD20} and \cite{DGEDJ18}. The experiment uses solid paraffin wax (or candle wax) as the fuel because of its low cost and high regression rate characteristics. The paraffin used is laboratory grade manufactured by the Carolina Biological Supply Company. The oxidizer is controlled using a solenoid valve and monitored using an Omega FMA $1744$a mass flow meter. The oxidizer flows through a $2.54~cm$ inner diameter stainless steel pipe with a length of $1.83~m$ to provide a fully developed flow \cite{DGEDJ18}. The oxidizer enters the combustion chamber shown in Fig. \ref{fig:cc} where the fuel sample is placed between two $3~mm$ thick borosilicate glasses for optical access with stainless steel plates on top and bottom. 

A CMOS based Nikon D$5600$ DSLR camera focused on the entire chamber is setup approximately $500~mm$ away, perpendicular to the chamber. A digital stopwatch is setup in the frame of the image to obtain image time to a millisecond accuracy. The fuel is ignited with ethylene vinyl acetate (EVA) and steel wool attached to a nichrome wire supplied with $12~V$, as performed by \cite{DGEDJ18}. The experimental time ranges from $5$-$10$ $s$ depending on the incoming oxidizer flux. Approximately $40$ continuous images are taken for each case during this time. The amount of images are limited by the flash's ability to recharge.

\begin{figure}[H]
  \centering
  \includegraphics[width=0.6\textwidth]{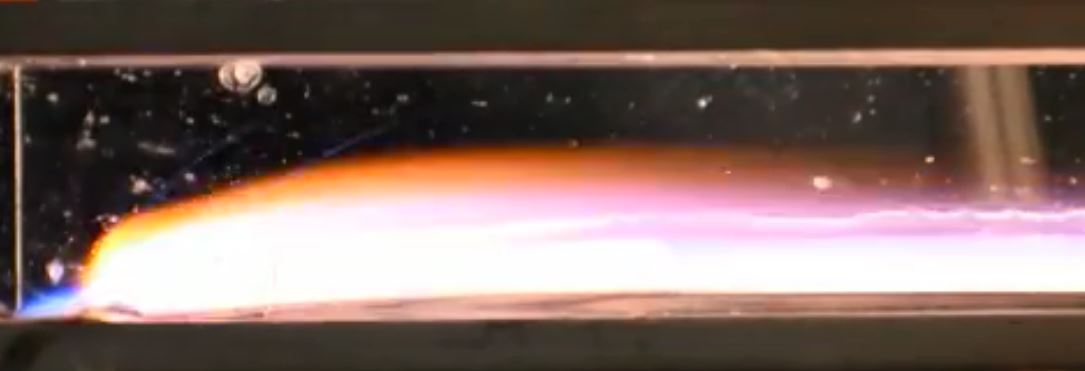}
  \caption{An example of the side image of the wax without using a high intensity flash.}
  \label{fig:flameEx}
\end{figure}

\begin{figure}[H]
	\centering
	\subfloat[]{%
		\includegraphics[width=0.8\textwidth]{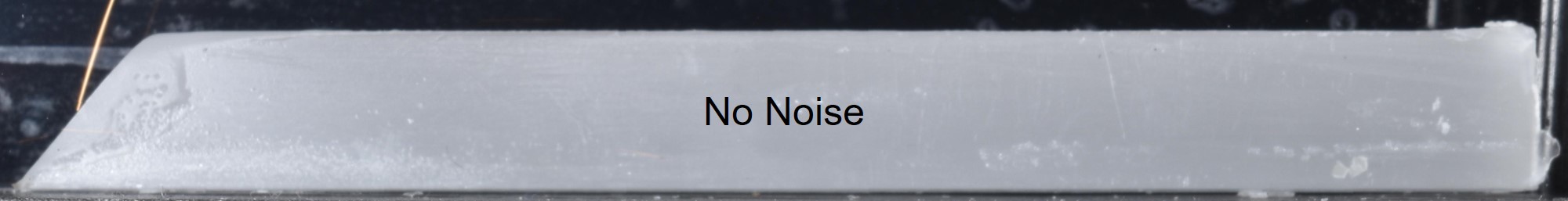}}\\
		\subfloat[\label{fig:noisec}]{%
		\includegraphics[width=0.8\textwidth]{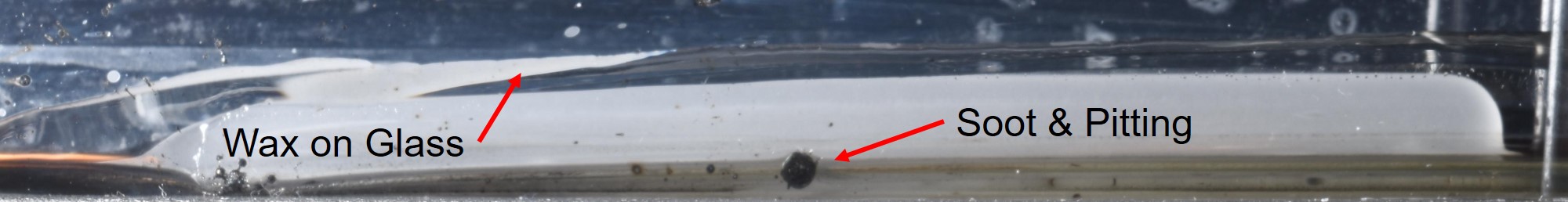}}\\
		\subfloat[\label{fig:noised}]{%
		\includegraphics[width=0.8\textwidth]{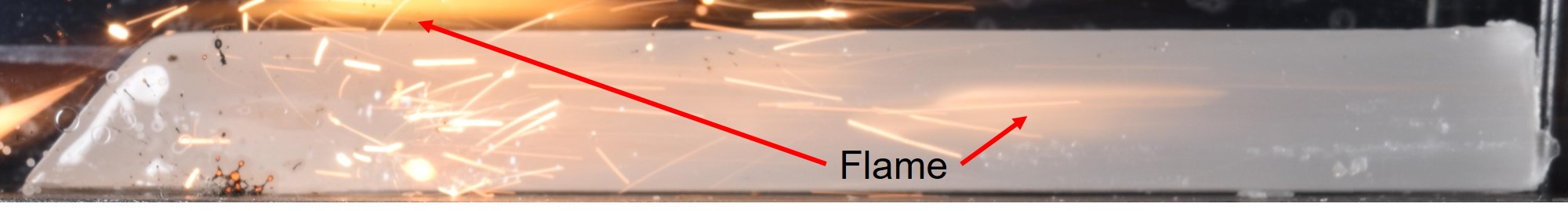}}\\
		\subfloat[\label{fig:noiseb}]{%
		\includegraphics[width=0.8\textwidth]{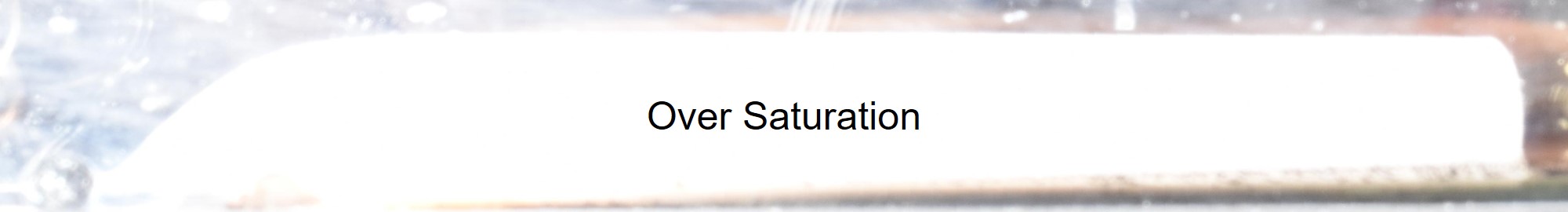}}
	\caption{Examples of side wax profile images using a high intensify flash from \cite{BAKSD20} with (a) no noise and noise from (b) wax on the glass, soot, pitting, (c) flame, and (d) over saturation pointed out.}
	\label{fig:noise}
\end{figure}

The slab burner experiments are conducted at four different oxidizer fluxes oxidizer fluxes ($5.91,$ $9.58$, $18.59$, $22.19$ $[kg/m^2-s]$). As shown in \cref{fig:flameEx}, the wax profile is blocked by flame that brightens with increasing oxidizer flux. To enable viewing the fuel through the flame, a high intensity flash is used to saturate the flame and illuminate the fuel surface, however a flame ghost image remains. \Cref{fig:noise} shows a series of images captured with the high intensity flash. Although the flash allows for imaging of the regressing fuel, it contains noise. With increasing oxidizer flux, the flame brightens and can exceed the intensity of flash, leading to flame traces in the images (\cref{fig:noisec}). The flash intensity can be increased but it results in over-saturation (\cref{fig:noised}). Apart from these issues, the optical windows are also prone to noise introduced by molten wax splashing on the glasses and soot pitting (\cref{fig:noiseb}). These issues, especially flame interference and wax on the glass make it difficult to automate the accurate tracing of the wax profile resulting error in regression rate estimates. Thus, a neural network enables automating the tracing of the fuel profile in the noisy images. 

\section{Methods of Image Processing}
\label{sec:meth}
\begin{figure}[H]
	\centering
	\subfloat[]{%
		\includegraphics[width=0.48\textwidth]{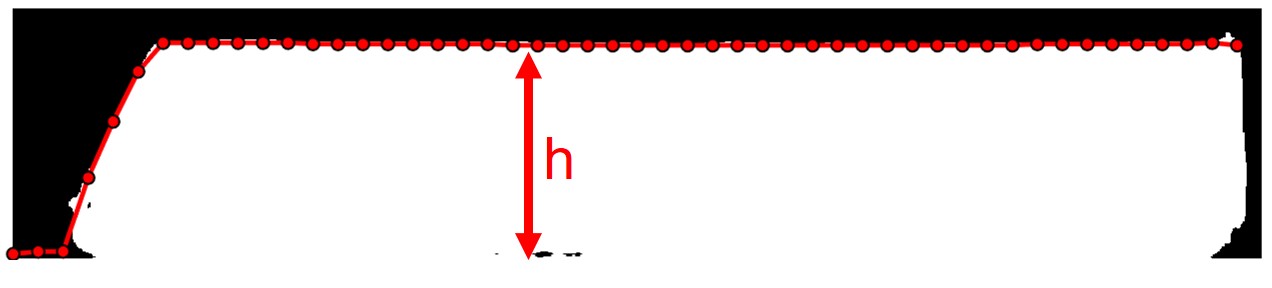}}
	\subfloat[]{%
		\includegraphics[width=0.5\textwidth]{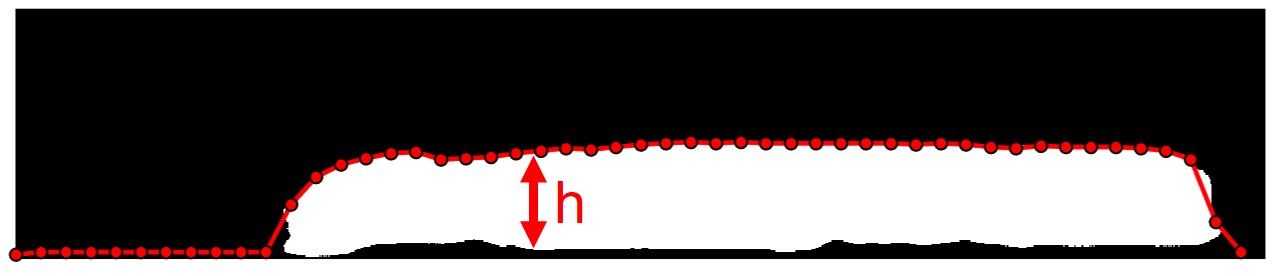}}
	\caption{An example of the masks of the regressing fuel with the profile height tracking.}
	\label{fig:height}
\end{figure}

The regression rates are calculated by tracking the height of the wax over time in successive images (from the fuel masks), fitting a cubic polynomial curve, and finally taking its derivative, consistent with \cite{BAKSD20}. The binary images are processed by starting from the top of the image and stepping down each pixel vertically while searching for the profile height. The profile is constrained to only regress or remain constant, i.e. the height of the wax can only decrease or remain constant. If the height increases in time, that data is neglected. Figure \ref{fig:height} shows representative binary masks of the regressing fuel with the profile tracking in red lines with circles, the profile height (h) for each column of pixels is used in the regression rate estimation. The accurate segmented binary masks are essential to estimate accurate regression rates. Four image processing methods are investigated and compared for the accuracy in estimating the regression rates. The methods of the image processing techniques include using (i) threshold, (ii) Threshold Last Image Subtraction (TLIS), (iii) spatial filtering, and (iv) U-net. Thresholding represents the least complex image processing method and has little to no ability to filter out noise. TLIS is a method that removes all of the noise but removes significant amounts of data in doing so. The spatial filtering method removes all of the small clusters of noise but amplifies large noise clusters. The U-net model artificially chooses the best filters to apply, including image processing techniques that are not readily available, providing the best results, if trained well enough. The resulting segmentation masks from each technique are compared with each other.

\subsection{Threshold}
\label{sec:T}
\begin{figure}[H]
  \centering
  \subfloat[]{%
		\includegraphics[width=0.7\textwidth]{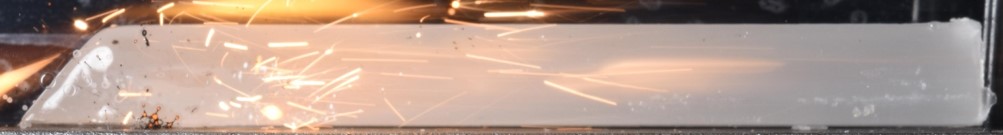}}\\
	\subfloat[]{%
		\includegraphics[width=0.7\textwidth]{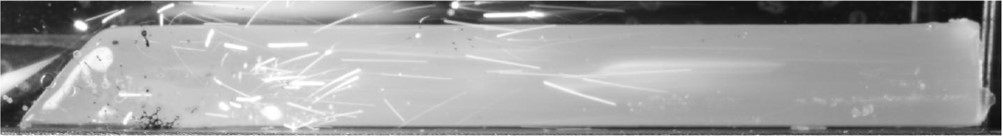}}\\
	\subfloat[]{%
		\includegraphics[width=0.7\textwidth]{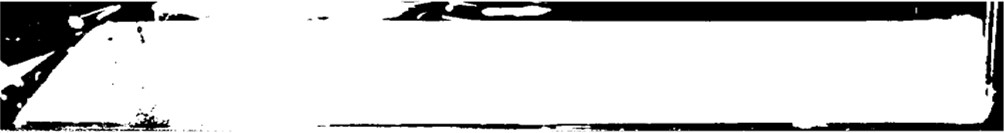}}
  \caption{An example of the process of the thresholding method starting from top to bottom, (a) original, (b) grayscale, and (c) segmented images.}
  \label{fig:Thres}
\end{figure}

One simple and straightforward way to convert the experimental images into a binary segmentation mask is by using a pixel intensity threshold. Performed in \cite{M} with the image processing toolbox, a threshold is chosen by using Otsu's method that uses discriminant analysis to pick the optimal threshold for the best separation of objects from their background \cite{Otsu1979}. The resulting images are converted to grayscale and each pixel falls below or above the threshold, converting its pixel value to a ``$0$" or ``$1$", respectively. The grayscale images are converted by using the weighted sum formula $0.2989 R + 0.5870 G + 0.1140 B$  \cite{RGB2011}, where $R$, $G$, and $B$ are the corresponding red, green, and blue channels, respectively (MATLAB's standard method to convert to grayscale). Pixels that have a value of ``$0$" correspond to background, and pixels that have a value of ``$1$" correspond to the fuel area. An example of this process is shown in \cref{fig:Thres}. The thresholding method is very easy to quickly apply and provides good results if there is little to no noise. However, if there is significant noise (such as the data used in this paper), this method will classify it as fuel and cause high error regression rate estimates.

\subsection{Threshold Last Image Subtraction (TLIS)}
\label{sec:TLIS}

\begin{figure}[H]
  \centering
    \subfloat[]{%
		\includegraphics[width=0.7\textwidth]{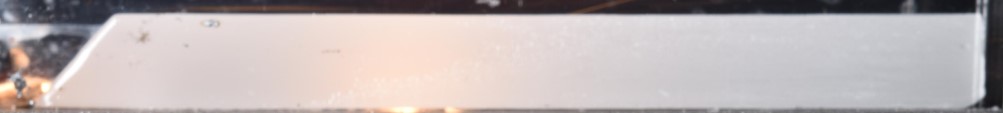}}\\
		\subfloat[]{%
		\includegraphics[width=0.7\textwidth]{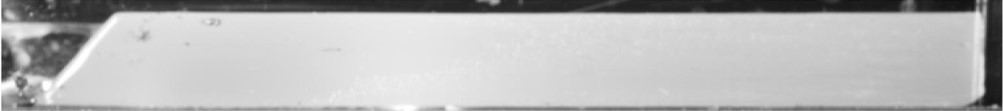}}\\
		\subfloat[]{%
		\includegraphics[width=0.7\textwidth]{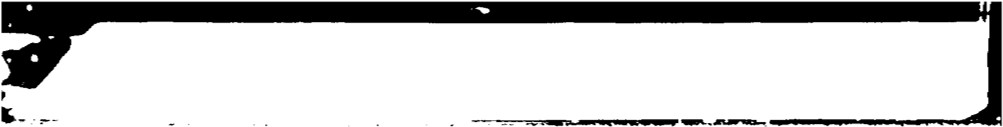}}\\
		\subfloat[]{%
		\includegraphics[width=0.7\textwidth]{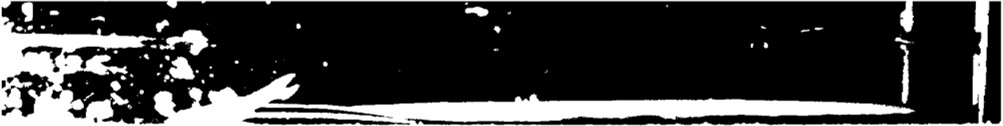}}\\
		\subfloat[]{%
		\includegraphics[width=0.7\textwidth]{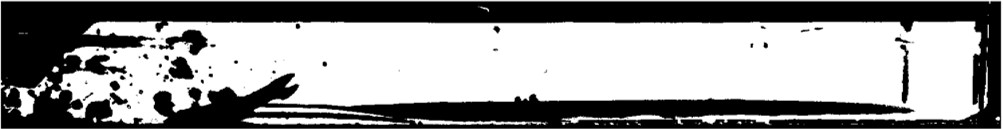}}\\
  \caption{An example of the process of the TLIS method starting from top to bottom, (a) original, (b) grayscale, (c) original segmented, (d) last segmented, and (e) original segmented minus the last segmented images.}
  \label{fig:TLIS}
\end{figure}

The TLIS technique reduces the noise that affects the mask data. The last image in the series of images is assumed to have the maximum noise originating from pitting/soot/wax on the glass. Thus, in TLIS this last image is subtracted from the rest of the images' binary masks. TLIS is helpful because it removes the locations with high amounts of noise, to minimize the errors in the calculation of regression rate. An example of this process is shown in \cref{fig:TLIS}. TLIS is very simple and relatively easy to apply and does not classify any noise as fuel. However, this method reduces the amount of data significantly and causes error in the regression rate estimates.

\subsection{Spatial Filtering}
\begin{figure}[H]
  \centering
  \subfloat[]{%
		\includegraphics[width=0.7\textwidth]{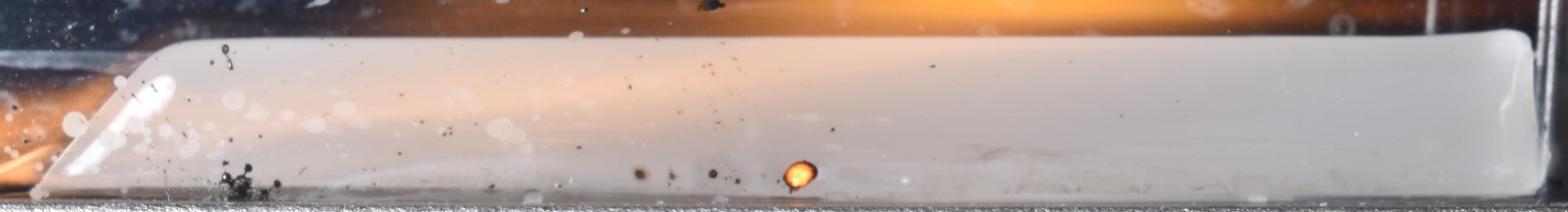}}\\
		\subfloat[]{%
		\includegraphics[width=0.7\textwidth]{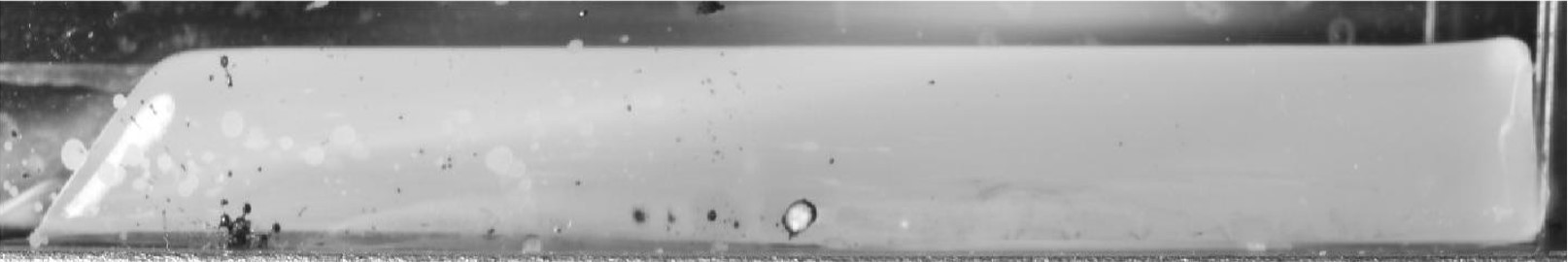}}\\
		\subfloat[]{%
		\includegraphics[width=0.7\textwidth]{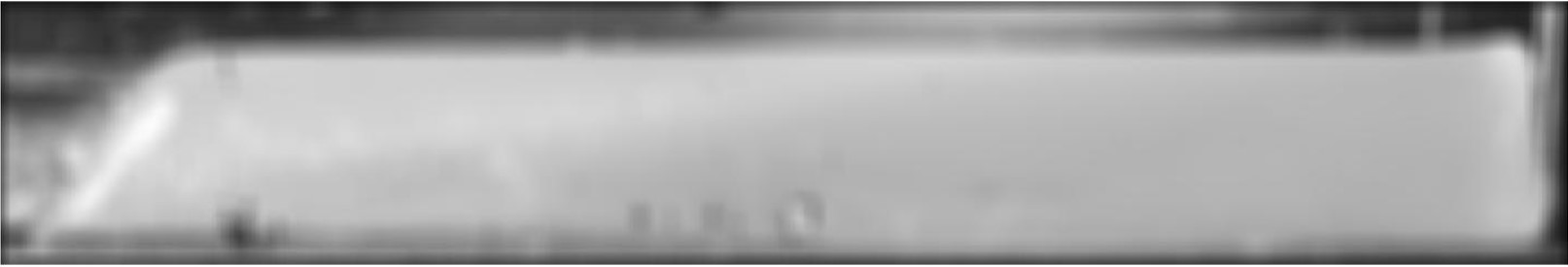}}\\
		\subfloat[]{%
		\includegraphics[width=0.7\textwidth]{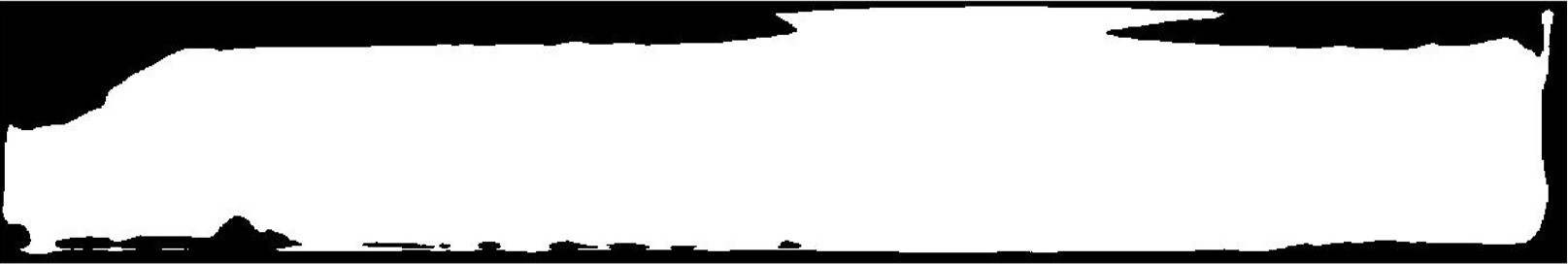}}\\
  \caption{An example of the process of the spatial filtering method starting from top to bottom, (a) original, (b) grayscale, (c) spatially filtered, and (d) segmented images.}
  \label{fig:Spat}
\end{figure}

Another image processing method that helps reduce the amount of noise that affects the data, specifically by filtering out the small data clusters, is spatial filtering. The images are first reduced in resolution by $50$\% ($\sim300$ by $2100$) and then a $30$ by $30$ ($\sim0.1$\% of the original image) sized square block sliding average filter is applied. The data is reduced to enable, effectively, a larger window without being unreasonably computationally expensive. A larger window could have been chosen, but it is at the cost of more computational resources and at the increasing cost of the definition of the fuel boundaries. An example of this process is shown in \cref{fig:Spat}.

\subsection{U-net with Monte-Carlo Dropout}
\begin{figure}[H]
  \centering
  \includegraphics[scale=0.6]{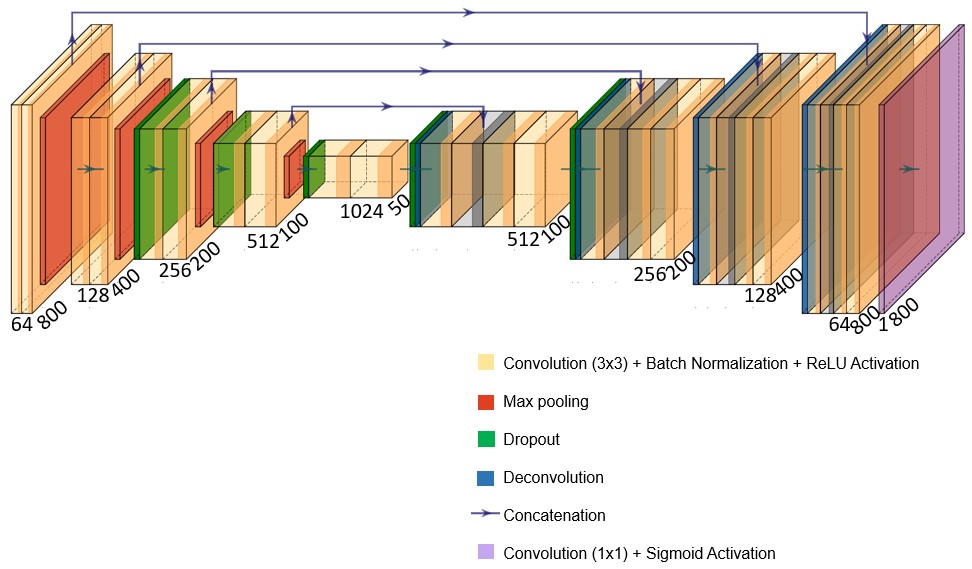}
  \caption{U-net neural network architecture with dropout in the intermediate layers based on \cite{RFB15} and \cite{devries2018leveraging}. The input images are $800$ by $800$ by $3$, their size is reduced and their depth increased in the contraction path of the U-net, and the opposite process takes place in the expansion path to arrive at the output binary segmentation mask of size $800$ by $800$ by $1$.}
  \label{fig:unet}
\end{figure}

The base U-net architecture was created using the Keras library \cite{chollet2015keras}. For additional details on U-net architecture, readers are referred to \cite{RFB15}. U-net is the choice of network for this study because it is capable of handling a low number of training data and still have a high success rate \cite{RFB15}. \Cref{fig:unet} shows a schematic of the U-net architecture used in this study. The deep network has a data reduction down-branch (contraction path) to locate the important features, followed by a data dilation up-branch (expansion path) to return to higher resolutions. U-net initially extracts the most important features from the image as resolution reduces and depth increases in the contraction path. Then, to acquire location information, the image is reconstructed by up-sampling to return to higher resolutions \cite{RFB15}. The contraction path consists of four sequences of two convolutions ($3$ by $3$ kernels and `same' padding so that the output image has the same height/width as the input image), followed by batch normalization, ReLU activation, and a max pooling operation ($2$ by $2$ kernel and strides equal to $2$). The expansion path consists of four sequences of upsampling ($2$ by $2$ kernels and `same' padding), followed by a concatenation with the feature map from the corresponding level in the contracting path, two convolutions ($3$ by $3$ kernels and `same' padding), batch normalization, and lastly a ReLU activation. Finally, U-net includes a last convolution operation ($1$ by $1$ kernel, `same' padding) to arrive at a gray-scale image of equal height and width as the original image (i.e., the fuel segmentation mask). The selected layer settings are consistent with those in \cite{RFB15} except for the extra padding in the convolutions to prevent the loss of border pixels, which is common practice \cite{Chauhan2018}. Additionally, batch normalization is included (which was not included in in the original U-net paper) to reduce covariant shift, as suggested in \cite{IS15}.

To capture model uncertainty information (i.e. how confident the U-net is for each predicted pixel in the segmentation masks), the process outlined in \cite{devries2018leveraging} utilizing Monte-Carlo Dropout is followed to generate the associated uncertainty map for each predicted segmentation mask. There are other approaches to quantify uncertainty in CNNs (e.g., using Bayesian networks that use distributions for the model weights \cite{Abdar2021}), but Monte Carlo Dropout is selected since the dropout layers can be easily implemented as extensions to the original U-net architecture. First, the U-net is trained with a $50$\% dropout probability in the intermediate layers to reduce overfitting according to \cite{JMLR:v15:srivastava14a}. The dropout process randomly silences neurons in the intermediate layers of the U-net by zero-ing the weights according to the set probability. Then at inference, the trained model is sampled $20$ times \cite{devries2018leveraging}, and the predicted segmentation masks are computed as the mean masks of the sampled model instances. The uncertainty map is then calculated as the entropy of the averaged probability vectors for the two class dimensions (``fuel" and ``noise"), as described in \cite[][p. 309]{Hastie2009}: 
\begin{equation}
  U = -\sum_{c = 1}^{2}p_{c}\log(p_{c}) = -[p_{_{fuel}}\log(p_{_{fuel}})+p_{_{noise}}\log(p_{_{noise}})]
\end{equation}
where $U$ is the uncertainty map, $p_{_{fuel}}$ is the predicted probability that the pixel corresponds to fuel, and $p_{_{noise}}$ is the predicted probability that the pixel corresponds to noise / background.

In order to increase robustness and reduce biases, the training data is initially augmented. Using the Albumentations library \cite{A}, each training image is rotated and flipped in random directions and magnitudes. The most common loss function for binary image segmentation - binary cross entropy loss function, is used here in combination with a sigmoid activation layer. The binary cross entropy loss function is defined as,
\begin{equation}
  \ell(x,y) = \overline{\{l_1,....,l_N\}^T},
\label{eq:bce}
\end{equation}
where $\ell$ is the loss function, the overbar corresponds to the mean all $l_n$, and
\begin{equation}
  l_n = y_n\cdot \log \sigma (x_n) + (1-y_n) \cdot \log(1-\sigma(x_n))
\end{equation}
where, $y_n$ is the prediction, $\sigma$ is the sigmoid activation function, and $x_n$ is the ground truth. Finally, the Adam optimization algorithm is used to update the weights in the kernels. The algorithm is chosen for its ability to handle noisy data \cite{KB17, LH17} instead of stochastic gradient descent that was used in the original U-net paper.

\section{Results and Discussion}
\label{sec:results}

\subsection{Training on all available images}
\label{sec:all}
As an initial test, the $150$ RGB images are split into the training set ($80$\% or $120$ images) and the testing and validation set ($20$\% or $30$ images). The images are randomly assigned to the training or validation sets, so that information from all four fluxes are used both in training the U-net to correctly capture the experiment phenomena and validate the U-net model to test whether it can reliably be used for fuel regression calculation. To train and validate the U-net, the original experiment images and manually traced masks are resized to $800$ by $800$ pixels to limit computational expense. An aspect ratio of $1$ ensures valid matrix operations during the convolution and deconvolution operations in the deeper layers on the U-net. The $150$ RGB experimental images follow a naming convention based on the oxidizer flux and the time of the experiment they are captured shown in \cref{tab:names}.
\begin{table}[h]
\begin{center}
\caption{\label{tab:names}Image naming conventions with the letters corresponding to different oxidizer fluxes and the numbers corresponding to the chronological order that the images were captured at.}
\begin{tabular}{ |c|c|c|c| } 
\hline
Oxidizer flux, $G [kg/m^2-s]$& First captured image & Last captured image & Total images \\
\hline
 $5.91$  & A1 & A37 & 37\\ 
9.58 & B1 & B36 &36 \\ 
18.59 & C1 & C39 & 39 \\ 
22.19 & D1 & D38 & 38\\
\hline
\end{tabular}
\end{center}
\end{table}

For the neural network parameters, the batch size was set to $5$ to account for possible model degradation in its ability to generalize due to the small training set as shown by \cite{keskarbatch} and the starting learning rate is $10^{-3}$, the default setting for the Adam optimizer (the learning rate is adapted during the training process). For this initial training, the U-net required $6$ $GB$ of memory, $3$ cores, and runtime of $20.5$ hours. 

\begin{figure}[H]
  \centering
  \subfloat[\label{fig:initialresultsa}Original]{%
		\includegraphics[width=0.217\textwidth]{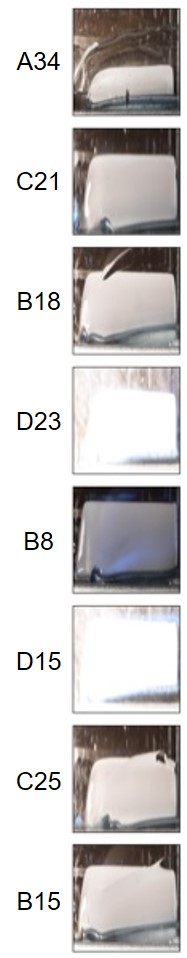}}\hspace{1cm}
		\subfloat[\label{fig:initialresultsb}Ground Truth]{%
		\includegraphics[width=0.14\textwidth]{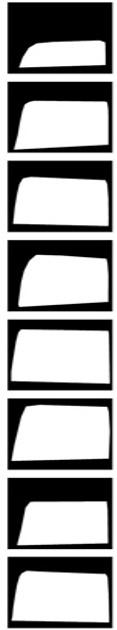}}\hspace{1cm}
		\subfloat[\label{fig:initialresultsc}Prediction]{%
		\includegraphics[width=0.14\textwidth]{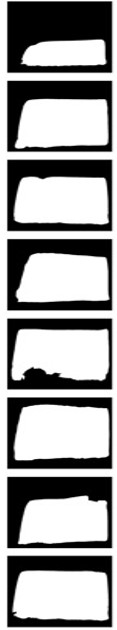}}\hspace{1cm}
		\subfloat[\label{fig:initialresultsd}Uncertainty]{%
		\includegraphics[width=0.1415\textwidth]{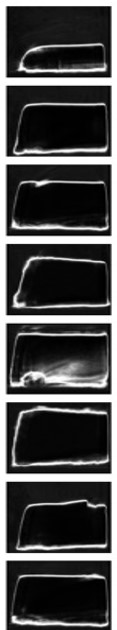}}\hspace{1cm}
  \caption{Selected few of (a) the original images from the testing/validation set of various oxidizer fluxes, (b) the ground truth generated by the manual process, (c) the prediction from the trained neural network, and (d) the associated uncertainty map.}
  \label{fig:initialresults}
\end{figure}

A selected few examples of original resized images of the testing set are shown in \cref{fig:initialresultsa}, the associated ground truth binary masks manually traced after the resolution reduction are shown in \cref{fig:initialresultsb}, the neural network's predictions are shown in \cref{fig:initialresultsc}, and the associated uncertainty maps are shown in \cref{fig:initialresultsd}.

The U-net model trained from images of all fluxes visually predicted the segmentation masks successfully. For example, for images A$34$, B$18$, and B$15$, U-net correctly traces the fuel mask without including the melted wax on the glass of the image. It also correctly predicts the mask for the over-saturated images D$15$ and D$23$. Some abnormal geometries (e.g., B$8$ and C$25$) are found to be the hardest to predict correctly. The uncertainty maps showed the largest level of uncertainty to be on the fuel boundaries and around abnormal geometries in the predicted masks (e.g., B$8$). The model uncertainties inside and outside the identified fuel mask border are very small, showing that once U-net identified the border of the fuel masks with some uncertainty, it correctly learned to classify the pixels inside and outside the border. Lastly, U-net performed well for fuel mask predictions in all stages of the experiment. Reliable prediction of the fuel masks over time is an important feature if U-net is to be used for fuel regression rate measurements because measurement requires tracking of the fuel mask over the entire burn duration.

\begin{figure}[H]
	\centering
	\subfloat[]{%
		\includegraphics[width=0.33\textwidth]{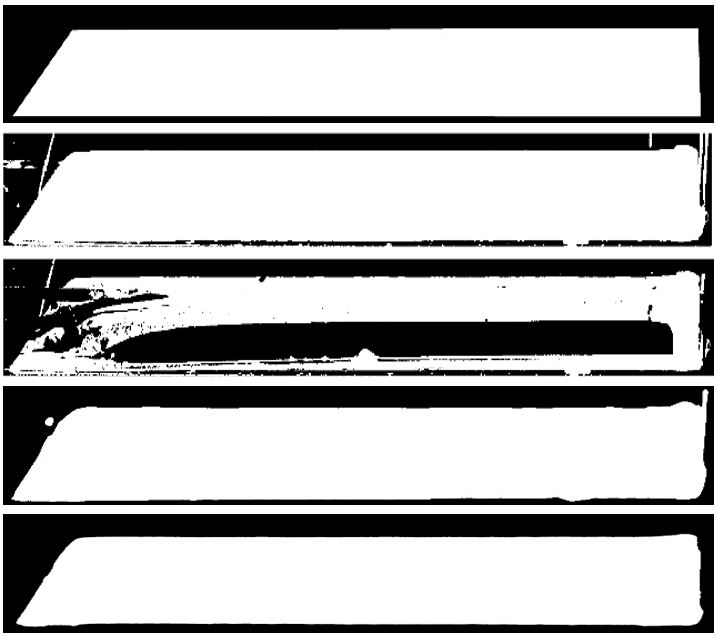}}
	\subfloat[]{%
		\includegraphics[width=0.33\textwidth]{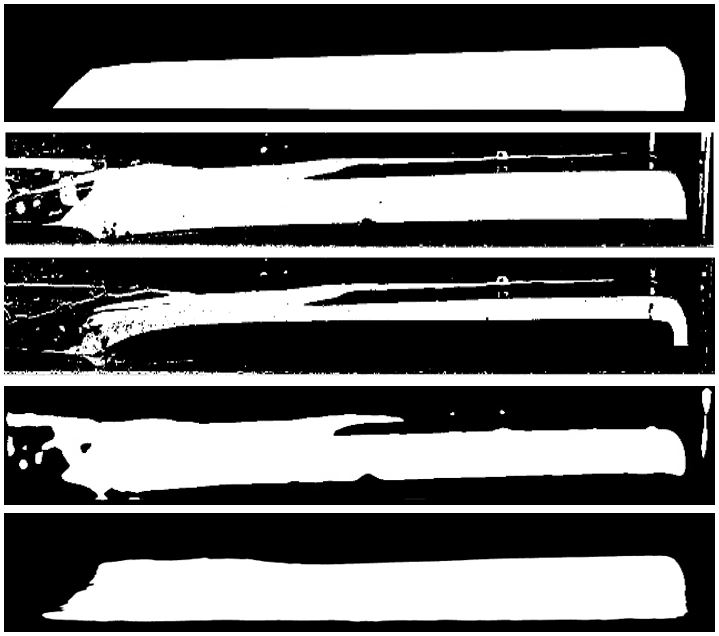}}
	\subfloat[]{%
		\includegraphics[width=0.33\textwidth]{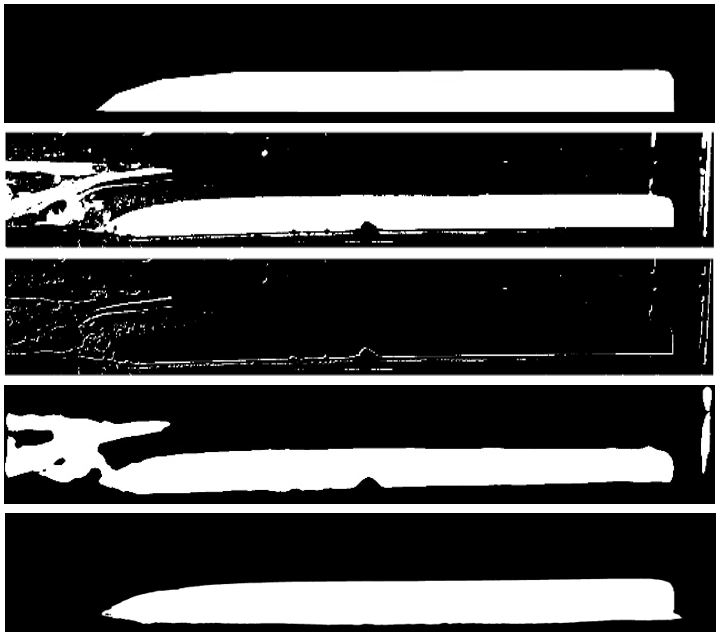}}
	\caption{From the top to bottom, ground truth, threshold, TLIS, spatial filtering, and the U-net binary masks for (a) early, (b) mid, and (c) late times for an oxidizer flux of $5.91$ $[kg/m^2-s]$.}
	\label{fig:5}
\end{figure}

\begin{figure}[H]
	\centering
	\subfloat[]{%
		\includegraphics[width=0.33\textwidth]{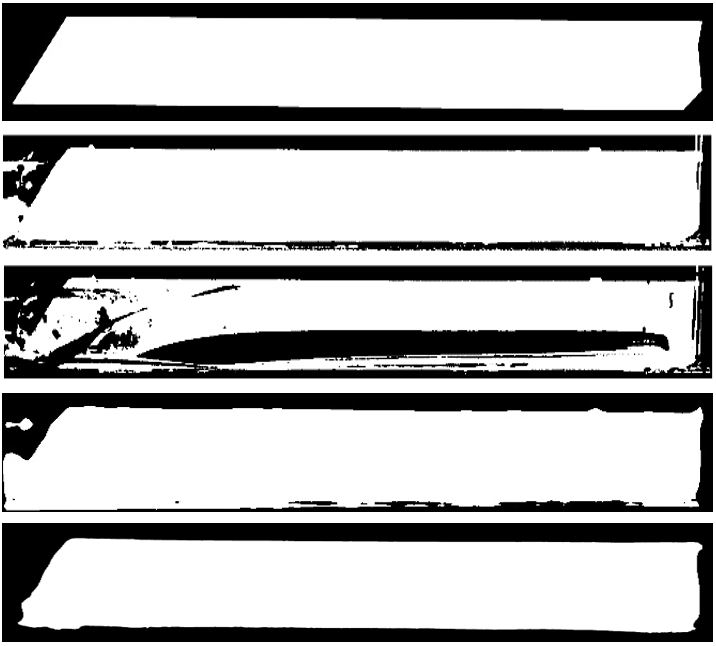}}
	\subfloat[]{%
		\includegraphics[width=0.33\textwidth]{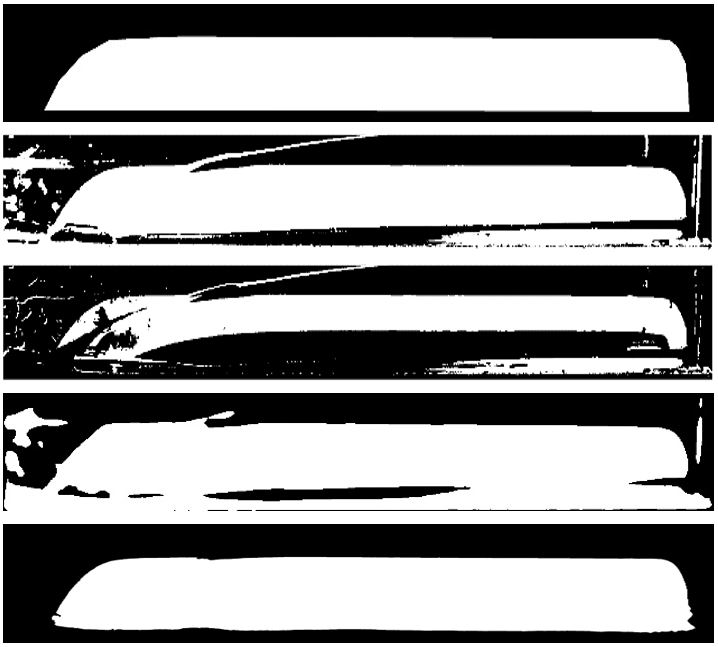}}
	\subfloat[]{%
		\includegraphics[width=0.33\textwidth]{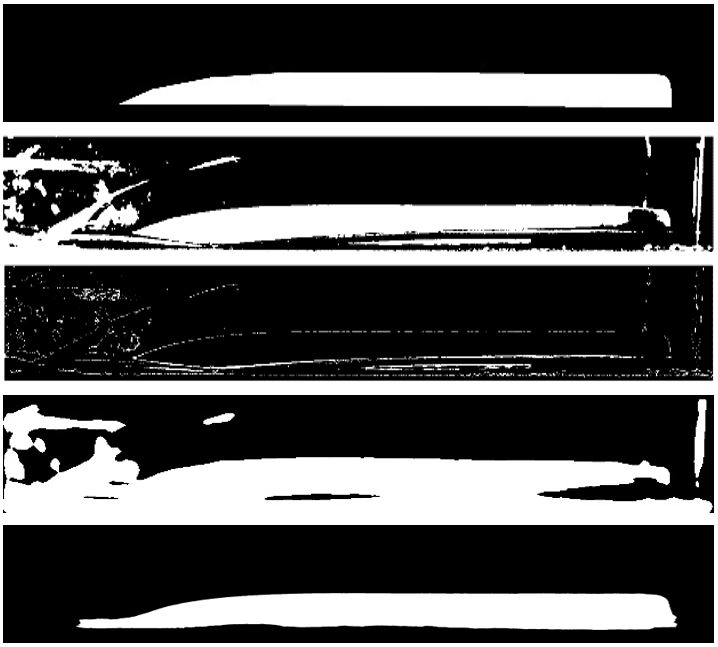}}
	\caption{From the top to bottom, ground truth, threshold, TLIS, spatial filtering, and the U-net binary masks for (a) early, (b) mid, and (c) late times for an oxidizer flux of $9.58$ $[kg/m^2-s]$.}
	\label{fig:9}
\end{figure}

\begin{figure}[H]
	\centering
	\subfloat[]{%
		\includegraphics[width=0.33\textwidth]{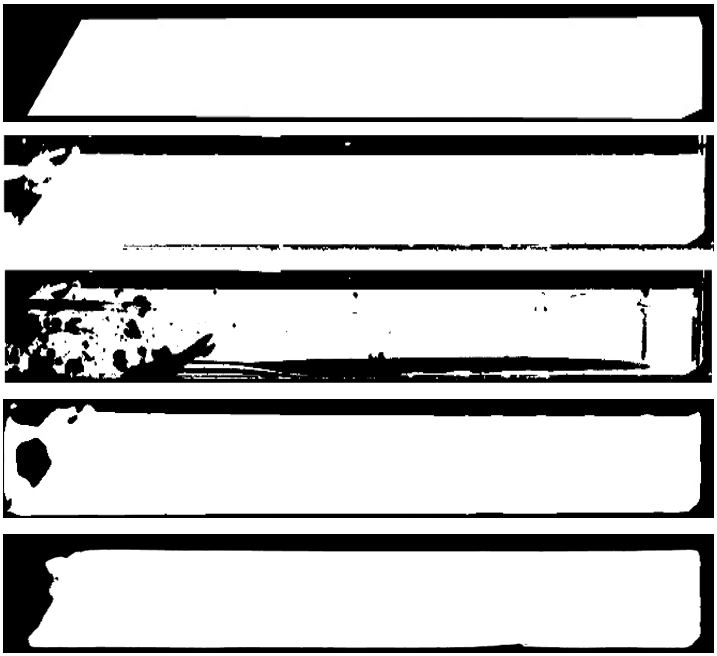}}
	\subfloat[]{%
		\includegraphics[width=0.33\textwidth]{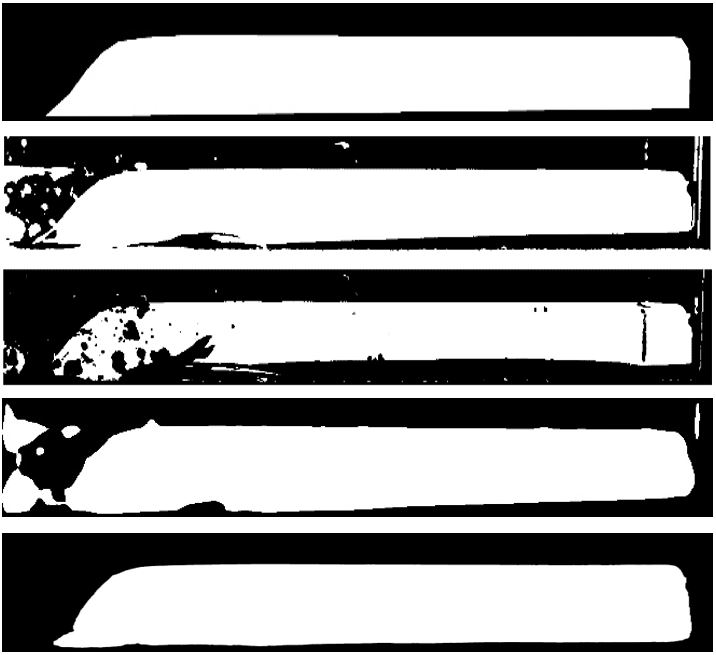}}
	\subfloat[]{%
		\includegraphics[width=0.33\textwidth]{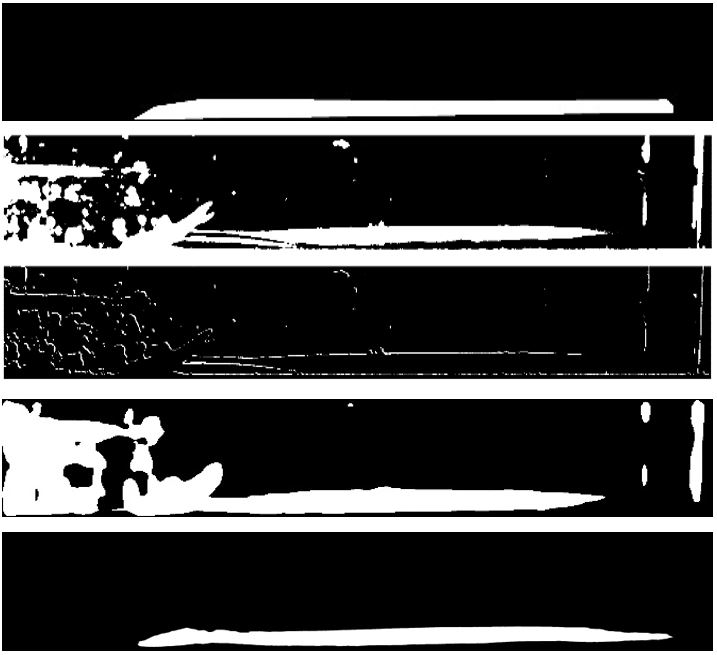}}
	\caption{From the top to bottom, ground truth, threshold, TLIS, spatial filtering, and the U-net binary masks for (a) early, (b) mid, and (c) late times for an oxidizer flux of $18.59$ $[kg/m^2-s]$.}
	\label{fig:18G}
\end{figure}

\begin{figure}[H]
	\centering
	\subfloat[]{%
		\includegraphics[width=0.33\textwidth]{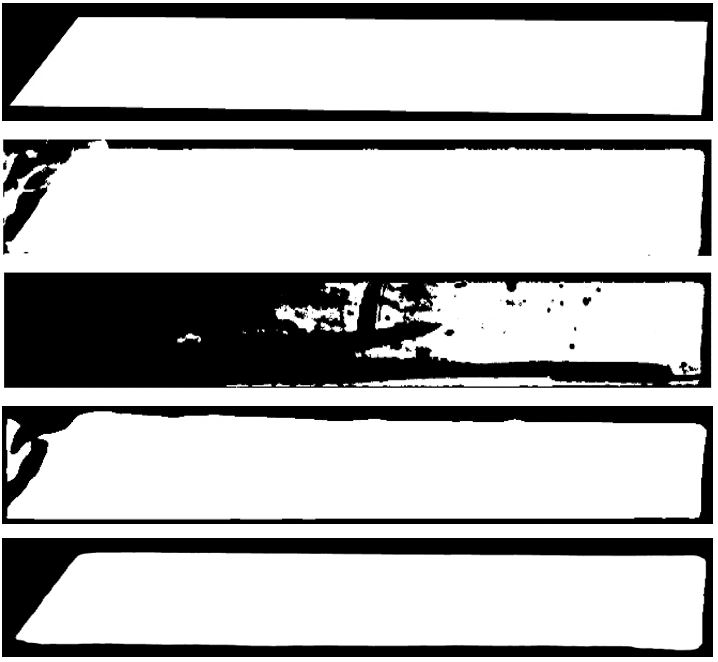}}
	\subfloat[]{%
		\includegraphics[width=0.33\textwidth]{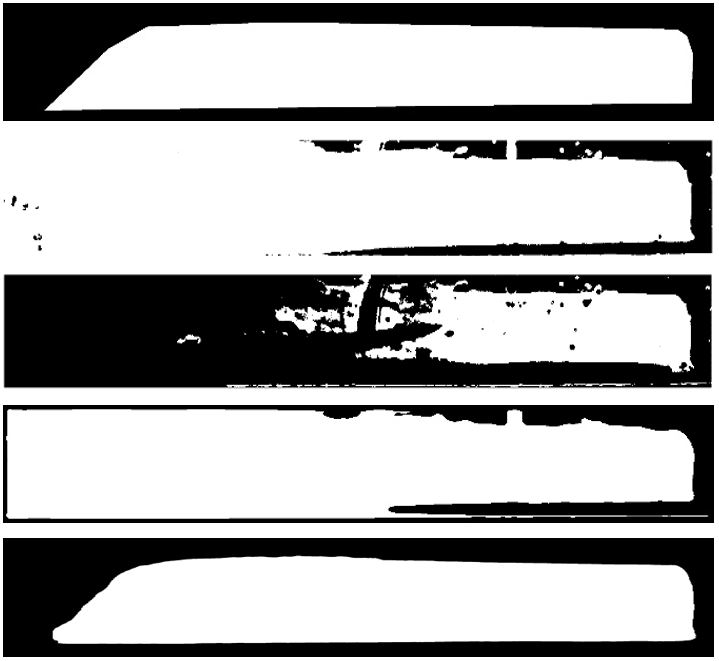}}
	\subfloat[]{%
		\includegraphics[width=0.33\textwidth]{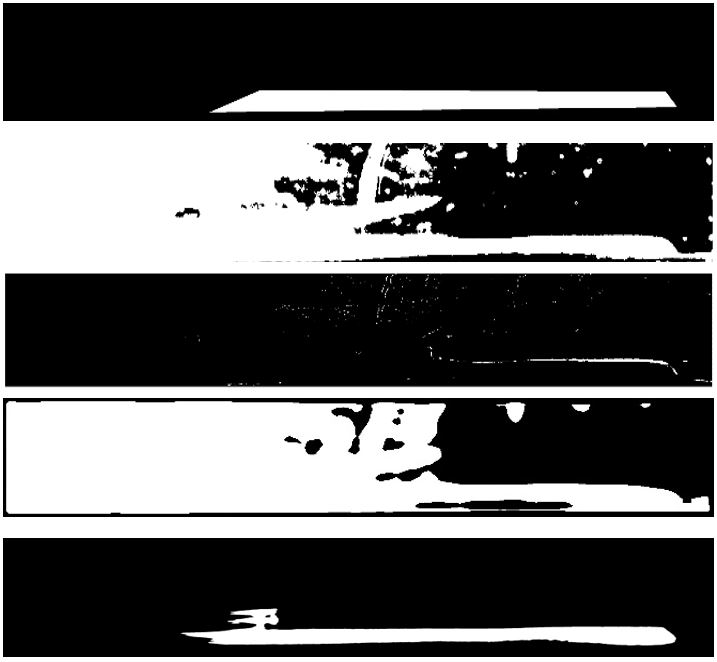}}
	\caption{From the top to bottom, ground truth, threshold, TLIS, spatial filtering, and the U-net binary masks for (a) early, (b) mid, and (c) late times for an oxidizer flux of $22.19$ $[kg/m^2-s]$.}
	\label{fig:22}
\end{figure}

Predictions from the oxidizer fluxes of $5.91$, $9.58$, $18.59$, and $22.19$ $[kg/m^2-s]$ using threshold, TLIS, spatial filtering, and U-net at early ($t\approx 0s$), mid ($t\approx 5s$), and late ($t\approx 10s$) times are presented in \cref{fig:5,fig:9,fig:18G,fig:22}. Visually, the U-net predictions to perform the best and the simple image processing methods especially struggle with the oxidizer flux of $22.19$ $[kg/m^2-s]$ case in \cref{fig:22} most likely due to the over-saturated data seen uniquely in only the $22.19$ $[kg/m^2-s]$ case. The threshold, TLIS, and spatial filtering all have some advantages and do provide reasonable predictions, however, they do not provide the level of reliable filtering that U-net provides. 

\begin{figure}[H]
	\centering
	\subfloat[]{%
		\includegraphics[width=0.4\textwidth]{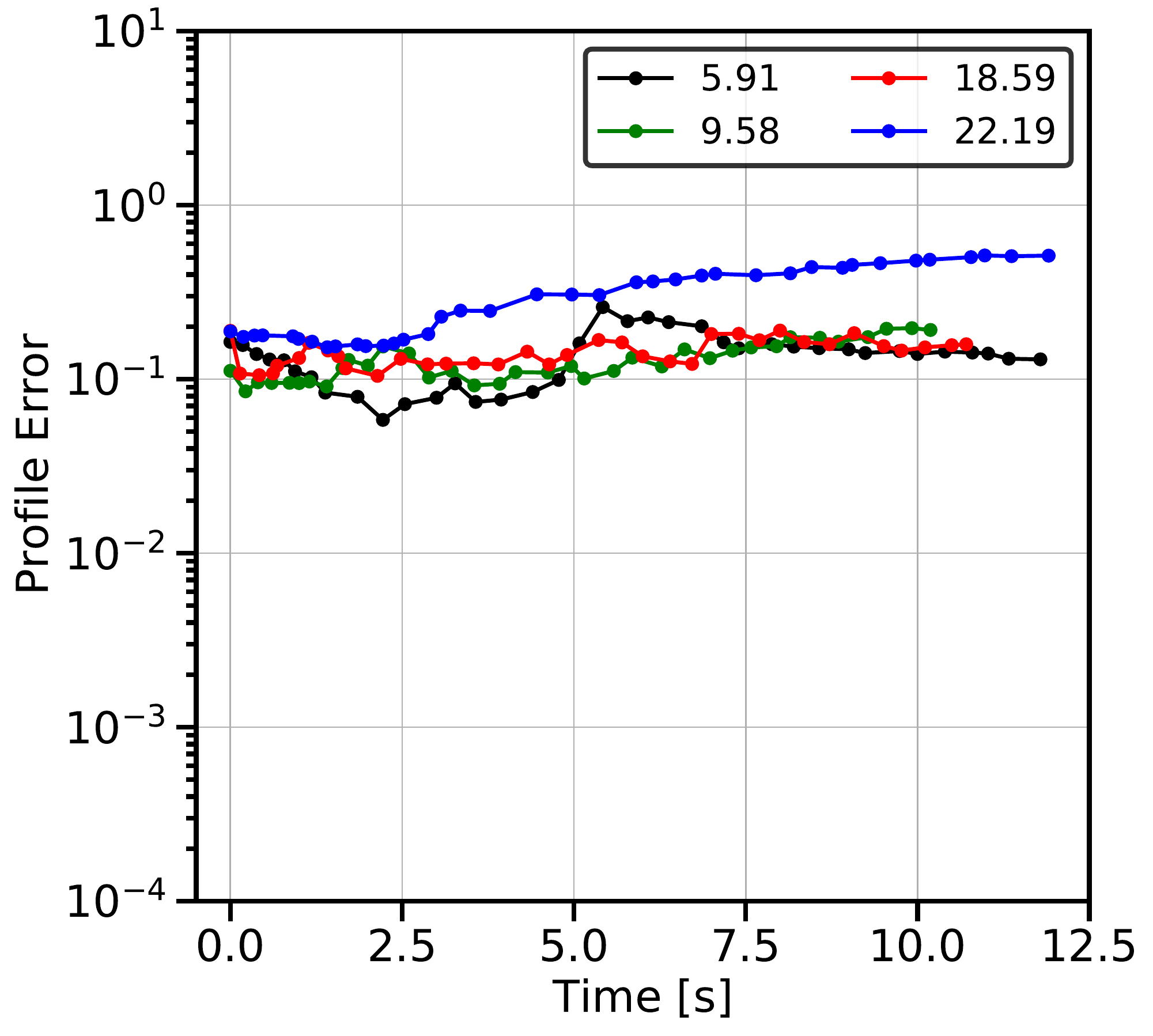}}
	\subfloat[]{%
		\includegraphics[width=0.4\textwidth]{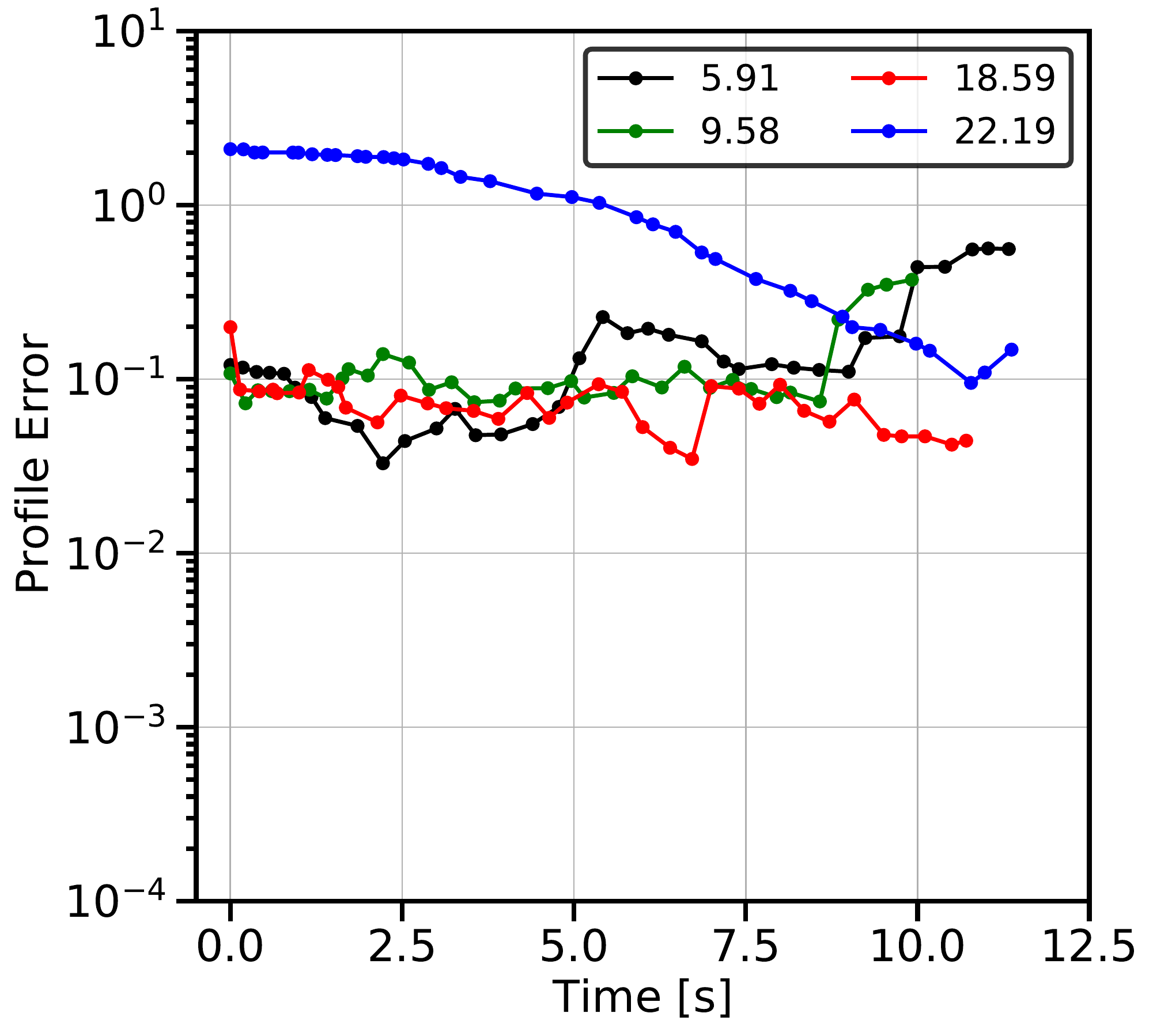}} \\
	\subfloat[]{%
		\includegraphics[width=0.4\textwidth]{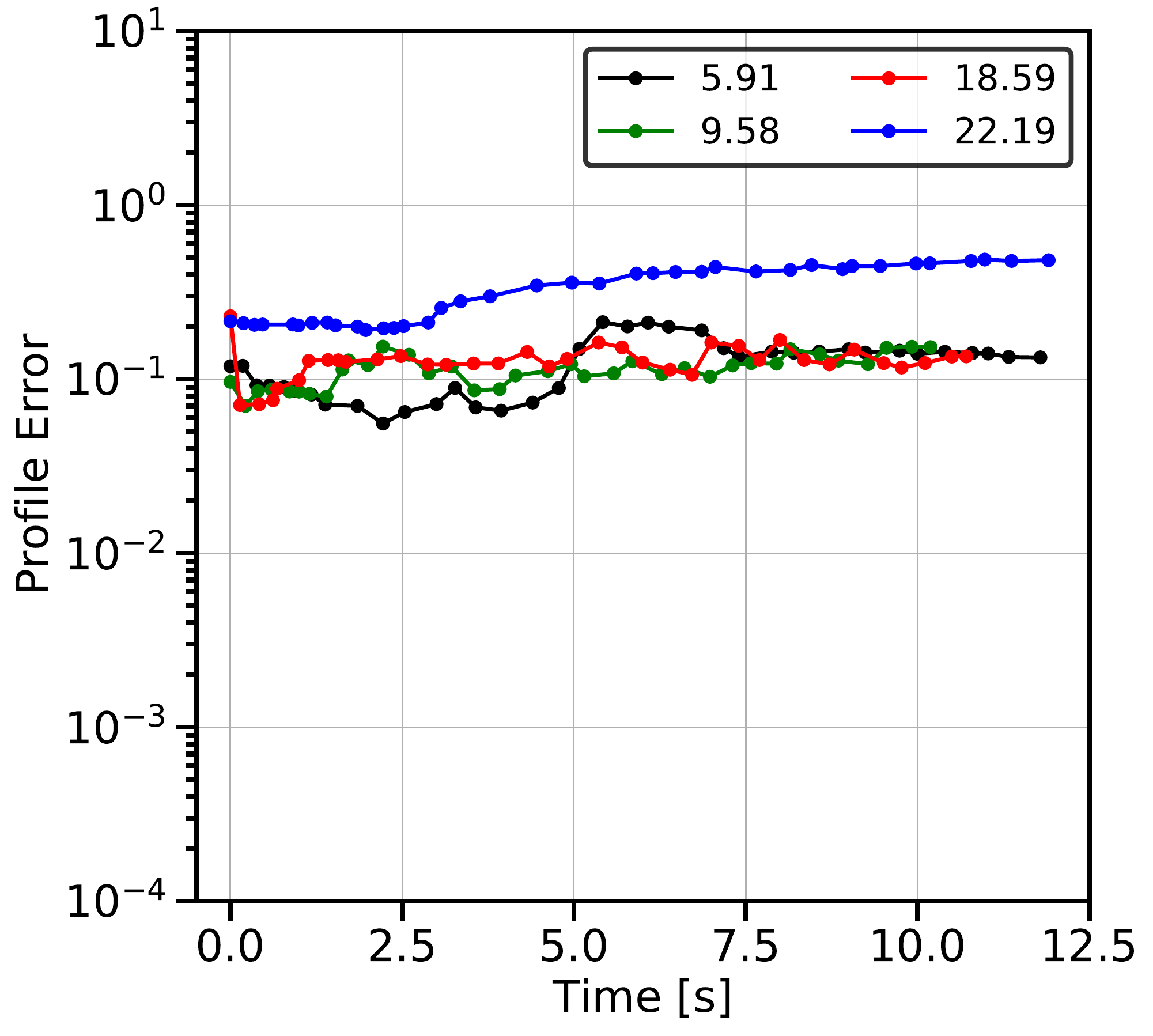}}
	\subfloat[\label{fig:mlerror}]{%
		\includegraphics[width=0.4\textwidth]{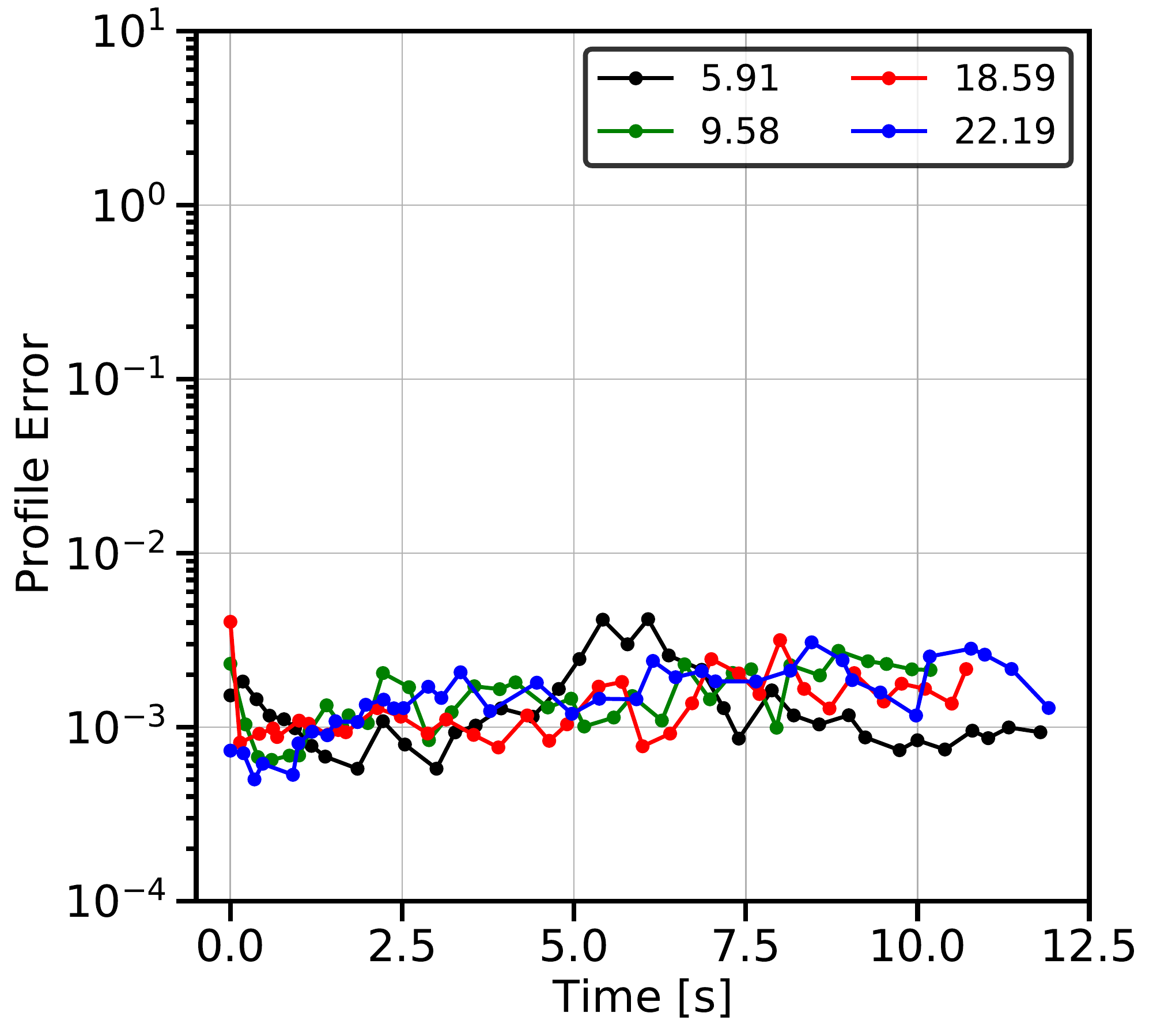}}
	\caption{Spatial error between the ground truth mask profiles and the profiles from (a) the threshold, (b) TLIS, (c) spatial filtering, and (d) U-net.}
	\label{fig:errors}
\end{figure}

Figure \ref{fig:errors} shows the total absolute spatial error, $(\sum |h - h_{truth}|/h_{truth})/width_{image}$ between the ground truth mask profiles and the profiles from the image processing techniques for each point in time. The U-net results are resized into the native resolution of the ground truth images in order to compare the profiles. Overall, the U-net produced the lowest maximum error out of all four image processing methods. The U-net architecture is the only approach that performed the most consistent out of the other image processing methods with error below $10^{-2}$ for all fluxes. Specifically, in the $22.19$ $[kg/m^2-s]$ case, the U-net filtered the noise caused by over-saturation correctly, whereas, the other image processing methods generated significant errors, especially the TLIS method with errors over $10^0$.

\begin{figure}[H]
  \centering
  \subfloat[\label{fig:results}]{\includegraphics[scale=0.32]{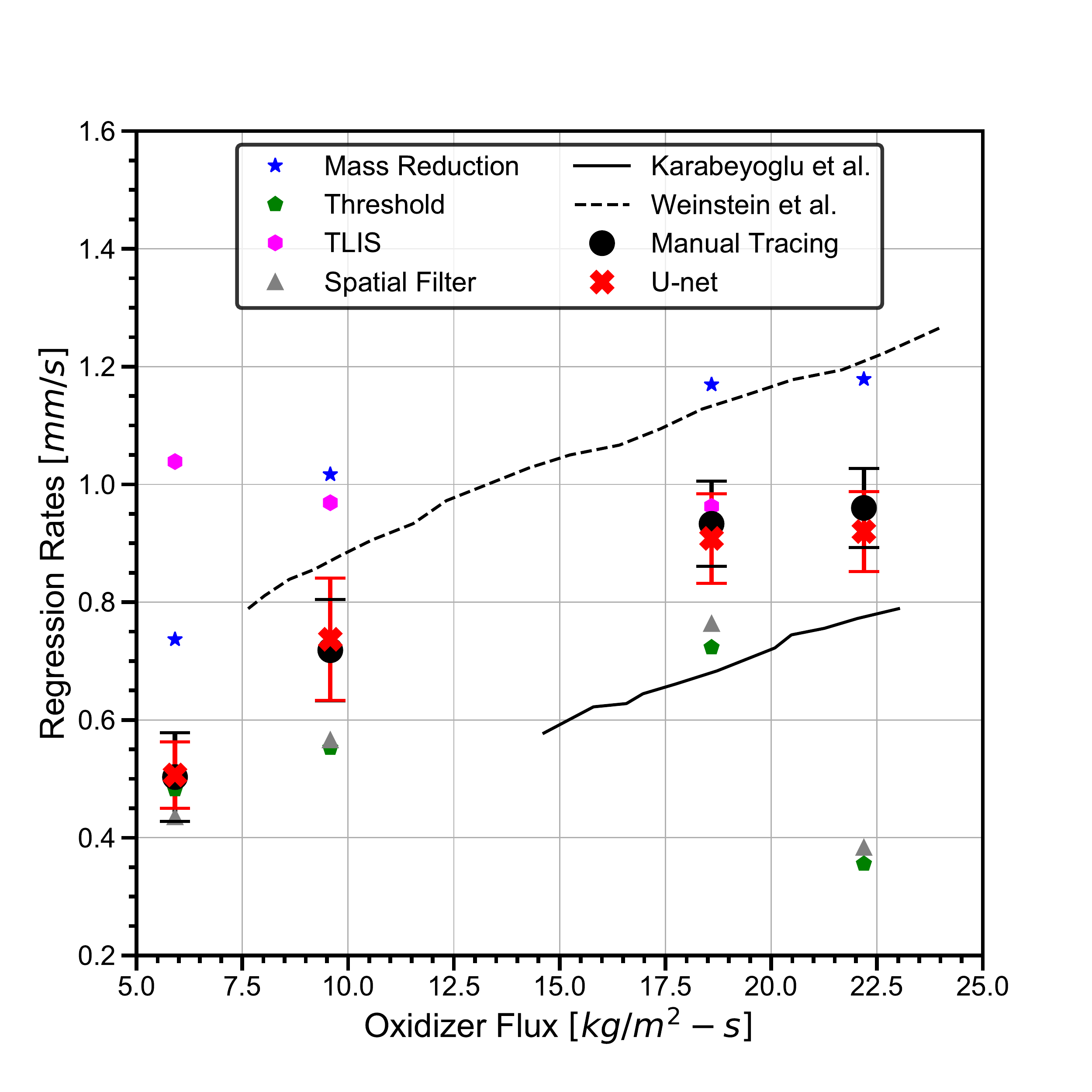}}
  \subfloat[\label{fig:error}]{\includegraphics[scale=0.32]{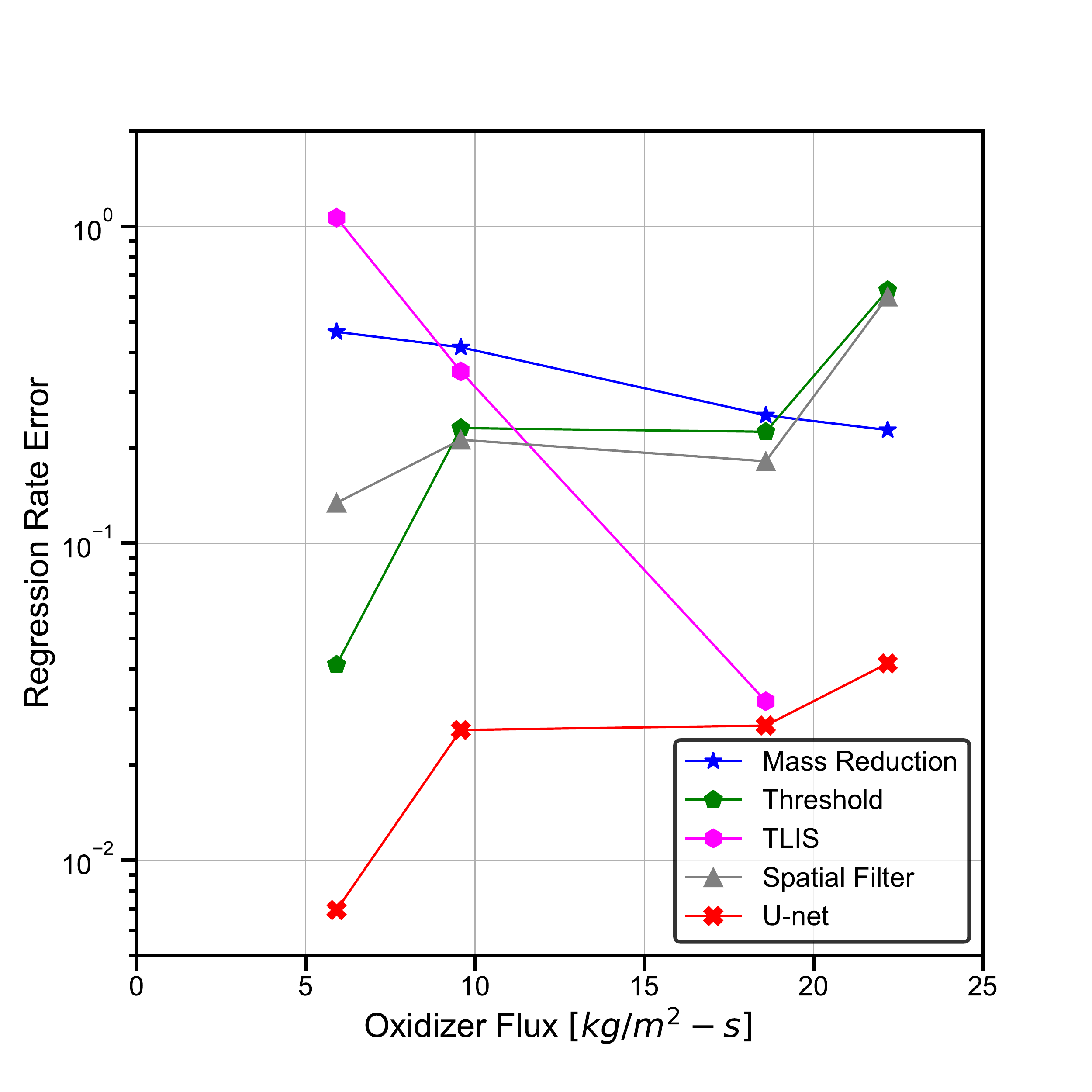}}
  \caption{(a) Comparison to published data from \cite{BAKSD20} and (b) the absolute error for the imaging processing techniques assuming the manual data from \cite{BAKSD20} as truth (TLIS did not give a feasible result for $22.19$ $[kg/m^2-s]$ and therefore is not shown).}
\end{figure}

\Cref{fig:results} shows the comparison between the regression rate obtained from the different image processing techniques and the ground truth (manually drawn data from \cite{BAKSD20}). The results from the manual process are in black circles and the results from U-net are shown by the red circles. The regression rate error bars for the U-net approach are calculated by first getting an average threshold value for each mask by summing the entire uncertainty in the map and then normalizing by the number of pixels in the mask. If a pixel's uncertainty falls below this threshold the pixel becomes a ``$0$ (noise)" and if it falls above, it becomes a ``$1$ (fuel)." This process creates another binary mask which is added and subtracted to the original mean binary masks to get two new sets of regression rate data. The maximum deviation from the original regression rate from the two new regression rates is then calculated and this value is used for the error bars. The regression rate measurements inferred from the measurement fuel mass before and after the burn are labeled as ``Mass Reduction". The results from the other image processing methods are also plotted. The TLIS method did not provide feasible results for the data for the oxidizer flux of $22.19$ $[kg/m^2-s]$ resulting in no data for this method at this oxidizer flux. The black dashed and solid lines are regression rates obtained by \cite{Weinstein2013} and \cite{Karabeyoglu2001}, respectively. \Cref{fig:error} shows the total absolute error, $|\dot{r}_{prediction} - \dot{r}_{truth}|/\dot{r}_{truth}$, assuming the manual tracing, the black circles in \cref{fig:results}, from \cite{BAKSD20} as truth. TLIS performs poorly for $5.91$ and $9.58$ $[kg/m^2-s]$ fluxes, but performs well for $18.59$ $[kg/m^2-s]$ with an error below $10\%$. As for the $22.19$ $[kg/m^2-s]$ flux, the TLIS results do not provide a feasible regression rate and therefore is not shown. The threshold and spatial filtering have approximately the same accuracy as the mass reduction method. Almost all of the regression rate data points from all of the baseline image processing methods do not fall in the error bars for the manual tracing. The U-net regression rate estimates agree well and are all inside the manual tracing error bars with regression rate error for all fluxes below $10\%$. The U-net error bars as well encompass the manual tracing data points. U-net is the only method that consistently predicts the regression rate for all oxidizer fluxes.

\subsection{U-net Cross-Validation}
\begin{figure}[H]
  \centering
  \includegraphics[scale=0.5]{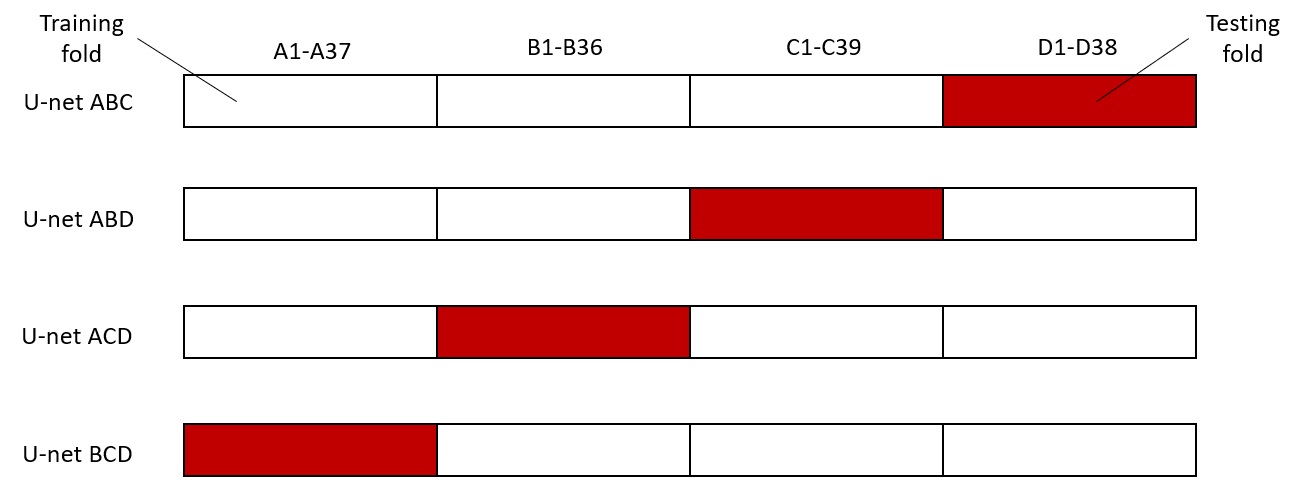}
  \caption{Leave-one-out cross-validation (LOOCV) process. In total, four different U-net models were built by using images from $3$ out of $4$ oxidizer fluxes as the training set each time, and images from the left-out oxidizer flux as the testing set. Image naming convention follows \cref{tab:names}.}
  \label{fig:crossval}
\end{figure}

The results of the U-net architecture presented in the previous section come from training the model with data from all available oxidizer fluxes and showed that U-net accurately predicts the fuel regression rate. However, the image quality varies significantly across the four oxidizer fluxes (e.g., images from $22.19$ $[kg/m^2-s]$ are over-saturated as shown in \cref{fig:initialresults}). Therefore, to further validate the U-net architecture and its capability to make accurate predictions on data from different fluxes, a Leave-one-out cross-validation (LOOCV) \cite{MR763576} approach is implemented. To complete the cross-validation process, the dataset is initially split into $4$ folds, one for each oxidizer flux. Then, a U-net model is built by using $3$ of the folds as the training set  and the last fold as the testing set \cite{Arlot2010} (i.e., train with images from $3$ out of $4$ oxidizer fluxes and test on images from the left-out flux as shown in \cref{fig:crossval}). 

\begin{figure}[H]
  \centering
  \includegraphics[scale=0.3]{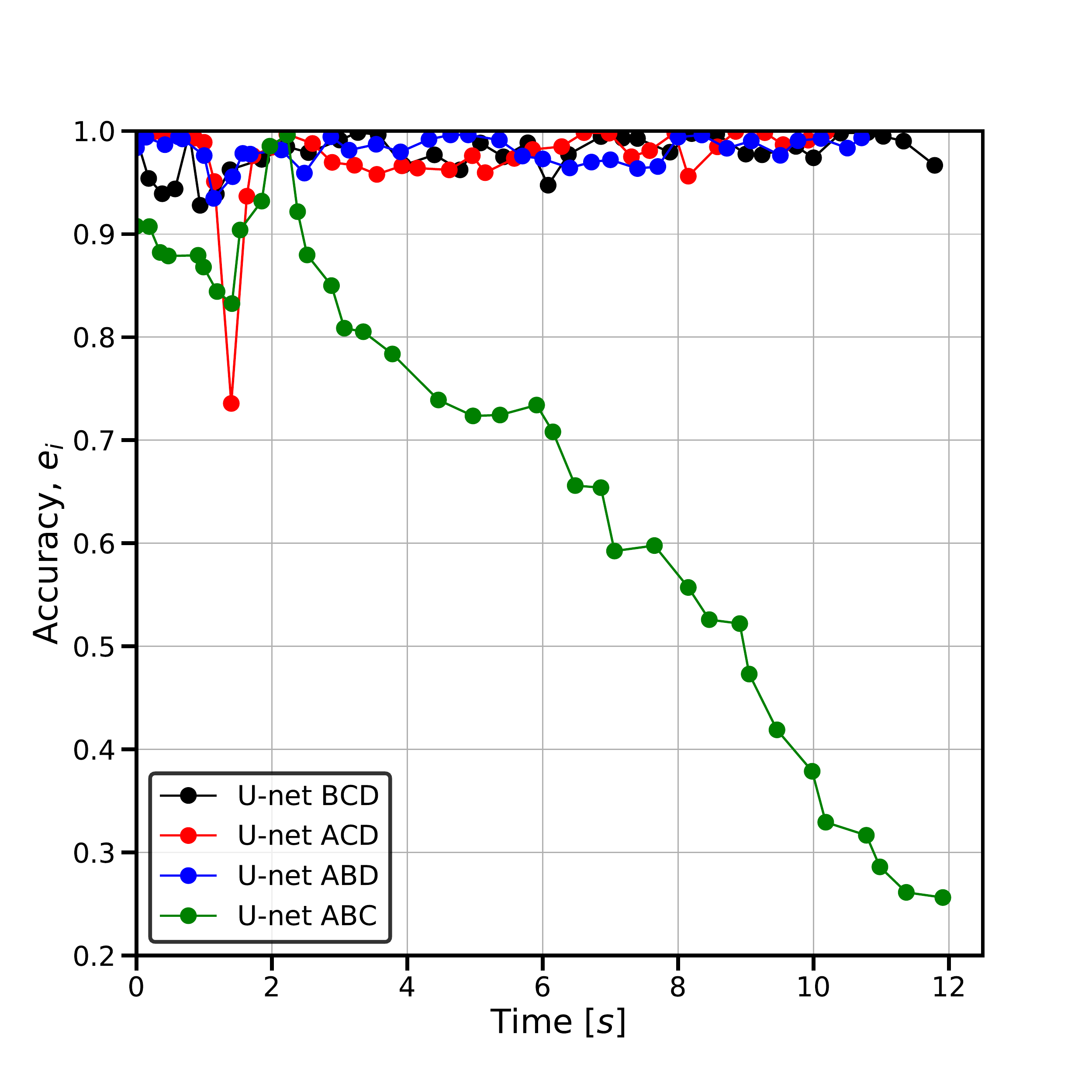}
  \caption{Leave-one-out cross-validation (LOOCV) accuracy measure for all images in the testing oxidizer flux, corresponding to different experiment burn times. }
  \label{fig:crossvalresults}
\end{figure}

Each of the U-net models created during the cross-validation are evaluated based on an accuracy metric that represents the number of correctly classified pixels as fuel or noise for all images in the corresponding testing fold. For each testing image, a confusion matrix is created:

\begin{table}[h]
\begin{center}
\caption{\label{tab:confusionmatrix}Confusion matrix for output of U-net predictions. $n$ represents the corresponding number of pixels.}
\begin{tabular}{ |c|c|c| } 
\hline
    & Predicted: Fuel& Predicted: Noise \\
\hline
 Truth: Fuel  & $n_1$ & $n_2$ \\ 
Truth: Noise & $n_3$ & $n_4$  \\ 

\hline
\end{tabular}
\end{center}
\end{table}
\noindent The accuracy measure is the ratio of correctly classified pixels by the model in the particular testing set, over the total number of pixels:

\begin{equation}
  e_i = \frac{n_{correct}}{{n_{total}}} = \frac{n_1+n_4}{\sum_{i=1}^4 n_i}
\end{equation}
\\
Following the validation process, the results are shown in \cref{fig:crossvalresults}.

The U-net architecture predicts the experimental images for the oxidizer fluxes $5.91$, $9.58$, and $18.59$ $[kg/m^2-s]$ correctly with an accuracy $>95\%$. However, the model that is trained without the over-saturated images of the $22.19$ $[kg/m^2-s]$ oxidizer flux (U-net ABC) fails to make correct predictions past the initial phases of the experiment. The result indicates that when U-net has not learned features of images with high noise or over-saturation, it should not be used for predicting fuel segmentation masks. Including images with high noise, improves the predictive power of the architecture and in those cases, U-net is a viable option for predicting fuel segmentation masks for fuel regression rate measurement. 

The U-net ACD does not predict one particular image well, B$8$. The image includes large amounts of noise from the bottom plate of the chamber and the model fails to classify that area as fuel. Even when trained from data on all four fluxes, noise in B$8$ is not captured correctly by the U-net, as shown previously in \cref{fig:initialresults}. 

After training and validating the U-net architecture using all the available images from the experiments, two practical applications of fuel regression rate measurement are investigated in the following sections: (A) the availability of experimental images from only one oxidizer flux and (B) the availability of only grayscale images. These two scenarios may be frequent in fuel regression measurement research either due to laboratory equipment constraints, process data that has not been manually processed, or an inability to repeat or collect measurements. 

\subsection{Application A: Oxidizer Flux Independency}

\begin{figure}[H]
	\centering
	\subfloat[]{%
		\includegraphics[width=0.4\textwidth]{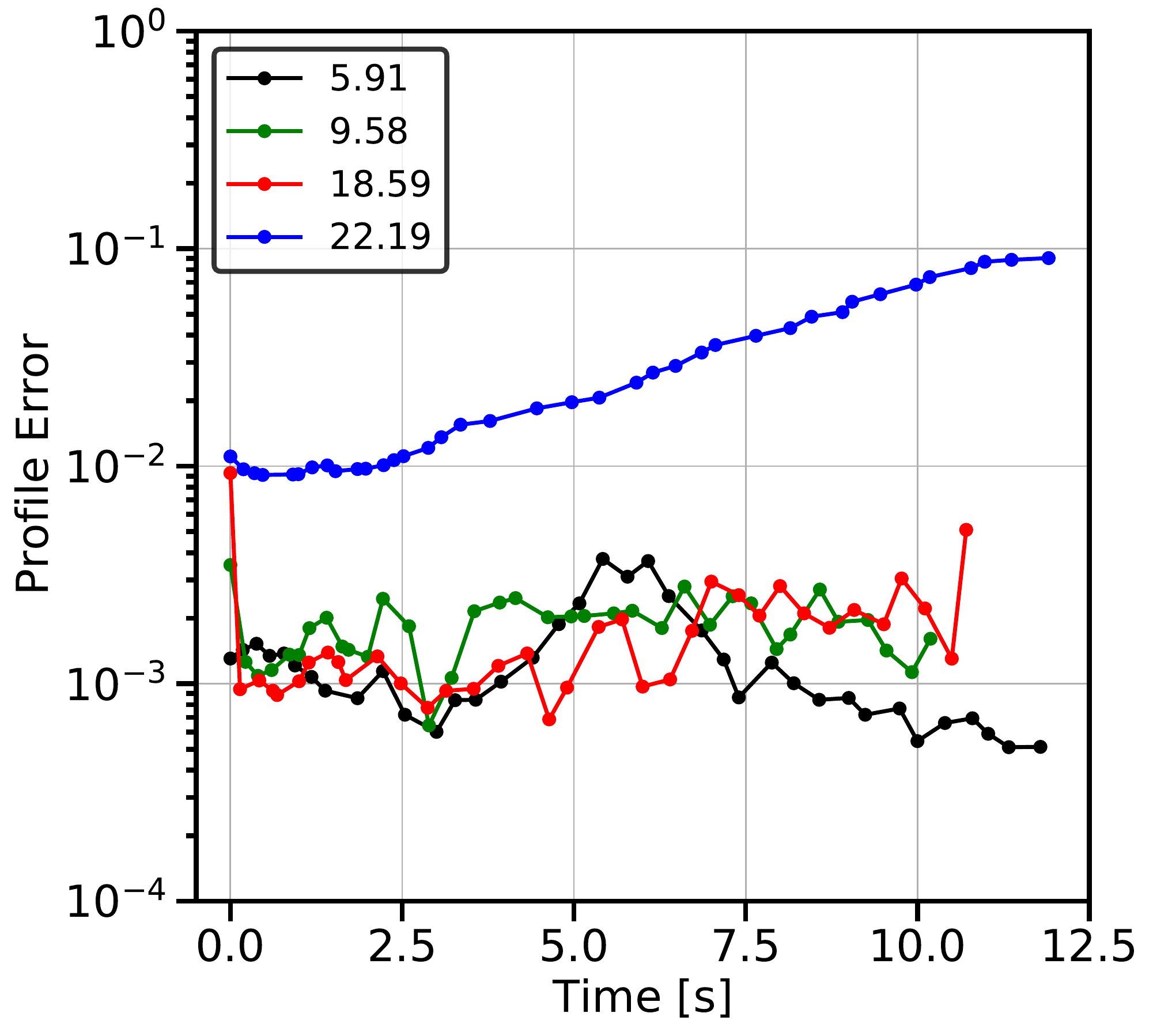}}
	\subfloat[]{%
		\includegraphics[width=0.4\textwidth]{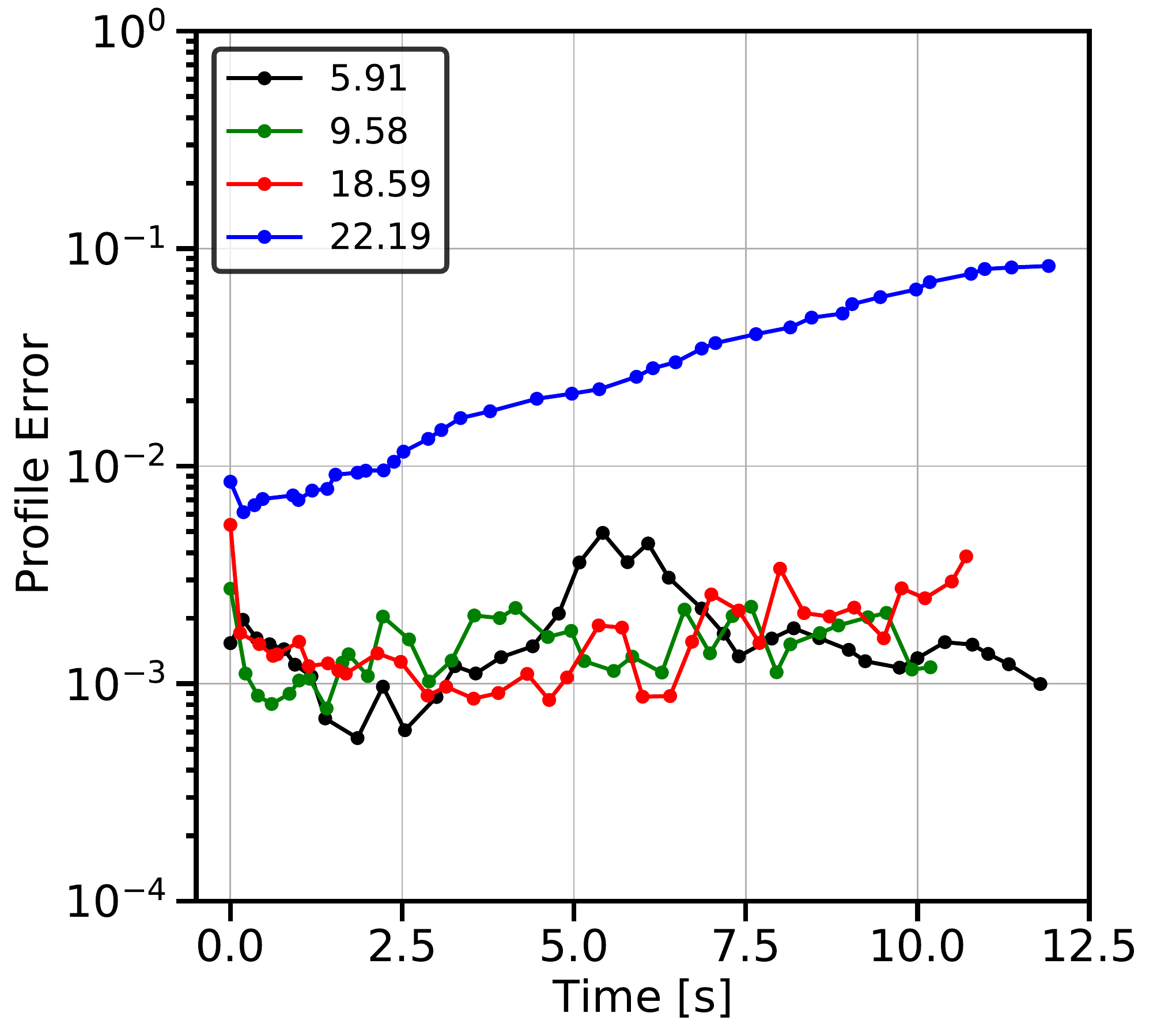}} \\
	\subfloat[]{%
		\includegraphics[width=0.4\textwidth]{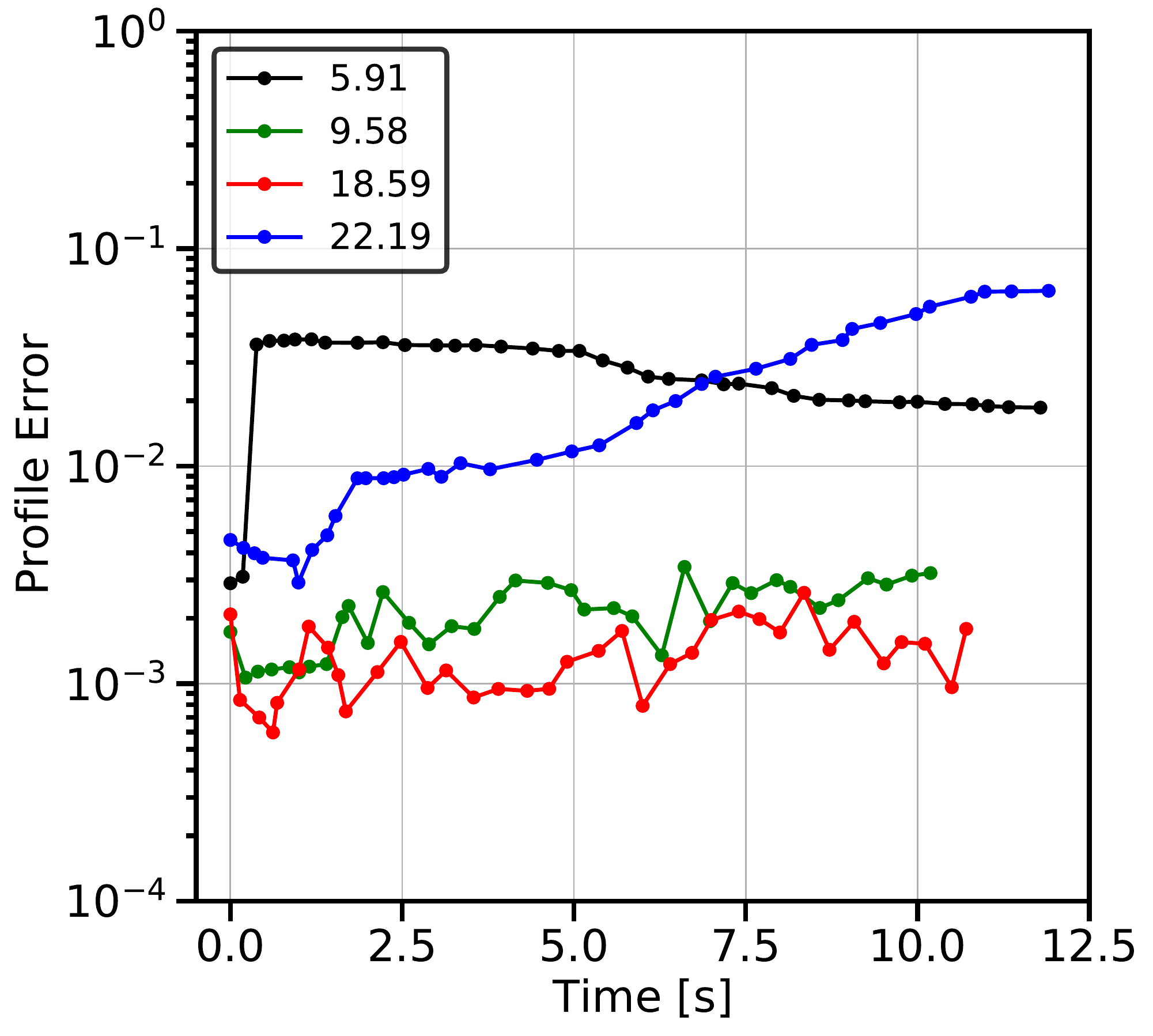}}
	\subfloat[\label{fig:errors_OXIml}]{%
		\includegraphics[width=0.4\textwidth]{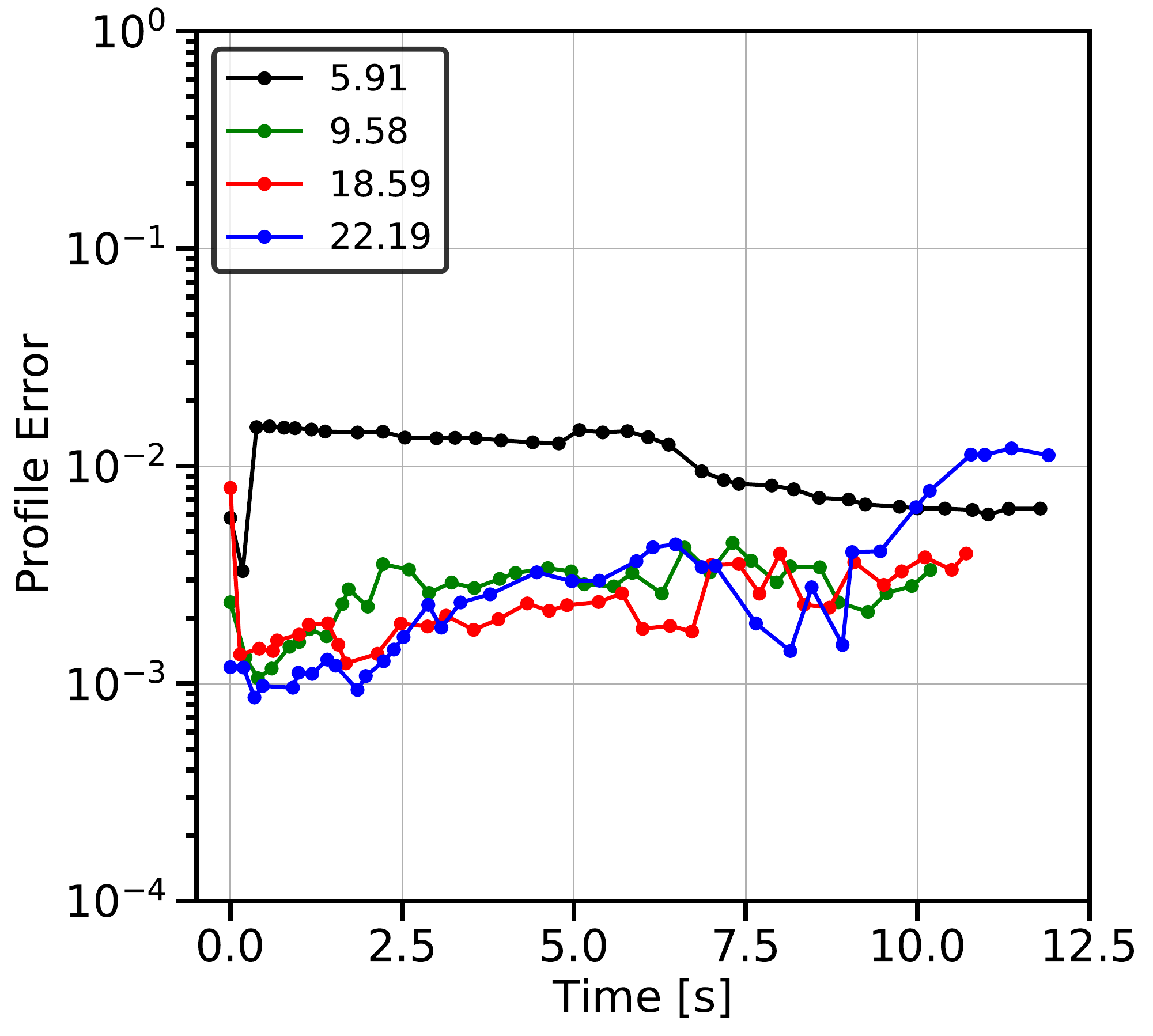}}
	\caption{Spatial error between the ground truth mask profiles and the predicted profiles from U-net (a) trained on only $5.91$, (b) trained on only $9.58$, (c) trained on only $18.59$, and (d) trained on only $22.19$ $[kg/m^2-s]$ RGB images.}
	\label{fig:errors_OXI}
\end{figure}

The ability of the neural network to make satisfactory fuel mask predictions independent of the oxidizer flux is investigated by training the network on only one flux data set and testing on the remaining three. This application is particularly useful to explore when the data for other oxidizer fluxes is to be processed without having access to the manually traced fuel masks. For these cases, due to the low number of training images and to avoid overfitting, an additional stop criterion is implemented: model training should stop when the model does not improve at predicting the remaining three fluxes for $20$ consecutive epochs. In total, four separately trained neural networks are produced and are used to predict all of the other oxidizer fluxes. \Cref{fig:errors_OXI} shows the total absolute spatial error, $(\sum |h - h_{truth}|/h_{truth})/width_{image}$ between the ground truth mask profiles and the profiles from each network's prediction for each point in time. Overall, each network outperforms each of the basic image processing techniques. The networks trained with the images from fluxes $5.91$, $9.58$, and $18.59$ $[kg/m^2-s]$, predict the $22.19$ $[kg/m^2-s]$ case with the largest error ($>10^{-2}$), particularly during the later stages of the fuel burn, most likely due to the presence of over-saturated images. However, the model trained on the data from the $22.19$ $[kg/m^2-s]$ flux (\cref{fig:errors_OXIml}) is overall better at capturing the profile for all the fluxes, as the profile error for each does not exceed $10^{-2}$ in any case.

\begin{figure}[H]
  \centering
  \subfloat[\label{fig:results_OXIa}]{\includegraphics[scale=0.32]{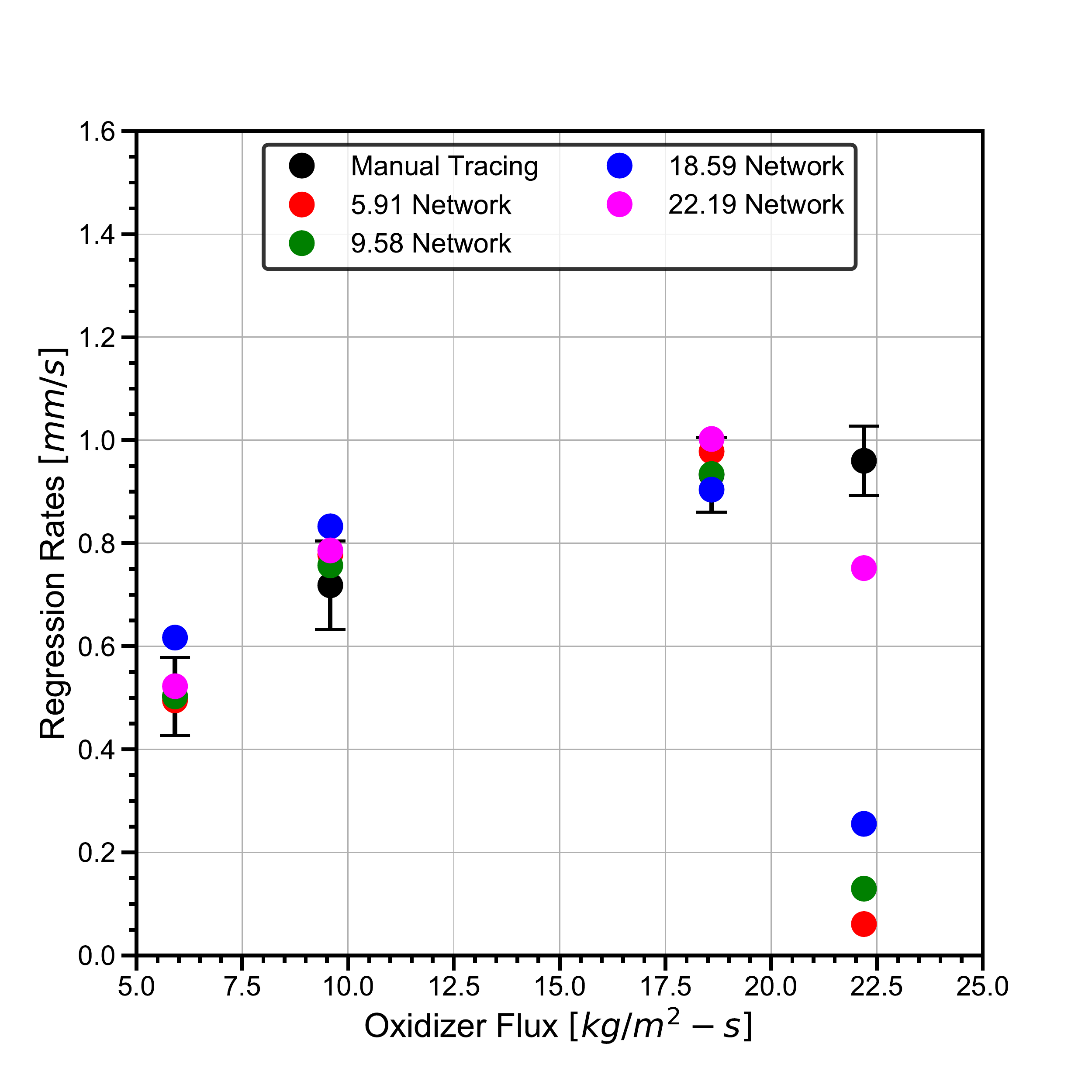}}
  \subfloat[\label{fig:results_OXIb}]{\includegraphics[scale=0.32]{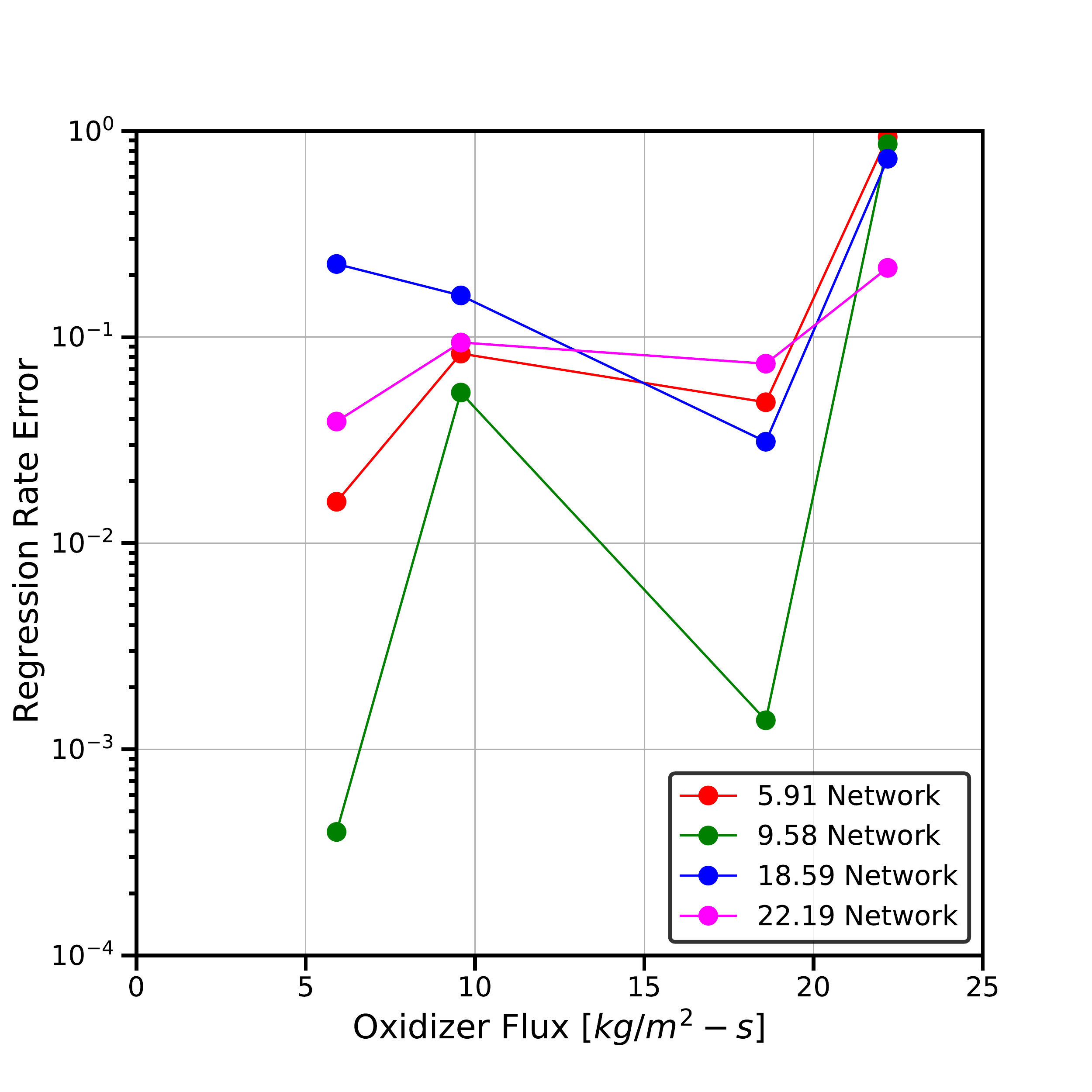}}\\
  \subfloat[\label{fig:results_OXIc}]{\includegraphics[scale=0.32]{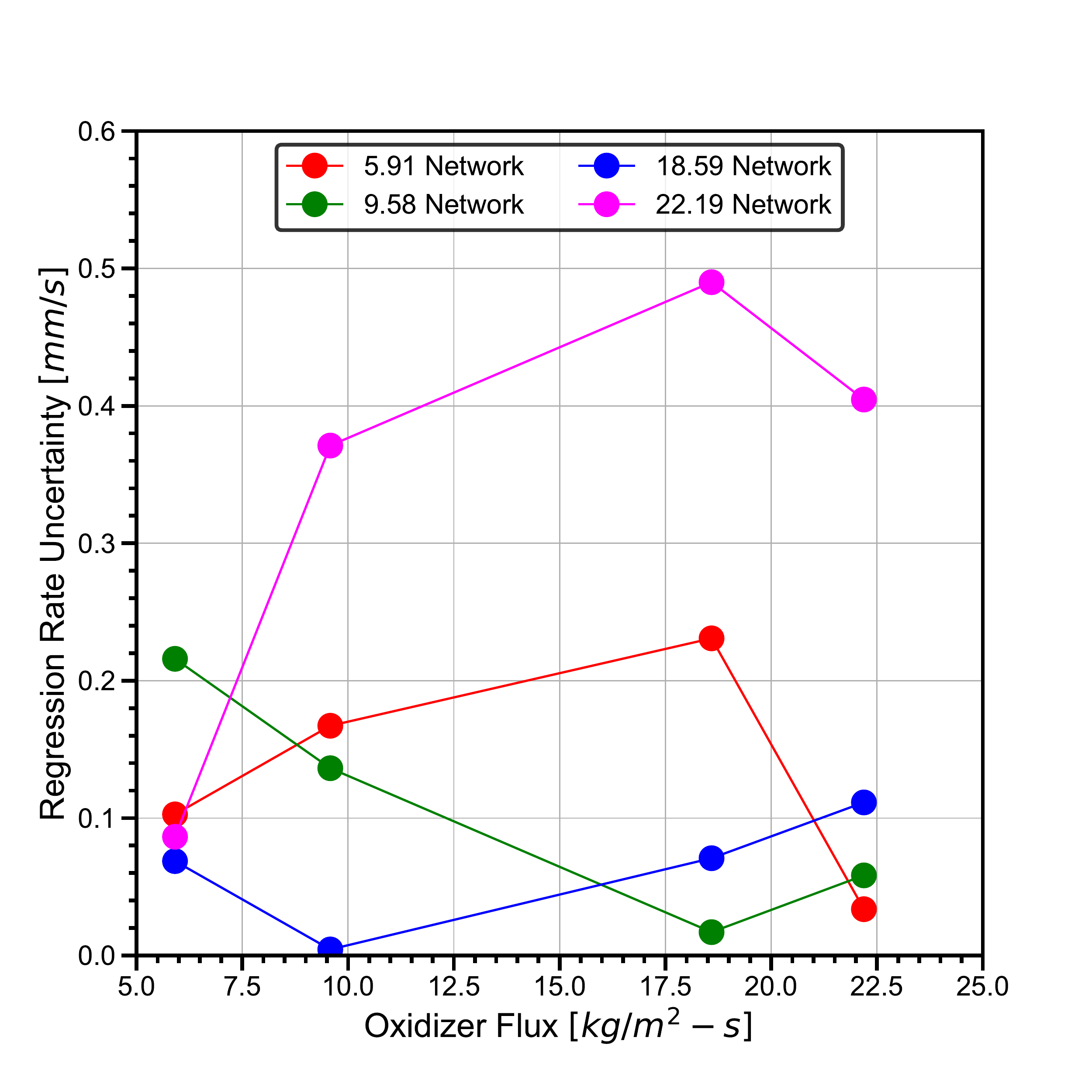}}
  \caption{(a) Regression rate measurements based on fuel mask predictions from the individual oxidizer flux U-net networks and comparison to published data from \cite{BAKSD20}, (b) the absolute error assuming the manual data from \cite{BAKSD20} as truth, and (c) the corresponding uncertainty in the fuel profile.}
\end{figure}

\Cref{fig:results_OXIa} shows the measurement of regression rates from the predicted fuel masks for each of the four networks. \Cref{fig:results_OXIb} shows the total absolute error, $|\dot{r}_{prediction} - \dot{r}_{truth}|/\dot{r}_{truth}$, assuming the manual tracing (represented as the black circles) from \cite{BAKSD20} as truth. \Cref{fig:results_OXIc} shows the uncertainty for each network, derived from the uncertainty map of each predicted fuel mask profile. The networks for $5.91$, $9.58$, and $18.59$ $[kg/m^2-s]$ predict the regression rate within first order accuracy for each other, but do not have satisfactory performance at predicting the regression rate for the $22.19$ $[kg/m^2-s]$ case. The network trained with the $22.19$ $[kg/m^2-s]$ images, has approximately a first order accuracy for all of the oxidizer fluxes as discussed before, however, it also has the overall highest uncertainty in the model. With monochrome data, it is possible to get good regression rate predictions, however order to get sufficiently accurate results the networks needs more training data. In such scenarios, the use of a high speed camera should provide enough data to train the network and obtain improved regression rate estimates.

\subsection{Application B: Monochrome Study}
\begin{figure}[H]
	\centering
	\subfloat[]{%
		\includegraphics[width=0.4\textwidth]{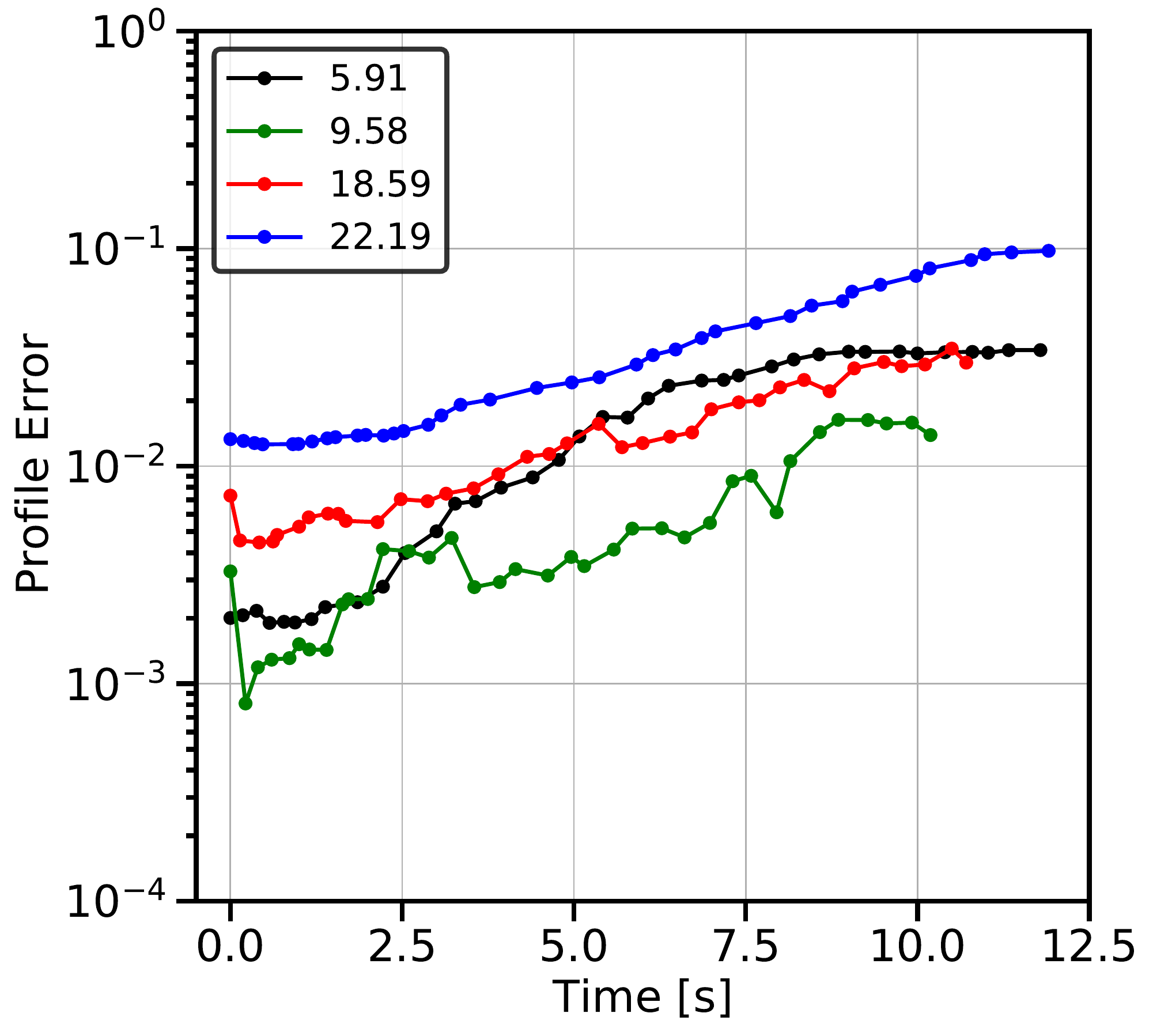}}
	\subfloat[]{%
		\includegraphics[width=0.4\textwidth]{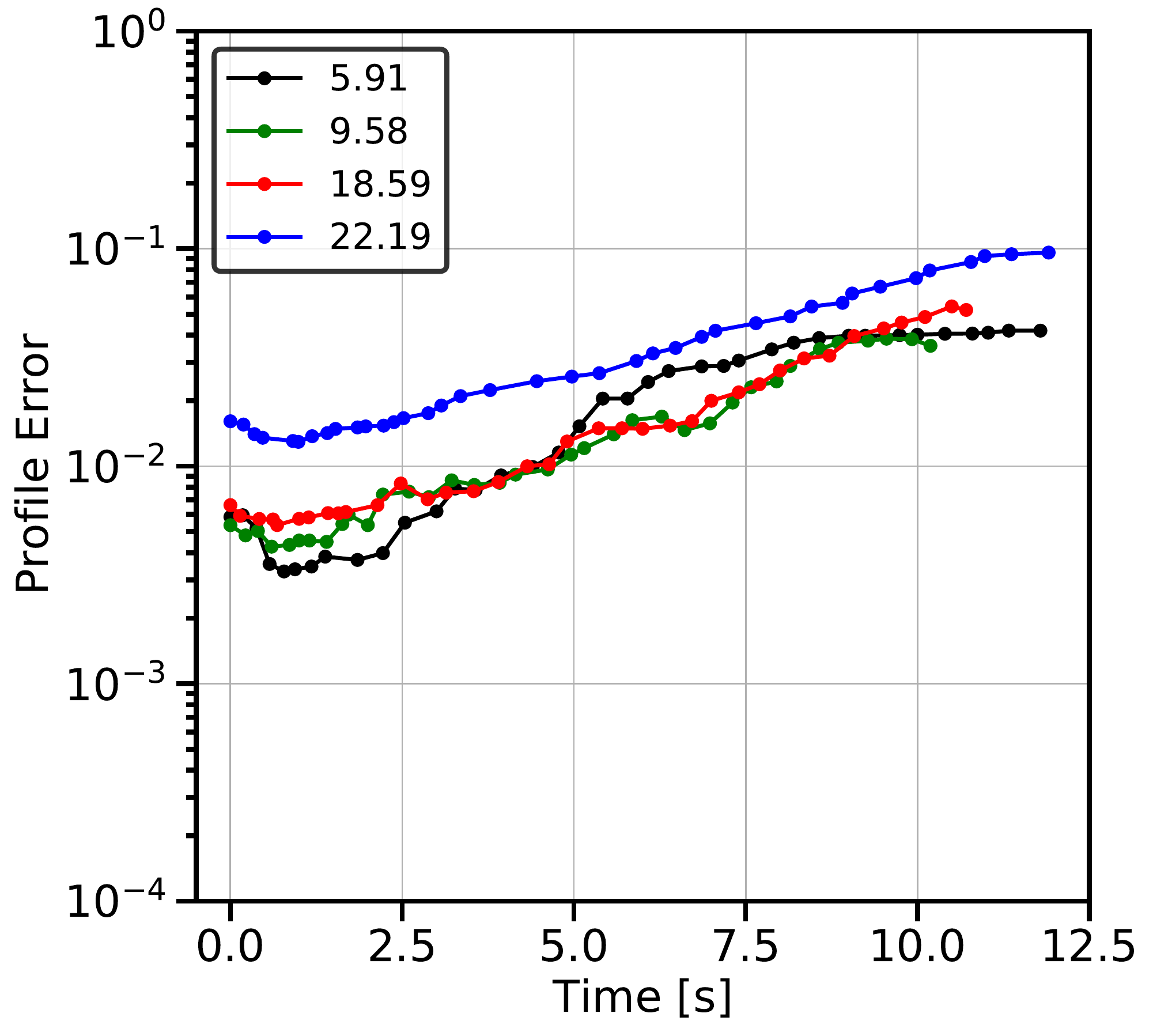}} \\
	\subfloat[]{%
		\includegraphics[width=0.4\textwidth]{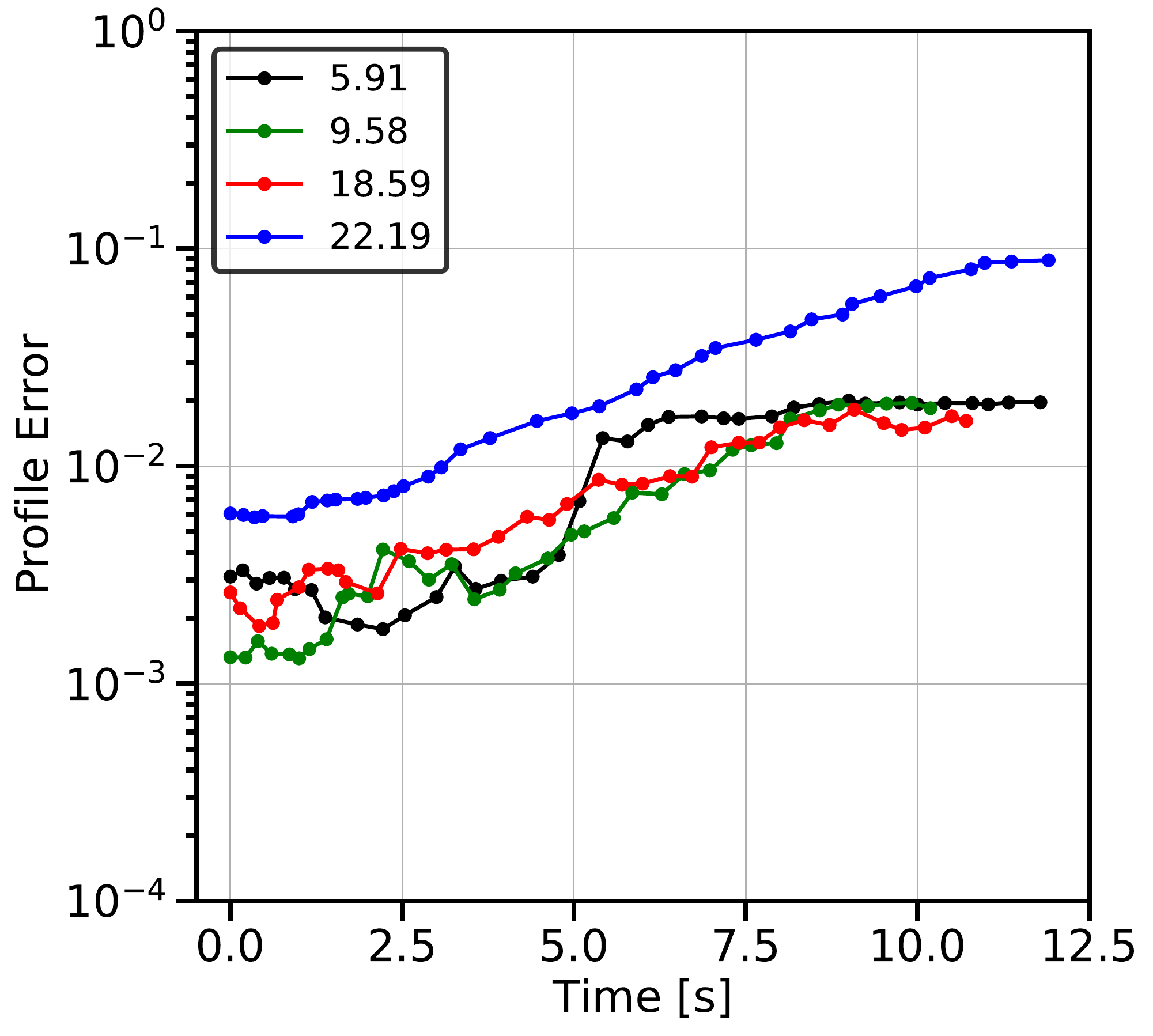}}
	\subfloat[\label{fig:errors_OXImlg}]{%
		\includegraphics[width=0.4\textwidth]{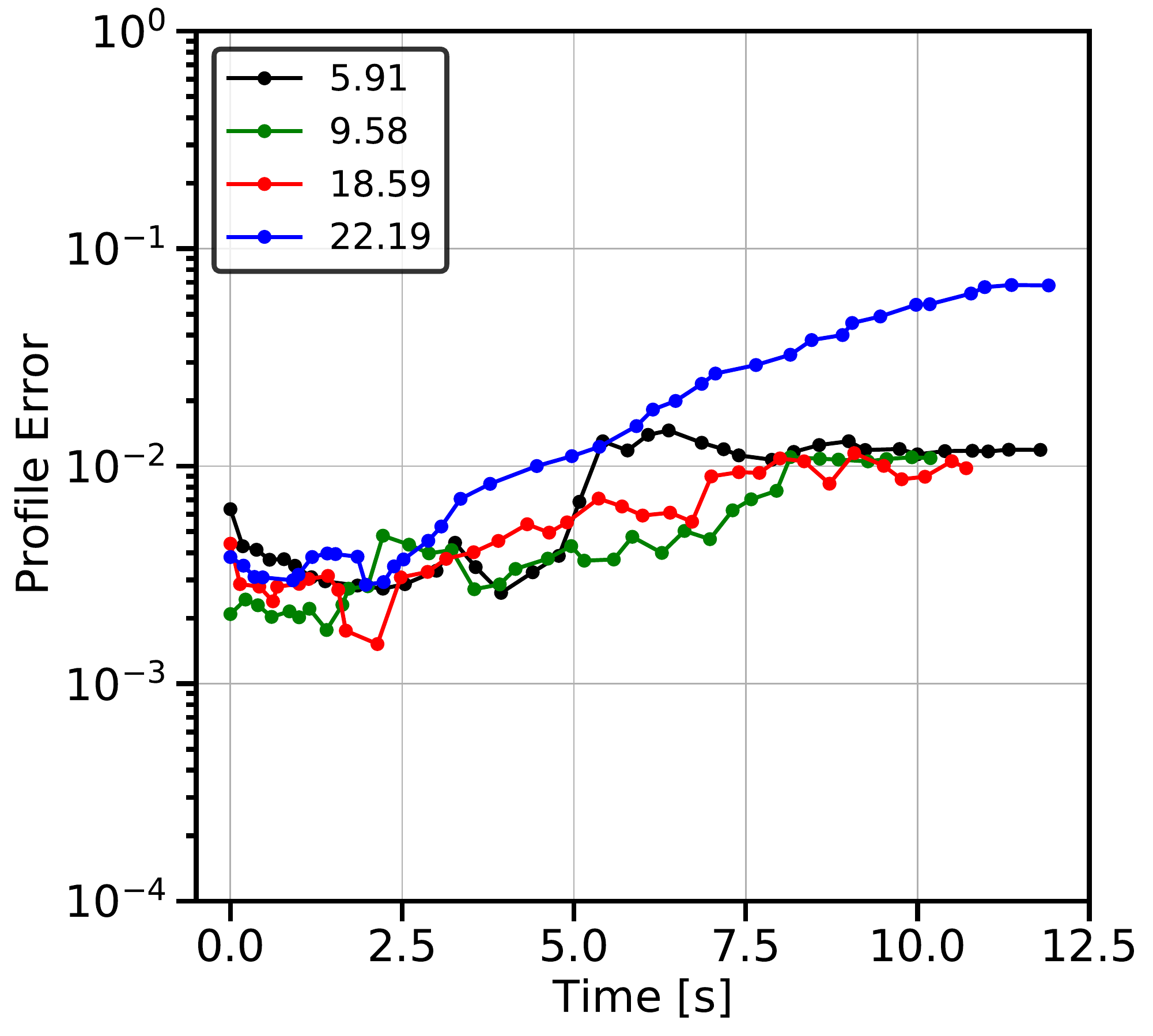}}
	\\
	\subfloat[]{%
		\includegraphics[width=0.4\textwidth]{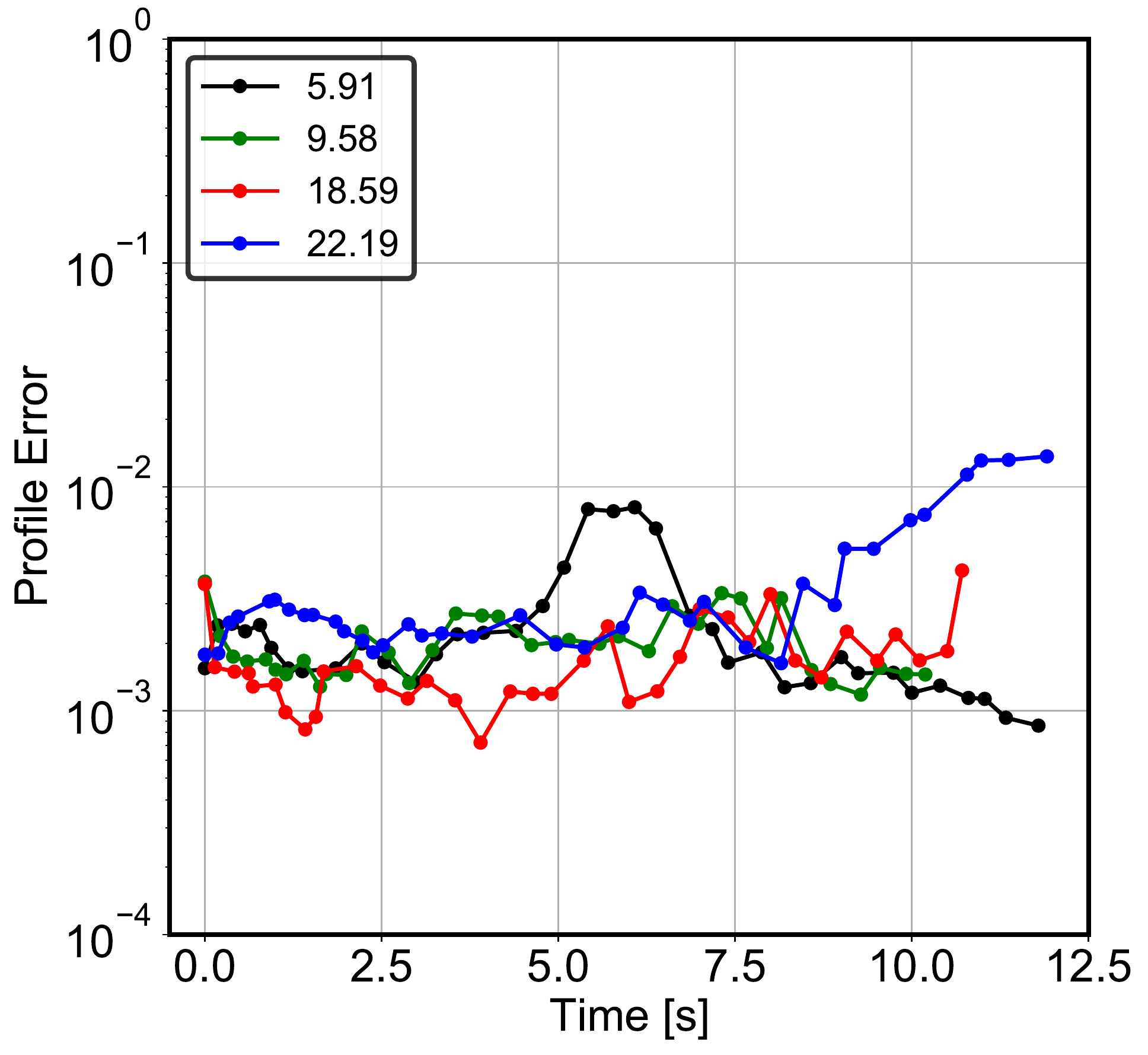}}
	\caption{Spatial error between the ground truth mask profiles and the predicted profiles from U-net (a) trained on only $5.91$, (b) trained on only $9.58$, (c) trained on only $18.59$, and (d) trained on only $22.19$ $[kg/m^2-s]$ grayscale images.}
	\label{fig:errors_OXIg}
\end{figure}

\begin{figure}[H]
  \centering
  \subfloat[\label{fig:image22}]{\includegraphics[width=0.3\textwidth]{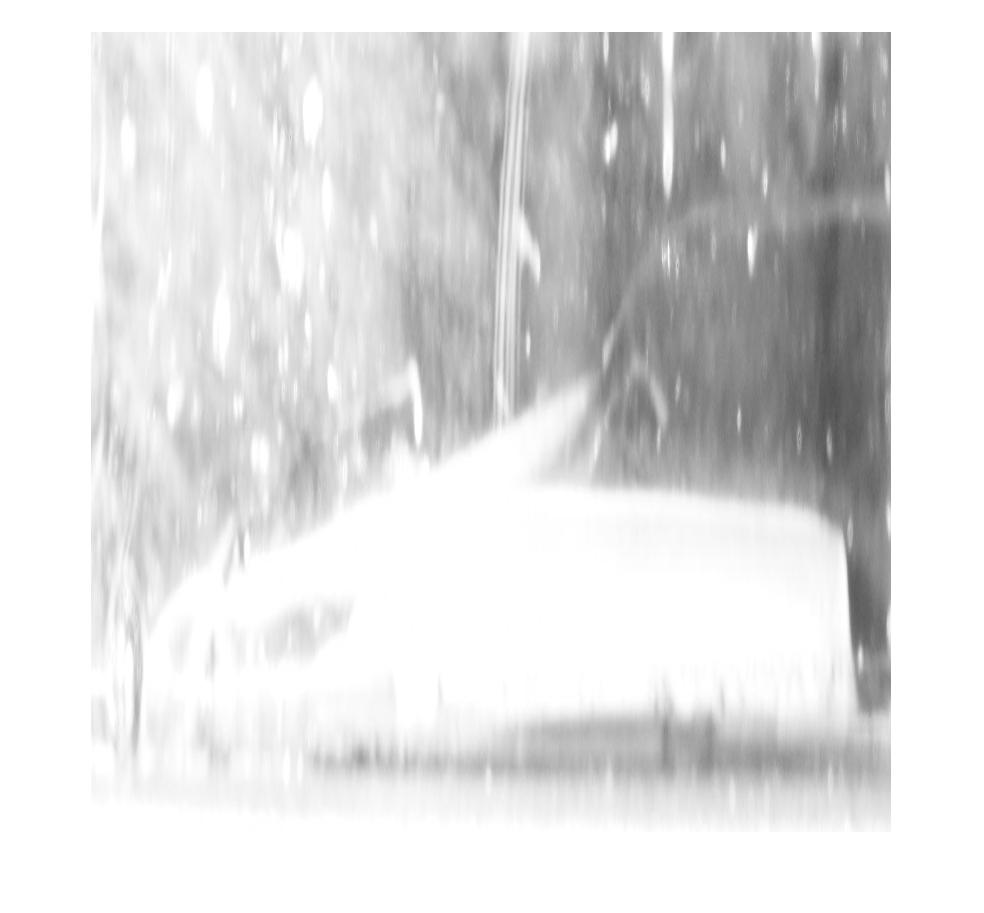}}
  \subfloat[\label{fig:mask22}]{\includegraphics[width=0.3\textwidth]{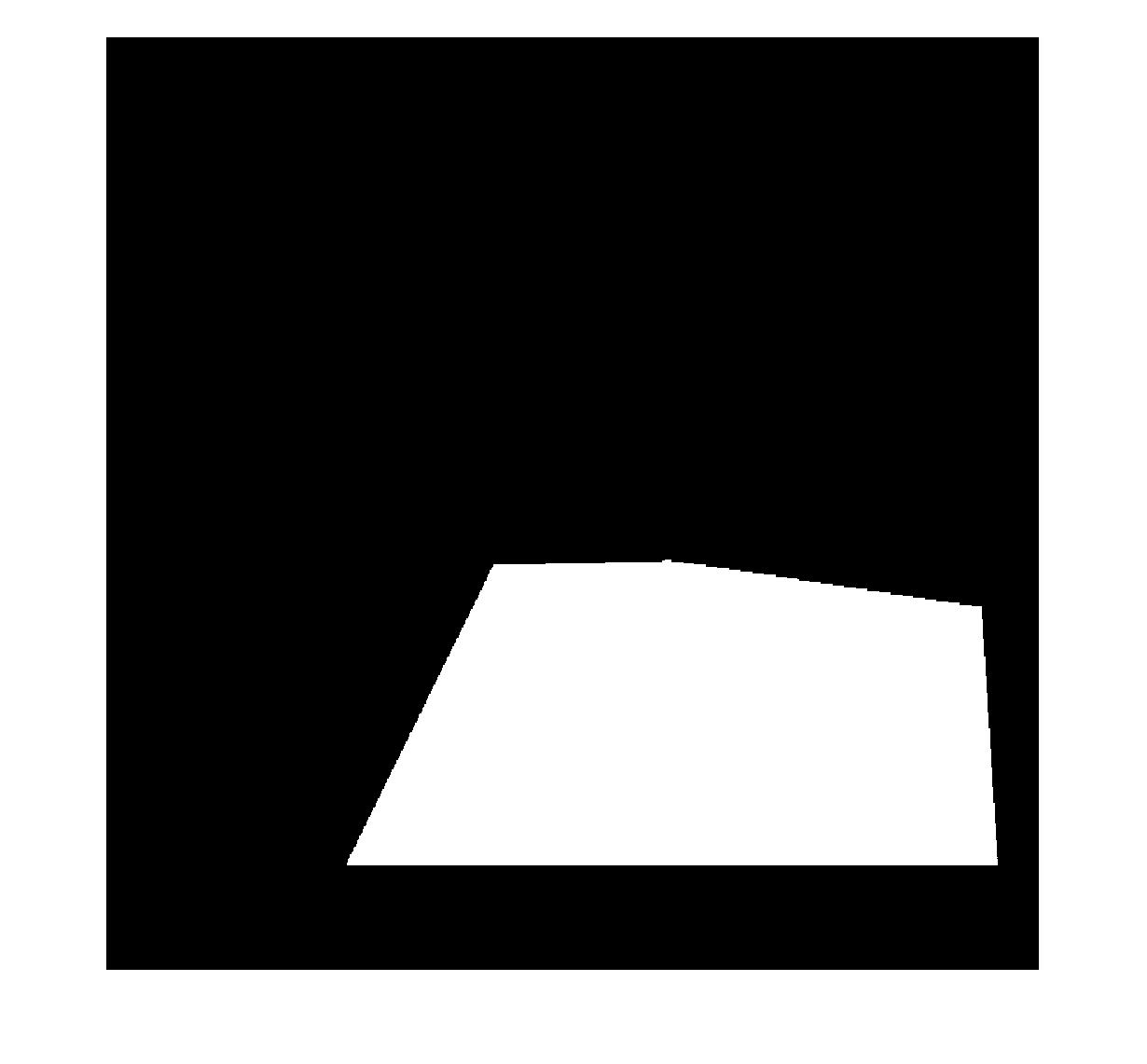}}
  \subfloat[\label{fig:pred22}]{\includegraphics[width=0.3\textwidth]{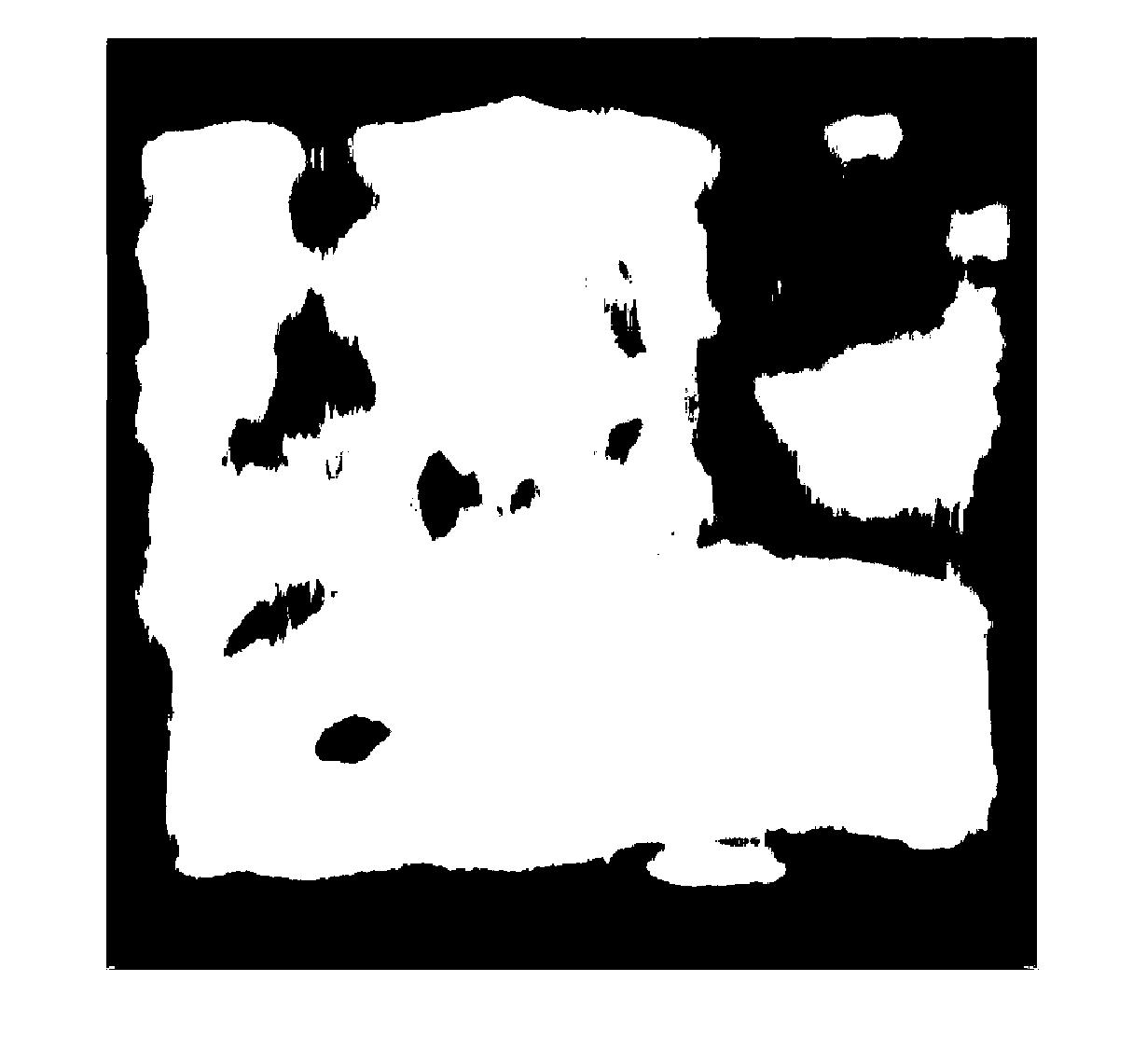}}
  \caption{An example of data from the $22.19$ $[kg/m^2-s]$ network predicting the $22.19$ $[kg/m^2-s]$ case where (a) is the resized grayscale image, (b) is the ground truth, and (c) is the fuel mask prediction.}
  \label{fig:22_22}
\end{figure}

To further explore the limits of satisfactory fuel mask predictions with the U-net, a study is performed on how the network performs when trained on lower depth image data. This study is useful in applications where only a monochrome camera is available, which costs a fraction compared to a color high speed camera. In this case, the experimental RGB images are first converted to grayscale images before using them as the training set. Similar to the individual flux cases, to avoid overfitting due to the small training set, the additional stop criterion to halt model training if it has not improved at prediction for $20$ consecutive epochs is implemented. \Cref{fig:errors_OXIg} shows the total absolute spatial error, $(\sum |h - h_{truth}|/h_{truth})/width_{image}$ between the ground truth masks profiles and the profiles from each network's prediction for each point in time. The networks trained with the $5.19$, $9.58$, and $18.59$ $[kg/m^2-s]$ grayscale images, have the highest error at correctly predicting fuel masks for the $22.19$ $[kg/m^2-s]$ case particularly at the later stages of fuel burn. The network trained with the $22.19$ $[kg/m^2-s]$ grayscale images, also has the highest error when predicting fuel masks for the $22.19$ $[kg/m^2-s]$ case and after inspection of the results (an example of the predicted mask is shown in \cref{fig:22_22}), there is significant noise most likely caused by the over-saturation in the experimental images for the $22.19$ $[kg/m^2-s]$ flux that are problematic for the network predictions. The network that is trained on all fluxes randomly chosen (similar to \cref{sec:all}) has the lowest error $<10^{-2}$ for all fluxes and experimental times.  

\begin{figure}[H]
  \centering
  \subfloat[\label{fig:results_OXIag}]{\includegraphics[scale=0.32]{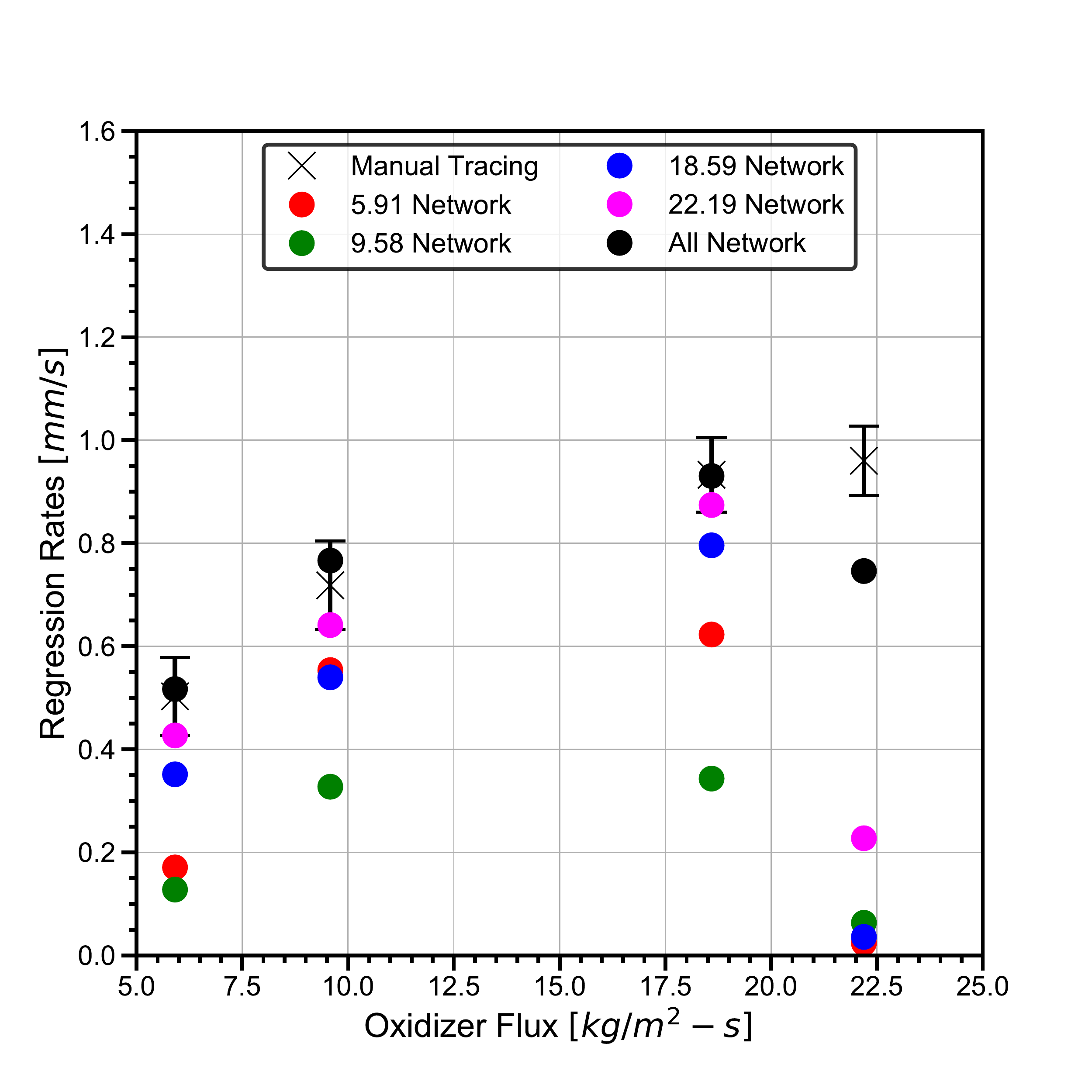}}
  \subfloat[\label{fig:results_OXIbg}]{\includegraphics[scale=0.32]{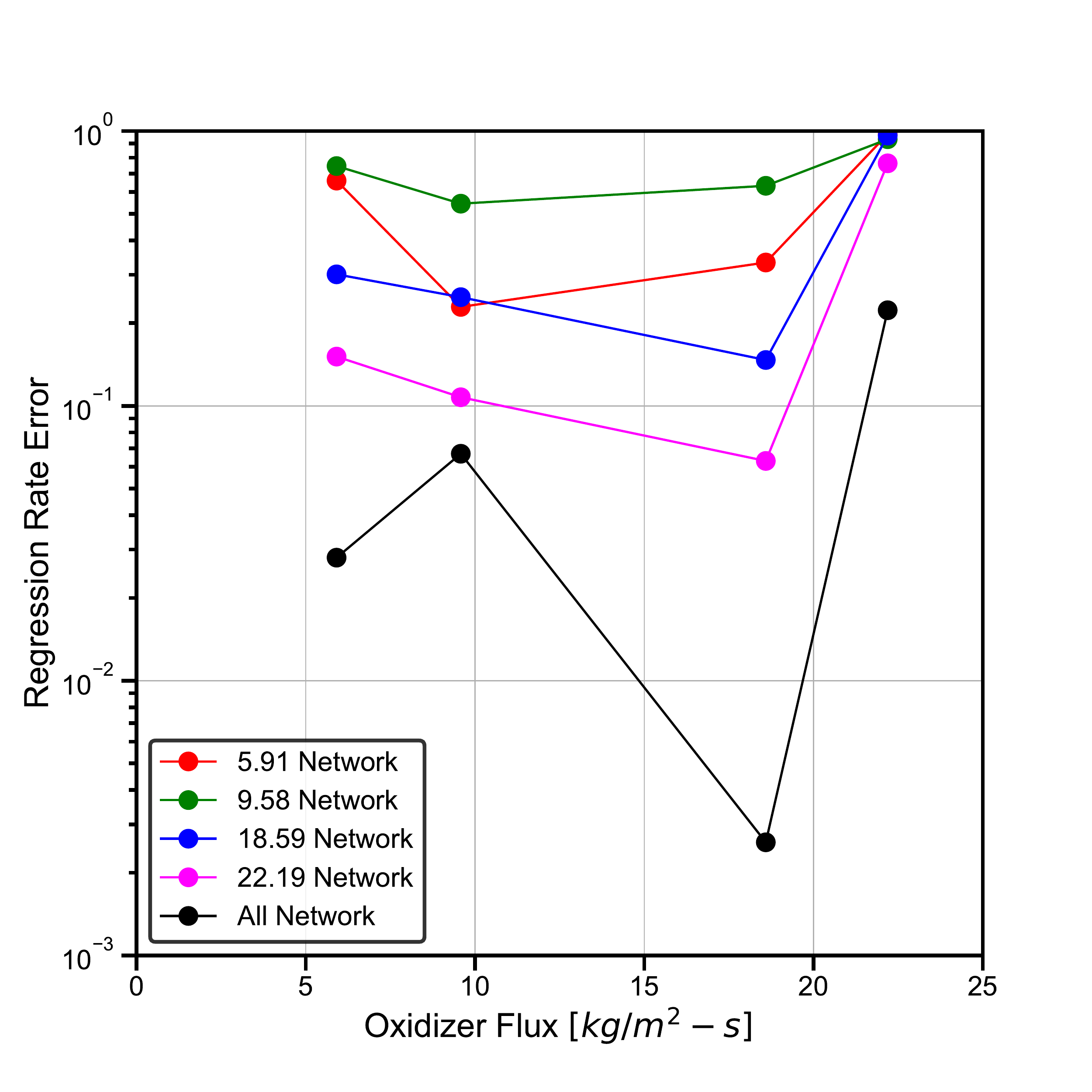}}\\
  \subfloat[\label{fig:results_OXIcg}]{\includegraphics[scale=0.32]{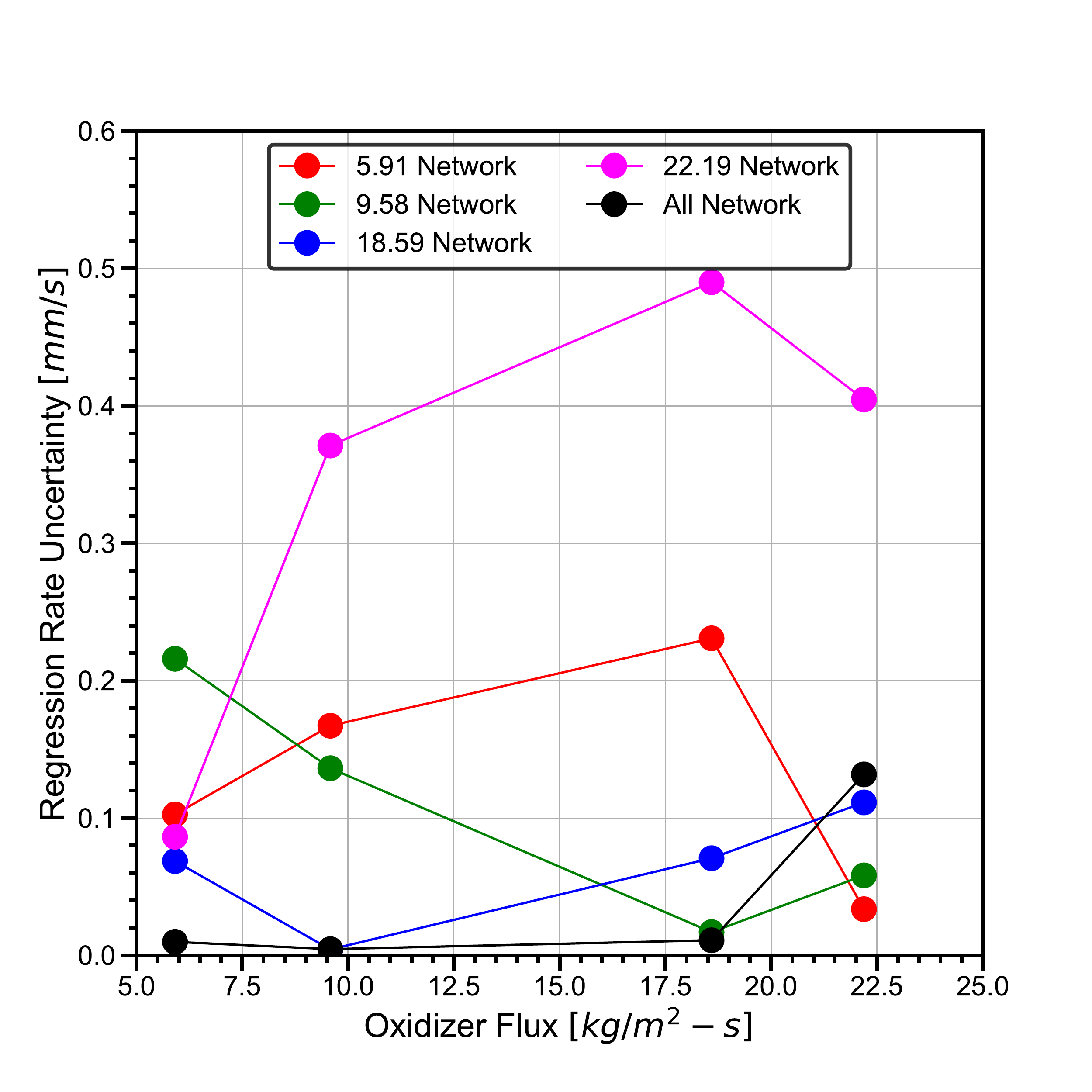}}
  \caption{(a) Grayscale network regression rate results comparison to published data from \cite{BAKSD20}, (b) the absolute error assuming the manual data from \cite{BAKSD20} as truth, and (c) the corresponding uncertainty.}
\end{figure}

\Cref{fig:results_OXIag} shows the comparison between the regression rate obtained from the networks trained on grayscale images from different oxidizer fluxes and the ground truth (manually drawn data from \cite{BAKSD20}). \Cref{fig:results_OXIbg} shows the total absolute error, $|\dot{r}_{prediction} - \dot{r}_{truth}|/\dot{r}_{truth}$, assuming the manual tracing (represented as the black circles), from \cite{BAKSD20} as truth. \Cref{fig:results_OXIcg} shows the uncertainty for each network. The results for the networks trained on grayscale images from $5.91$, $9.58$, and $18.59$ $[kg/m^2-s]$ all have over first order error and do not provide good regression rate results consistently. The network trained on $22.19$ $[kg/m^2-s]$ provides the best results out of the four oxidizer flux independently trained networks, however falls short on predicting the regression rate for its own data, likely due to the noise seen in \cref{fig:pred22}. Even the network trained on all the oxidizer fluxes has better results predicting fuel masks for $5.91$, $9.58$, and $18.59$ $[kg/m^2-s]$ with error less than $10^{-1}$ but still has difficulty predicting the $22.19$ $[kg/m^2-s]$ case. The networks trained with lower depth data, as expected, have higher error in predicting the regression rate compared to the ones trained with RGB data.

\section{Conclusion}
\label{sec:con}
In this study, for the first time, a CNN architecture, specifically, U-net is applied to measure the solid fuel regression rate in a $2$D slab burner hybrid rocket combustion experiment. The slab burner experiment consists of gaseous oxygen flowing through a rectangular combustion chamber with optical access on both sides. Paraffin wax is ignited in the chamber and is photographed continuously with a high intensity flash. The resulting images are high in noise caused by interference from wax, soot, pitting, and flame. Four different image segmentation techniques are applied and resulting regression rate error estimates are quantified. The U-net neural network is found to perform the best and the regression rate estimates are in very good agreement with the values from literature. The uncertainty in regression rate originating from the U-net architecture is quantified using a Monte Carlo Dropout method. The U-net architecture training is independent of the oxidizer mass flux and provides good estimates, even for noisy data. The U-net network performs well in situations where the data can be sparse, such as with monochrome images but the data must have a minimum level of detail to which features can be detected and patterns can be found. The proposed technique automates manual fuel mask segmentation and enables accurate regression rate estimates.



\section*{Acknowledgments}
\label{sec:Acknowledgment}
Funding: This work was supported by the United States Department of Energy’s (DoE) National Nuclear Security Administration (NNSA) under the Predictive Science Academic Alliance Program III (PSAAP III) at the University at Buffalo, under contract number DE-NA$0003961$. \\

Computational resources: The authors acknowledge the Tufts University High Performance Compute Cluster (https://it.tufts.edu/high-performance-computing) which was used for training and evaluating the U-net models.

\clearpage
\bibliography{bib}

\begin{thebibliography}{54}
\expandafter\ifx\csname natexlab\endcsname\relax\def\natexlab#1{#1}\fi
\providecommand{\url}[1]{\texttt{#1}}
\providecommand{\href}[2]{#2}
\providecommand{\path}[1]{#1}
\providecommand{\DOIprefix}{doi:}
\providecommand{\ArXivprefix}{arXiv:}
\providecommand{\URLprefix}{URL: }
\providecommand{\Pubmedprefix}{pmid:}
\providecommand{\doi}[1]{\href{http://dx.doi.org/#1}{\path{#1}}}
\providecommand{\Pubmed}[1]{\href{pmid:#1}{\path{#1}}}
\providecommand{\bibinfo}[2]{#2}
\ifx\xfnm\relax \def\xfnm[#1]{\unskip,\space#1}\fi
\bibitem[{Abdar et~al.(2021)Abdar, Pourpanah, Hussain, Rezazadegan, Liu,
  Ghavamzadeh, Fieguth, Cao, Khosravi, Acharya, Makarenkov and
  Nahavandi}]{Abdar2021}
\bibinfo{author}{Abdar, M.}, \bibinfo{author}{Pourpanah, F.},
  \bibinfo{author}{Hussain, S.}, \bibinfo{author}{Rezazadegan, D.},
  \bibinfo{author}{Liu, L.}, \bibinfo{author}{Ghavamzadeh, M.},
  \bibinfo{author}{Fieguth, P.}, \bibinfo{author}{Cao, X.},
  \bibinfo{author}{Khosravi, A.}, \bibinfo{author}{Acharya, U.R.},
  \bibinfo{author}{Makarenkov, V.}, \bibinfo{author}{Nahavandi, S.},
  \bibinfo{year}{2021}.
\newblock \bibinfo{title}{A review of uncertainty quantification in deep
  learning: Techniques, applications and challenges}.
\newblock \bibinfo{journal}{Inf. Fusion} \bibinfo{volume}{76},
  \bibinfo{pages}{243--297}.
\newblock \DOIprefix\doi{10.1016/j.inffus.2021.05.008}.
\bibitem[{Arlot and Celisse(2010)}]{Arlot2010}
\bibinfo{author}{Arlot, S.}, \bibinfo{author}{Celisse, A.},
  \bibinfo{year}{2010}.
\newblock \bibinfo{title}{{A survey of cross-validation procedures for model
  selection}}.
\newblock \bibinfo{journal}{Stat. Surv.} \bibinfo{volume}{4},
  \bibinfo{pages}{40 -- 79}.
\newblock \DOIprefix\doi{10.1214/09-SS054}.
\bibitem[{Bochkov and Kataeva(2021)}]{BK21}
\bibinfo{author}{Bochkov, V.S.}, \bibinfo{author}{Kataeva, L.Y.},
  \bibinfo{year}{2021}.
\newblock \bibinfo{title}{{wUUNet}: Advanced fully convolutional neural network
  for multiclass fire segmentation}.
\newblock \bibinfo{journal}{Symmetry} \bibinfo{volume}{13},
  \bibinfo{pages}{98}.
\newblock \DOIprefix\doi{10.3390/sym13010098}.
\bibitem[{Budzinski et~al.(2020)Budzinski, Aphale, Ismael, Surina and
  DesJardin}]{BAKSD20}
\bibinfo{author}{Budzinski, K.}, \bibinfo{author}{Aphale, S.S.},
  \bibinfo{author}{Ismael, E.K.}, \bibinfo{author}{Surina, G.},
  \bibinfo{author}{DesJardin, P.E.}, \bibinfo{year}{2020}.
\newblock \bibinfo{title}{Radiation heat transfer in ablating boundary layer
  combustion theory used for hybrid rocket motor analysis}.
\newblock \bibinfo{journal}{Combust. Flame} \bibinfo{volume}{217},
  \bibinfo{pages}{248--261}.
\newblock \DOIprefix\doi{10.1016/j.combustflame.2020.04.011}.
\bibitem[{Buslaev et~al.(2020)Buslaev, Iglovikov, Khvedchenya, Parinov,
  Druzhinin and Kalinin}]{A}
\bibinfo{author}{Buslaev, A.}, \bibinfo{author}{Iglovikov, V.I.},
  \bibinfo{author}{Khvedchenya, E.}, \bibinfo{author}{Parinov, A.},
  \bibinfo{author}{Druzhinin, M.}, \bibinfo{author}{Kalinin, A.A.},
  \bibinfo{year}{2020}.
\newblock \bibinfo{title}{Albumentations: fast and flexible image
  augmentations}.
\newblock \bibinfo{journal}{Information} \bibinfo{volume}{11},
  \bibinfo{pages}{125}.
\bibitem[{Cai et~al.(2013)Cai, Zeng, Li, Tian and Yu}]{Cai2013}
\bibinfo{author}{Cai, G.}, \bibinfo{author}{Zeng, P.}, \bibinfo{author}{Li,
  X.}, \bibinfo{author}{Tian, H.}, \bibinfo{author}{Yu, N.},
  \bibinfo{year}{2013}.
\newblock \bibinfo{title}{Scale effect of fuel regression rate in hybrid rocket
  motor}.
\newblock \bibinfo{journal}{Aerosp. Sci. Technol.} \bibinfo{volume}{24},
  \bibinfo{pages}{141--146}.
\newblock \DOIprefix\doi{10.1016/j.ast.2011.11.001}.
\bibitem[{Carmicino and Sorge(2006)}]{Carmicino2006}
\bibinfo{author}{Carmicino, C.}, \bibinfo{author}{Sorge, A.R.},
  \bibinfo{year}{2006}.
\newblock \bibinfo{title}{Influence of a conical axial injector on hybrid
  rocket performance}.
\newblock \bibinfo{journal}{J. Propula. Power} \bibinfo{volume}{22},
  \bibinfo{pages}{984--995}.
\newblock \DOIprefix\doi{10.2514/1.19528}.
\bibitem[{Chauhan et~al.(2018)Chauhan, Ghanshala and Joshi}]{Chauhan2018}
\bibinfo{author}{Chauhan, R.}, \bibinfo{author}{Ghanshala, K.K.},
  \bibinfo{author}{Joshi, R.}, \bibinfo{year}{2018}.
\newblock \bibinfo{title}{Convolutional neural network ({CNN}) for image
  detection and recognition}, in: \bibinfo{booktitle}{2018 First Int. Conf.
  Secure Cyber Comput. Commun. ({ICSCCC})}, \bibinfo{publisher}{{IEEE}}.
\newblock \DOIprefix\doi{10.1109/icsccc.2018.8703316}.
\bibitem[{Chiaverini et~al.(2001)Chiaverini, Kuo, Peretz and
  Harting}]{Chiaverini2001}
\bibinfo{author}{Chiaverini, M.J.}, \bibinfo{author}{Kuo, K.K.},
  \bibinfo{author}{Peretz, A.}, \bibinfo{author}{Harting, G.C.},
  \bibinfo{year}{2001}.
\newblock \bibinfo{title}{Regression-rate and heat-transfer correlations for
  hybrid rocket combustion}.
\newblock \bibinfo{journal}{J. Propuls. Power} \bibinfo{volume}{17},
  \bibinfo{pages}{99--110}.
\newblock \DOIprefix\doi{10.2514/2.5714}.
\bibitem[{Chiaverini et~al.(2000)Chiaverini, Serin, Johnson, Lu, Kuo and
  Risha}]{Chiaverini2000}
\bibinfo{author}{Chiaverini, M.J.}, \bibinfo{author}{Serin, N.},
  \bibinfo{author}{Johnson, D.K.}, \bibinfo{author}{Lu, Y.C.},
  \bibinfo{author}{Kuo, K.K.}, \bibinfo{author}{Risha, G.A.},
  \bibinfo{year}{2000}.
\newblock \bibinfo{title}{Regression rate behavior of hybrid rocket solid
  fuels}.
\newblock \bibinfo{journal}{J. Propuls. Power} \bibinfo{volume}{16},
  \bibinfo{pages}{125--132}.
\newblock \DOIprefix\doi{10.2514/2.5541}.
\bibitem[{Chollet et~al.(2015)}]{chollet2015keras}
\bibinfo{author}{Chollet, F.}, et~al., \bibinfo{year}{2015}.
\newblock \bibinfo{title}{Keras}.
\newblock \bibinfo{howpublished}{\url{https://keras.io}}.
\bibitem[{Chowdhary and Acharjya(2020)}]{Chowdhary2020}
\bibinfo{author}{Chowdhary, C.L.}, \bibinfo{author}{Acharjya, D.},
  \bibinfo{year}{2020}.
\newblock \bibinfo{title}{Segmentation and feature extraction in medical
  imaging: A systematic review}.
\newblock \bibinfo{journal}{Procedia Comput. Sci.} \bibinfo{volume}{167},
  \bibinfo{pages}{26--36}.
\newblock \DOIprefix\doi{10.1016/j.procs.2020.03.179}.
\bibitem[{Devillers et~al.(2019)Devillers, Nugue, Chan-Hon-Tong, Le~Besnerais
  and Pichilloup}]{DNCBP19}
\bibinfo{author}{Devillers, R.}, \bibinfo{author}{Nugue, M.},
  \bibinfo{author}{Chan-Hon-Tong, A.}, \bibinfo{author}{Le~Besnerais, G.},
  \bibinfo{author}{Pichilloup, J.}, \bibinfo{year}{2019}.
\newblock \bibinfo{title}{Experimental analysis of aluminum-droplet combustion
  in solid-propellant conditions using deep learning}.
\newblock \bibinfo{journal}{Proc. 8th Eur. Conf. Aeronaut. Sp. Sci.}
  \DOIprefix\doi{10.13009/EUCASS2019-593}.
\bibitem[{DeVries and Taylor(2018)}]{devries2018leveraging}
\bibinfo{author}{DeVries, T.}, \bibinfo{author}{Taylor, G.W.},
  \bibinfo{year}{2018}.
\newblock \bibinfo{title}{Leveraging uncertainty estimates for predicting
  segmentation quality}.
\newblock \bibinfo{journal}{CoRR} \href{http://arxiv.org/abs/1807.00502}{{\tt
  arXiv:1807.00502}}.
\bibitem[{Dunn et~al.(2018)Dunn, Gustafson, Edwards, Dunbrack and
  Johansen}]{DGEDJ18}
\bibinfo{author}{Dunn, C.}, \bibinfo{author}{Gustafson, G.},
  \bibinfo{author}{Edwards, J.}, \bibinfo{author}{Dunbrack, T.},
  \bibinfo{author}{Johansen, C.}, \bibinfo{year}{2018}.
\newblock \bibinfo{title}{Spatially and temporally resolved regression rate
  measurements for the combustion of paraffin wax for hybrid rocket motor
  applications}.
\newblock \bibinfo{journal}{Aerosp. Sci. and Technol.} \bibinfo{volume}{72},
  \bibinfo{pages}{371--379}.
\newblock \DOIprefix\doi{10.1016/j.ast.2017.11.024}.
\bibitem[{Elmaz et~al.(2020)Elmaz, B\"{u}y\"{u}k{\c{c}}ak{\i}r, \"{O}zg\"{u}n
  Y\"{u}cel and Mutlu}]{EBYM20}
\bibinfo{author}{Elmaz, F.}, \bibinfo{author}{B\"{u}y\"{u}k{\c{c}}ak{\i}r, B.},
  \bibinfo{author}{\"{O}zg\"{u}n Y\"{u}cel}, \bibinfo{author}{Mutlu, A.Y.},
  \bibinfo{year}{2020}.
\newblock \bibinfo{title}{Classification of solid fuels with machine learning}.
\newblock \bibinfo{journal}{Fuel} \bibinfo{volume}{266},
  \bibinfo{pages}{117066}.
\newblock \DOIprefix\doi{10.1016/j.fuel.2020.117066}.
\bibitem[{Gallo et~al.(2021)Gallo, Mungiguerra and Savino}]{Gallo2021}
\bibinfo{author}{Gallo, G.}, \bibinfo{author}{Mungiguerra, S.},
  \bibinfo{author}{Savino, R.}, \bibinfo{year}{2021}.
\newblock \bibinfo{title}{New entrainment model for modelling the regression
  rate in hybrid rocket engines}.
\newblock \bibinfo{journal}{J. Propuls. Power}
  \DOIprefix\doi{10.2514/1.B38333}.
\bibitem[{Ghosh et~al.(2019)Ghosh, Pal, Jaiswal, Santosh, Das and
  Nasipuri}]{Ghosh2019}
\bibinfo{author}{Ghosh, S.}, \bibinfo{author}{Pal, A.},
  \bibinfo{author}{Jaiswal, S.}, \bibinfo{author}{Santosh, K.C.},
  \bibinfo{author}{Das, N.}, \bibinfo{author}{Nasipuri, M.},
  \bibinfo{year}{2019}.
\newblock \bibinfo{title}{{SegFast}-v2: Semantic image segmentation with less
  parameters in deep learning for autonomous driving}.
\newblock \bibinfo{journal}{Int. J. Mach. Learn. \& Cyber.}
  \bibinfo{volume}{10}, \bibinfo{pages}{3145--3154}.
\newblock \DOIprefix\doi{10.1007/s13042-019-01005-5}.
\bibitem[{Haas et~al.(2020)Haas, Schubert, Eickhoff and Pfeifer}]{HSEP20}
\bibinfo{author}{Haas, T.}, \bibinfo{author}{Schubert, C.},
  \bibinfo{author}{Eickhoff, M.}, \bibinfo{author}{Pfeifer, H.},
  \bibinfo{year}{2020}.
\newblock \bibinfo{title}{{BubCNN}: Bubble detection using faster {RCNN} and
  shape regression network}.
\newblock \bibinfo{journal}{Chem. Eng. Sci.} \bibinfo{volume}{216},
  \bibinfo{pages}{115467}.
\newblock \DOIprefix\doi{10.1016/j.ces.2019.115467}.
\bibitem[{Hastie et~al.(2009)Hastie, Tibshirani and Friedman}]{Hastie2009}
\bibinfo{author}{Hastie, T.}, \bibinfo{author}{Tibshirani, R.},
  \bibinfo{author}{Friedman, J.}, \bibinfo{year}{2009}.
\newblock \bibinfo{title}{The Elements of Statistical Learning}.
  \bibinfo{publisher}{Springer New York}.
\newblock p. \bibinfo{pages}{309}.
\newblock \DOIprefix\doi{10.1007/978-0-387-84858-7}.
\bibitem[{Hirata et~al.(2011)Hirata, Aso, Hayashida, Nakawatase, Tani,
  Morishita and Shimada}]{Hirata2011}
\bibinfo{author}{Hirata, Y.}, \bibinfo{author}{Aso, S.},
  \bibinfo{author}{Hayashida, T.}, \bibinfo{author}{Nakawatase, R.},
  \bibinfo{author}{Tani, Y.}, \bibinfo{author}{Morishita, K.},
  \bibinfo{author}{Shimada, T.}, \bibinfo{year}{2011}.
\newblock \bibinfo{title}{Improvement of regression rate and combustion
  efficiency of high density polyethylene fuel and paraffin fuel of hybrid
  rockets with multi-section swirl injection method}, in:
  \bibinfo{booktitle}{47th {AIAA}/{ASME}/{SAE}/{ASEE} Jt. Propuls. Conf.
  Exhib.}, \bibinfo{publisher}{AIAA}.
\newblock \DOIprefix\doi{10.2514/6.2011-5907}.
\bibitem[{Ioffe and Szegedy(2015)}]{IS15}
\bibinfo{author}{Ioffe, S.}, \bibinfo{author}{Szegedy, C.},
  \bibinfo{year}{2015}.
\newblock \bibinfo{title}{Batch normalization: Accelerating deep network
  training by reducing internal covariate shift}.
\newblock \bibinfo{journal}{Proc. 32nd Int. Conf. Mach. Learning}
  \bibinfo{volume}{37}, \bibinfo{pages}{448--456}.
\newblock \URLprefix \url{http://proceedings.mlr.press/v37/ioffe15.html}.
\bibitem[{ITU-R(2011)}]{RGB2011}
\bibinfo{author}{ITU-R}, \bibinfo{year}{2011}.
\newblock \bibinfo{title}{Studio encoding parameters of digital television for
  standard 4:3 and wide-screen 16:9 aspect ratios}.
\newblock \bibinfo{type}{techreport} \bibinfo{number}{Rec. ITU-R BT.601-7}.
\bibitem[{Jens et~al.(2016)Jens, Cantwell and Hubbard}]{Jens2016}
\bibinfo{author}{Jens, E.T.}, \bibinfo{author}{Cantwell, B.J.},
  \bibinfo{author}{Hubbard, G.S.}, \bibinfo{year}{2016}.
\newblock \bibinfo{title}{Hybrid rocket propulsion systems for outer planet
  exploration missions}.
\newblock \bibinfo{journal}{Acta Astronaut.} \bibinfo{volume}{128},
  \bibinfo{pages}{119--130}.
\newblock \DOIprefix\doi{10.1016/j.actaastro.2016.06.036}.
\bibitem[{Jiao et~al.(2020)Jiao, Huo, Hu and Tang}]{JHHT20}
\bibinfo{author}{Jiao, L.}, \bibinfo{author}{Huo, L.}, \bibinfo{author}{Hu,
  C.}, \bibinfo{author}{Tang, P.}, \bibinfo{year}{2020}.
\newblock \bibinfo{title}{Refined {UNet}: {UNet}-based refinement network for
  cloud and shadow precise segmentation}.
\newblock \bibinfo{journal}{Remote Sens.} \bibinfo{volume}{12},
  \bibinfo{pages}{2001}.
\newblock \DOIprefix\doi{10.3390/rs12122001}.
\bibitem[{Karabeyoglu et~al.(2003)Karabeyoglu, Zilliac, Cantwell, Zilwa and
  Castellucci}]{Karabeyoglu2003}
\bibinfo{author}{Karabeyoglu, A.}, \bibinfo{author}{Zilliac, G.},
  \bibinfo{author}{Cantwell, B.}, \bibinfo{author}{Zilwa, S.D.},
  \bibinfo{author}{Castellucci, P.}, \bibinfo{year}{2003}.
\newblock \bibinfo{title}{Scale-up tests of high regression rate liquefying
  hybrid rocket fuels}, in: \bibinfo{booktitle}{41st Aerosp. Sci. Meet.
  Exhib.}, \bibinfo{publisher}{AIAA}.
\newblock \DOIprefix\doi{10.2514/6.2003-1162}.
\bibitem[{Karabeyoglu et~al.(2001)Karabeyoglu, Cantwell and
  Altman}]{Karabeyoglu2001}
\bibinfo{author}{Karabeyoglu, M.}, \bibinfo{author}{Cantwell, B.},
  \bibinfo{author}{Altman, D.}, \bibinfo{year}{2001}.
\newblock \bibinfo{title}{Development and testing of paraffin-based hybrid
  rocket fuels}, in: \bibinfo{booktitle}{37th Jt. Propuls. Conf. Exhib.},
  \bibinfo{publisher}{AIAA}.
\newblock \DOIprefix\doi{10.2514/6.2001-4503}.
\bibitem[{Karabeyoglu(1998)}]{Karabeyoglu98}
\bibinfo{author}{Karabeyoglu, M.A.}, \bibinfo{year}{1998}.
\newblock \bibinfo{title}{Transient Combustion in Hybrid Rockets}.
\newblock Ph.D. thesis. Stanford University.
\bibitem[{Keskar et~al.(2016)Keskar, Mudigere, Nocedal, Smelyanskiy and
  Tang}]{keskarbatch}
\bibinfo{author}{Keskar, N.S.}, \bibinfo{author}{Mudigere, D.},
  \bibinfo{author}{Nocedal, J.}, \bibinfo{author}{Smelyanskiy, M.},
  \bibinfo{author}{Tang, P.T.P.}, \bibinfo{year}{2016}.
\newblock \bibinfo{title}{On large-batch training for deep learning:
  Generalization gap and sharp minima}.
\newblock \bibinfo{journal}{CoRR} \href{http://arxiv.org/abs/1609.04836}{{\tt
  arXiv:1609.04836}}.
\bibitem[{Kingma and Ba(2017)}]{KB17}
\bibinfo{author}{Kingma, D.P.}, \bibinfo{author}{Ba, J.}, \bibinfo{year}{2017}.
\newblock \bibinfo{title}{Adam: A method for stochastic optimization}.
\newblock \href{http://arxiv.org/abs/1412.6980}{{\tt arXiv:1412.6980}}.
\bibitem[{Knuth et~al.(2002)Knuth, Chiaverini, Sauer and Gramer}]{Knuth2002}
\bibinfo{author}{Knuth, W.H.}, \bibinfo{author}{Chiaverini, M.J.},
  \bibinfo{author}{Sauer, J.A.}, \bibinfo{author}{Gramer, D.J.},
  \bibinfo{year}{2002}.
\newblock \bibinfo{title}{Solid-fuel regression rate behavior of vortex hybrid
  rocket engines}.
\newblock \bibinfo{journal}{J. Propul. Power} \bibinfo{volume}{18},
  \bibinfo{pages}{600--609}.
\newblock \DOIprefix\doi{10.2514/2.5974}.
\bibitem[{Korting et~al.(1987)Korting, Sch\"{o}yer and Timnat}]{Korting1987}
\bibinfo{author}{Korting, P.}, \bibinfo{author}{Sch\"{o}yer, H.},
  \bibinfo{author}{Timnat, Y.}, \bibinfo{year}{1987}.
\newblock \bibinfo{title}{Advanced hybrid rocket motor experiments}.
\newblock \bibinfo{journal}{Acta Astronaut.} \bibinfo{volume}{15},
  \bibinfo{pages}{97--104}.
\newblock \DOIprefix\doi{10.1016/0094-5765(87)90009-9}.
\bibitem[{Kumar and Ramakrishna(2014)}]{Kumar2014}
\bibinfo{author}{Kumar, R.}, \bibinfo{author}{Ramakrishna, P.},
  \bibinfo{year}{2014}.
\newblock \bibinfo{title}{Measurement of regression rate in hybrid rocket using
  combustion chamber pressure}.
\newblock \bibinfo{journal}{Acta Astronaut.} \bibinfo{volume}{103},
  \bibinfo{pages}{226--234}.
\newblock \DOIprefix\doi{10.1016/j.actaastro.2014.06.044}.
\bibitem[{Li et~al.(2020)Li, Shao and Hong}]{LSH21}
\bibinfo{author}{Li, J.}, \bibinfo{author}{Shao, S.}, \bibinfo{author}{Hong,
  J.}, \bibinfo{year}{2020}.
\newblock \bibinfo{title}{Machine learning shadowgraph for particle size and
  shape characterization}.
\newblock \bibinfo{journal}{Meas. Sci. Technol.} \bibinfo{volume}{32},
  \bibinfo{pages}{015406}.
\newblock \DOIprefix\doi{10.1088/1361-6501/abae90}.
\bibitem[{Li et~al.(2018)Li, Chen, Qi, Dou, Fu and Heng}]{LCDFH18}
\bibinfo{author}{Li, X.}, \bibinfo{author}{Chen, H.}, \bibinfo{author}{Qi, X.},
  \bibinfo{author}{Dou, Q.}, \bibinfo{author}{Fu, C.W.}, \bibinfo{author}{Heng,
  P.A.}, \bibinfo{year}{2018}.
\newblock \bibinfo{title}{H-{DenseUNet}: Hybrid densely connected {UNet} for
  liver and tumor segmentation from {CT} volumes}.
\newblock \bibinfo{journal}{{IEEE} Trans. on Med. Imaging}
  \bibinfo{volume}{37}, \bibinfo{pages}{2663--2674}.
\newblock \DOIprefix\doi{10.1109/tmi.2018.2845918}.
\bibitem[{Loshchilov and Hutter(2019)}]{LH17}
\bibinfo{author}{Loshchilov, I.}, \bibinfo{author}{Hutter, F.},
  \bibinfo{year}{2019}.
\newblock \bibinfo{title}{Decoupled weight decay regularization}
  \href{http://arxiv.org/abs/1711.05101}{{\tt arXiv:1711.05101}}.
\bibitem[{Ma et~al.(2020)Ma, Zhang, Haidn, Thuerey and Hu}]{MZHTH20}
\bibinfo{author}{Ma, H.}, \bibinfo{author}{Zhang, Y.}, \bibinfo{author}{Haidn,
  O.J.}, \bibinfo{author}{Thuerey, N.}, \bibinfo{author}{Hu, X.},
  \bibinfo{year}{2020}.
\newblock \bibinfo{title}{Supervised learning mixing characteristics of film
  cooling in a rocket combustor using convolutional neural networks}.
\newblock \bibinfo{journal}{Acta Astronaut.} \bibinfo{volume}{175},
  \bibinfo{pages}{11--18}.
\newblock \DOIprefix\doi{10.1016/j.actaastro.2020.05.021}.
\bibitem[{MATLAB(2021)}]{M}
\bibinfo{author}{MATLAB}, \bibinfo{year}{2021}.
\newblock \bibinfo{title}{version 9.10.0.1613233 (R2021a)}.
\newblock \bibinfo{organization}{The Mathworks, Inc.}.
  \bibinfo{address}{Natick, Massachusetts}.
\bibitem[{Minaee et~al.(2021)Minaee, Boykov, Porikli, Plaza, Kehtarnavaz and
  Terzopoulos}]{MBPPKT20}
\bibinfo{author}{Minaee, S.}, \bibinfo{author}{Boykov, Y.Y.},
  \bibinfo{author}{Porikli, F.}, \bibinfo{author}{Plaza, A.J.},
  \bibinfo{author}{Kehtarnavaz, N.}, \bibinfo{author}{Terzopoulos, D.},
  \bibinfo{year}{2021}.
\newblock \bibinfo{title}{Image segmentation using deep learning: A survey}.
\newblock \bibinfo{journal}{IEEE Trans. Pattern Analysis Mach. Intelligence}
  \DOIprefix\doi{10.1109/TPAMI.2021.3059968}.
\bibitem[{Otsu(1979)}]{Otsu1979}
\bibinfo{author}{Otsu, N.}, \bibinfo{year}{1979}.
\newblock \bibinfo{title}{A threshold selection method from gray-level
  histograms}.
\newblock \bibinfo{journal}{{IEEE} Trans. Sys., Man, Cybern.}
  \bibinfo{volume}{9}, \bibinfo{pages}{62--66}.
\newblock \DOIprefix\doi{10.1109/tsmc.1979.4310076}.
\bibitem[{Perrotta et~al.(2017)Perrotta, Parry and Neves}]{PPN17}
\bibinfo{author}{Perrotta, F.}, \bibinfo{author}{Parry, T.},
  \bibinfo{author}{Neves, L.C.}, \bibinfo{year}{2017}.
\newblock \bibinfo{title}{Application of machine learning for fuel consumption
  modelling of trucks}, in: \bibinfo{booktitle}{2017 {IEEE} Int. Conf. Big
  Data}, \bibinfo{publisher}{{IEEE}}. pp. \bibinfo{pages}{3810--3815}.
\newblock \DOIprefix\doi{10.1109/bigdata.2017.8258382}.
\bibitem[{Picard and Cook(1984)}]{MR763576}
\bibinfo{author}{Picard, R.R.}, \bibinfo{author}{Cook, R.D.},
  \bibinfo{year}{1984}.
\newblock \bibinfo{title}{Cross-validation of regression models}.
\newblock \bibinfo{journal}{J. Am. Stat. Assoc.} \bibinfo{volume}{79},
  \bibinfo{pages}{575--583}.
\newblock \DOIprefix\doi{10.1080/01621459.1984.10478083}.
\bibitem[{Qamar et~al.(2020)Qamar, Jin, Zheng, Ahmad and Usama}]{QJZAU20}
\bibinfo{author}{Qamar, S.}, \bibinfo{author}{Jin, H.}, \bibinfo{author}{Zheng,
  R.}, \bibinfo{author}{Ahmad, P.}, \bibinfo{author}{Usama, M.},
  \bibinfo{year}{2020}.
\newblock \bibinfo{title}{A variant form of 3d-{UNet} for infant brain
  segmentation}.
\newblock \bibinfo{journal}{Future Generation Comput. Systs.}
  \bibinfo{volume}{108}, \bibinfo{pages}{613--623}.
\newblock \DOIprefix\doi{10.1016/j.future.2019.11.021}.
\bibitem[{Ronneberger et~al.(2015)Ronneberger, Fischer and Brox}]{RFB15}
\bibinfo{author}{Ronneberger, O.}, \bibinfo{author}{Fischer, P.},
  \bibinfo{author}{Brox, T.}, \bibinfo{year}{2015}.
\newblock \bibinfo{title}{U-net: Convolutional networks for biomedical image
  segmentation}, in: \bibinfo{booktitle}{Lecture Notes in Comput. Sci.}.
  \bibinfo{publisher}{Springer International Publishing}, pp.
  \bibinfo{pages}{234--241}.
\newblock \DOIprefix\doi{10.1007/978-3-319-24574-4_28}.
\bibitem[{R\"{u}ttgers et~al.(2019)R\"{u}ttgers, Petrarolo and Kobald}]{RPK20}
\bibinfo{author}{R\"{u}ttgers, A.}, \bibinfo{author}{Petrarolo, A.},
  \bibinfo{author}{Kobald, M.}, \bibinfo{year}{2019}.
\newblock \bibinfo{title}{Clustering of paraffin-based hybrid rocket fuels
  combustion data}.
\newblock \bibinfo{journal}{Exp. Fluids} \bibinfo{volume}{61}.
\newblock \DOIprefix\doi{10.1007/s00348-019-2837-8}.
\bibitem[{Shin et~al.(2005)Shin, Lee, Chang and Koo}]{Shin2005}
\bibinfo{author}{Shin, K.H.}, \bibinfo{author}{Lee, C.},
  \bibinfo{author}{Chang, S.Y.}, \bibinfo{author}{Koo, J.Y.},
  \bibinfo{year}{2005}.
\newblock \bibinfo{title}{The enhancement of regression rate of hybrid rocket
  fuel by various methods}, in: \bibinfo{booktitle}{43rd {AIAA} Aerospace. Sci.
  Meet. Exhib.}, \bibinfo{publisher}{AIAA}.
\newblock \DOIprefix\doi{10.2514/6.2005-359}.
\bibitem[{Srivastava et~al.(2014)Srivastava, Hinton, Krizhevsky, Sutskever and
  Salakhutdinov}]{JMLR:v15:srivastava14a}
\bibinfo{author}{Srivastava, N.}, \bibinfo{author}{Hinton, G.},
  \bibinfo{author}{Krizhevsky, A.}, \bibinfo{author}{Sutskever, I.},
  \bibinfo{author}{Salakhutdinov, R.}, \bibinfo{year}{2014}.
\newblock \bibinfo{title}{Dropout: A simple way to prevent neural networks from
  overfitting}.
\newblock \bibinfo{journal}{J. Mach. Learning Res.} \bibinfo{volume}{15},
  \bibinfo{pages}{1929--1958}.
\newblock \URLprefix \url{http://jmlr.org/papers/v15/srivastava14a.html}.
\bibitem[{Wang et~al.(2020)Wang, Xie, Xuan and Jiao}]{WXXJ20}
\bibinfo{author}{Wang, B.}, \bibinfo{author}{Xie, B.}, \bibinfo{author}{Xuan,
  J.}, \bibinfo{author}{Jiao, K.}, \bibinfo{year}{2020}.
\newblock \bibinfo{title}{{AI}-based optimization of {PEM} fuel cell catalyst
  layers for maximum power density via data-driven surrogate modeling}.
\newblock \bibinfo{journal}{Energy Convers. Manag.} \bibinfo{volume}{205},
  \bibinfo{pages}{112460}.
\newblock \DOIprefix\doi{10.1016/j.enconman.2019.112460}.
\bibitem[{Weinstein and Gany(2013)}]{Weinstein2013}
\bibinfo{author}{Weinstein, A.}, \bibinfo{author}{Gany, A.},
  \bibinfo{year}{2013}.
\newblock \bibinfo{title}{Testing and modeling liquefying fuel combustion in
  hybrid propulsion}, in: \bibinfo{booktitle}{Prog. in Propuls. Phys.},
  \bibinfo{publisher}{{EDP} Sciences}. pp. \bibinfo{pages}{99--112}.
\newblock \DOIprefix\doi{10.1051/eucass/201304099}.
\bibitem[{Xiao et~al.(2018)Xiao, Lian, Luo and Li}]{XLLL18}
\bibinfo{author}{Xiao, X.}, \bibinfo{author}{Lian, S.}, \bibinfo{author}{Luo,
  Z.}, \bibinfo{author}{Li, S.}, \bibinfo{year}{2018}.
\newblock \bibinfo{title}{Weighted res-{UNet} for high-quality retina vessel
  segmentation}, in: \bibinfo{booktitle}{2018 9th Int. Conf. Inf. Technol. Med.
  Education ({ITME})}, \bibinfo{publisher}{{IEEE}}. pp.
  \bibinfo{pages}{327--331}.
\newblock \DOIprefix\doi{10.1109/itme.2018.00080}.
\bibitem[{Yaqub et~al.(2020)Yaqub, Feng, Zia, Arshid, Jia, Rehman and
  Mehmood}]{brainsci}
\bibinfo{author}{Yaqub, M.}, \bibinfo{author}{Feng, J.}, \bibinfo{author}{Zia,
  M.S.}, \bibinfo{author}{Arshid, K.}, \bibinfo{author}{Jia, K.},
  \bibinfo{author}{Rehman, Z.U.}, \bibinfo{author}{Mehmood, A.},
  \bibinfo{year}{2020}.
\newblock \bibinfo{title}{State-of-the-art cnn optimizer for brain tumor
  segmentation in magnetic resonance images}.
\newblock \bibinfo{journal}{Brain Sci.} \bibinfo{volume}{10}.
\newblock \DOIprefix\doi{10.3390/brainsci10070427}.
\bibitem[{Zeng et~al.(2019)Zeng, Xie, Zhang and Lu}]{ZXZL19}
\bibinfo{author}{Zeng, Z.}, \bibinfo{author}{Xie, W.}, \bibinfo{author}{Zhang,
  Y.}, \bibinfo{author}{Lu, Y.}, \bibinfo{year}{2019}.
\newblock \bibinfo{title}{{RIC}-unet: An improved neural network based on unet
  for nuclei segmentation in histology images}.
\newblock \bibinfo{journal}{{IEEE} Access} \bibinfo{volume}{7},
  \bibinfo{pages}{21420--21428}.
\newblock \DOIprefix\doi{10.1109/access.2019.2896920}.
\bibitem[{Zhou et~al.(2021)Zhou, Huang, Dong, Xia and Wang}]{ZHDXW19}
\bibinfo{author}{Zhou, Y.}, \bibinfo{author}{Huang, W.}, \bibinfo{author}{Dong,
  P.}, \bibinfo{author}{Xia, Y.}, \bibinfo{author}{Wang, S.},
  \bibinfo{year}{2021}.
\newblock \bibinfo{title}{D-{UNet}: A dimension-fusion u shape network for
  chronic stroke lesion segmentation}.
\newblock \bibinfo{journal}{{IEEE}/{ACM} Trans. on Computational Biol. and
  Bioinformatics} \bibinfo{volume}{18}, \bibinfo{pages}{940--950}.
\newblock \DOIprefix\doi{10.1109/tcbb.2019.2939522}.
\bibitem[{Zilliac and Karabeyoglu(2006)}]{Zilliac2006}
\bibinfo{author}{Zilliac, G.}, \bibinfo{author}{Karabeyoglu, M.},
  \bibinfo{year}{2006}.
\newblock \bibinfo{title}{Hybrid rocket fuel regression rate data and
  modeling}, in: \bibinfo{booktitle}{42nd {AIAA}/{ASME}/{SAE}/{ASEE} Jt.
  Propuls. Conf. Exhib.}, \bibinfo{publisher}{AIAA}.
\newblock \DOIprefix\doi{10.2514/6.2006-4504}.

\end{thebibliography}

\end{document}